\documentclass[ reprint,twocolumn, prb,]{revtex4-1}
\usepackage{xr} 
\usepackage{graphicx}
\usepackage{dcolumn}
\usepackage{float}
\usepackage{bm}
\usepackage{hyperref}
\usepackage{amsfonts}
\usepackage{siunitx}
\usepackage{amsmath,amssymb}
\usepackage{afterpage}
\usepackage{feynmp}
\usepackage{amsmath,amsthm}
\usepackage{amssymb}
\usepackage{graphicx,color}
\usepackage{subfigure}
\usepackage{epstopdf}
 \usepackage{relsize}
\usepackage[toc,page]{appendix}
\usepackage[dvipsnames]{xcolor}
\usepackage{color}
\usepackage{natbib}
\usepackage{xcolor}
\usepackage[strict]{changepage}
\usepackage{placeins}
\usepackage{tikz}
\usepackage[page]{appendix}

\newcommand\cminus{\mathbin{\raisebox{-\height}{$-$}}}
\newcommand\ch{\cosh}
\newcommand\sh{\sinh}
\newcommand\eps{\varepsilon}
\newcommand\contdots{\raisebox{-\height}{$\vphantom{+}\dotsm$}}

\newcommand\Qn{Q_{n-\frac12}(\ch \mu)}

\newcommand{\bl}[1]{\textcolor{black}{#1}}

\definecolor{dgr}{rgb}{0.0, 0.5, 0.0}

\begin{document}
%
\noindent\fbox{
    \parbox{\textwidth}{
        This manuscript has been authored by UT-Battelle, LLC under Contract No. DE-AC05-00OR22725 
with the U.S. Department of Energy. The United States Government retains and the publisher, 
by accepting the article for publication, acknowledges that the United States Government retains 
a non-exclusive, paid-up, irrevocable, world-wide license to publish or reproduce the published 
form of this manuscript, or allow others to do so, for United States Government purposes. The 
Department of Energy will provide public access to these results of federally sponsored research
in accordance with the DOE Public Access Plan (http://energy.gov/downloads/doe-public-access-plan).
    }
}
\clearpage

\title{Poloidal and toroidal plasmons  and fields of multilayer nanorings}
\author{K.V. Garapati$^{1}$,  M. Salhi$^{2}$, S. Kouchekian$^{1}$, G. Siopsis$^{2}$, A. Passian$^{2,3}$}\email{passianan@ornl.gov}
\affiliation{$^1$ Department of Mathematics, University of South Florida, 33613, USA\\
$^2$ Department of Physics, University of Tennessee, Knoxville, Tennessee 37996-1200, USA\\
$^3$ Oak Ridge National Laboratory, Oak Ridge, Tennessee 37831-6123, USA}
\date{\today}

\begin{abstract}
Composite and janus type metallo-dielectric nanoparticles are increasingly considered as a means to control the spatial and temporal behavior of electromagnetic fields in diverse applications such as coupling to quantum emitters, achieve invisibility cloaks, and obtain quantum correlations between qubits.
We investigate the surface modes of a toroidal nano-structure and obtain the canonical plasmon dispersion relations and resonance modes for arbitrarily layered nanorings. 
Unlike particle plasmon eigenmodes in other geometries, the amplitudes of the eigenmodes of tori exhibit a distinct forward and backward coupling. 
\bl{We present the plasmon dispersion relations for  several relevant toroidal configurations in the quasistatic limit and obtain the dominant retarded dispersion relations of a single ring for comparison,} discuss mode complementarity and hybridization,  and introduce two new types of toroidal particles in the form of janus nanorings. 
The resonance  frequencies  for the first few dominant modes of a ring composed of plasmon supporting materials such as gold, silver, and aluminum 
are provided and compared to those for a silicon ring. A generalized Green's function is obtained for multilayer tori allowing for calculation of the scattering response to interacting fields. Employing the Green's function, the scalar electric potential distribution corresponding to  individual poloidal and toroidal modes 
 in response to  an arbitrarily polarized external field and the field of \bl{electrons is obtained. The results are applied to obtain the local density of states and decay rate of a dipole near the center of the torus.}
\end{abstract}
\maketitle
\section{Introduction}
Collective electronic effects in solid tori and toroidal shells by virtue of their topology exhibit unique electromagnetic response. 
Consequently, electronic excitation in nanorings has been of increasing importance in nano-optics~\cite{aizp:rev}, metamaterial~\cite{mari},  trapping of cold molecules~\cite{salhi} and are of great potential for achieving enhanced spectroscopies~\cite{mich, xu, nikoob, hayne, grand}, where  giant enhancement in the near field of the surface modes of similar nano-structures associated with the collective electronic excitation is being utilized. 
Ordered or structured embedded nanoparticles in metamaterial as in photonic bandgap material, and material with negative index of refraction rely strongly on the shape of the structures~\cite{vesel, pendry1}. Increasingly, the local curvature and geometry of the nanoparticles  are being capitalized  upon in tip and surface-enhanced spectroscopies~\cite{dier}, in spectral tuning of optical response~\cite{halas}, in optical antenna~\cite{bharad} and chemical and biological sensing~\cite{lee}.      
Moreover, during the last decade, the implications of the emerging field of transformation optics~\cite{leon,pendry} have driven the exploration of novel anisotropic materials and their realizations in spherical, cylindrical, and toroidal geometries within the optical regime. In these studies, to achieve the required anisotropy, mainly the homogenization principle of laminates~\cite{milton} has been resorted to, and the associated proof of principle of the modeled concepts has been attempted in the quasistatic regime~\cite{kadic,schit}.  In particular, in the case of toroidal structures, significant difficulties arise when employing such approaches due to the vectorial nature of the resulting three-dimensional electromagnetic problem.  Specifically, the degree of the underlying challenges can be appreciated in the approximation employed by Kadic et al in their work on 
plasmonic analogs of electromagnetic wormholes~\cite{kadica}, where the availability of effective properties of metallo-dielectric coatings would have been beneficial.

Employing advanced fabrication, both metallic and nonmetallic nanorings are being considered.
Metallic nanorings may be fabricated by various lithographic techniques such as vacuum evaporation and self-assembly. Aizpurua \emph{et al.} fabricated  gold nanorings on glass substrate using colloidal lithography and investigated their light scattering properties and found that the excitation wavelengths of metallic nanoparticles depend strongly on the particle geometry allowing one to choose and tailor the shape of a particle to achieve excitation spectra on demand~\cite{aizp:quant}.
Similarly, while Kong \emph{et al.} demonstrated the fabrication and application of zinc oxide nanorings in investigating fundamental physical phenomena, such as the Aharonov-Bohm oscillations in exciton luminescence besides possible use as nanoscale sensors, transducers, and resonators~\cite{kong}, Kuyucak \emph{et al.} used the toroidal domain as a realistic  model important for the conductance of ions in biological channels~\cite{kuk}. 
Many forms of toroidal structures are emerging, such as toroidal micelles of polystyrene-block-Poly(acrylic acid)~\cite{c:liu}, and toroidal carbon nanotubes~\cite{l:liu}. Similarly, various fabrication methods are being reported including  a strategy for generating monodisperse toroidal particles through solidification of droplets of polymer solution in a microchannel~\cite{b:wang}, and very recently, vortex ring-derived particles were used toward mass production of larger rings~\cite{an}. Other more elaborate configurations continue to be envisioned and studied as reported by Chen et al. in their theoretical work on the mechanical properties of connected carbon nanorings~\cite{n:chen}. \bl{Also in recent development~\cite{yi}, Janus particles have been employed as a new tool for imaging and sensing in biological systems.}

The expanding use of toroidal structures and fields is diverse. For example, virtually free from dissipation, all-dielectric metamaterials capable of supporting toroidal dipolar excitations  has also been proposed~\cite{basharin}. Similarly, the excitation of toroidal dipolar moment in a novel metal-dielectric-metal  nanostructure under the radially polarized light was investigated~\cite{bao}. A toroid-like nanostructure metamaterial comprising asymmetric double-bar magnetic resonators was designed to demonstrate the toroidal dipolar response in the optical regime~\cite{dong1}. Also the toroidal dipole response was numerically investigated on a torus-like structure obtained by rotating the planar double-ring structure into a multifold double-ring torus-like metamaterial~\cite{dong2}. Similar investigations have been performed in metamaterials with their unit cell including three magnetic resonators, metal-dielectric-metal, consisting of two Ag rods and a SiO$_2$ spacer~\cite{tang}. Experiments and numerical simulations were conducted to investigate the feasibility of tuning the optical response of a dimer type nanoantenna using plasmonic nanorings~\cite{panaretos}. Treating the charged ionic channels as deformable torus with a circular or elliptical cross section, analytical solution of the equilibrium ion distribution for a toroidal model of a ionic channel was discussed~\cite{enriquez}. Other examples include obtaining an exact solution for the electrostatic potential of a family of conducting charged toroids with no hole in the toroid~\cite{lekner}; obtaining the surface states of a toroidal topological insulator supporting both bound-states and charged zero-modes~\cite{fonseca}; and analysis of a uniformly magnetized torus and its derivative self-intersecting shapes~\cite{beleggia}.

Interestingly, while a plethora of applications are emerging, early treatment of the toroidal problem has been reported in the context of plasma modes and stability. Specifically, Tokamak confinement of fusion plasma typically entails the generation and magnetic containment of a ring of plasma~\cite{hinton}. Toroidal plasma stability and control have been a major focus in the field of hot fusion research~\cite{mukhov}. These early works that appear to have been unnoticed by recent investigators, most likely due to 
a length-scale difference of at least $10^6$, provide much of the necessary mathematical developments. 
However, to date in the context of collective electronic effects at the toroidal surfaces, that is, surface plasmon excitation or particle plasmons, an accurate and comprehensive account on the toroidal response is not available. Furthermore, there are discrepancies in what has been reported, obscuring the application of the reported results to materials of current interest in nanoscience. 
The interior and exterior fields of an isolated dielectric ring subject to a uniform axial electrostatic field was first obtained by Love~\cite{love:jmp}. In a follow on article,  Love also obtained the modes of a cold plasma~\cite{love:plasma}. Very recently, Di Biasio \emph{et al.} calculated the polarizibility of various dielectric materials including a ring-shaped dielectric~\cite{biasio}. 
The  quasi-static modes of a cold plasma obtained by Love were claimed by Avramov \emph{et al.} to account for only half of the existing modes when taking into consideration the parity of the modes under coordinate inversion~\cite{avramov}. A more recent attempt to obtain the plasmon dispersion relations for a gold  nanoring was made by Mary \emph{et al.} However, the reported solutions  were obtained by making an assumption that does not conform to a diagonalization or perturbation scheme to solve the resulting transcendental dispersion equation~\cite{mary}. 
Interestingly, employing a theorem established by Lord Kelvin, F. H. Safford, obtained the solution to the toroidal problem by using a geometric inversion and transformation from conical into toroidal surfaces~\cite{saff}. Similarly, an alternative separation scheme, where the coordinate dependence of the boundary conditions is altered, was discussed by M. Andrews~\cite{andrew}. While Love considered the Greens function to obtain the potential distribution of a torus in a static field, Scharstein and Wilson employed the static thin-wire kernel approximation in an integral equation to calculate the excitation of a conducting ring~\cite{schars}. 
Unlike the Laplace equation, the Helmholtz equation is not separable in the toroidal systems. Furthermore, the case of a multiply coated toroids have not previously been treated.  
While not straightforward, numerical techniques, in principle, can be employed to obtain the dispersion relations in the fully retarded case, both analytical and semi-analytical approaches in the quasi-electrostatic limit have proved to be useful. Tsuchimoto \emph{et al.} employed the boundary element method to solve an eigenvalue problem derived from the Helmholtz equation to obtain the dispersion relations of a cold plasma for both circular and oval cross sections~\cite{tsuch}. Using the Hertz vector to obtain a scalar Helmholtz equation, Janaki and Dasgupta attempted a partial separation for the azimuthal solutions followed by expansion to obtain approximate eigenmodes of a toroidal cavity~\cite{janaki}. \\

\bl{Motivated by the above introduction, we proceed to present the organization of this article.}
In section II, we begin by formulating the general case of a $k$-layered toroidal structure ($k=1,2, \cdots$) and obtain the corresponding \bl{plasmon} dispersion relations. These results, being reported for the first time, warrant an in-depth investigation~\cite{pap1}  due to a three-term difference equation that arises and its relation to continued fractions and infinite determinants. We will simply utilize the results of this investigation without giving a full account, which is out of the scope of this article.     
We will then specialize, in subsections II.A-II.B, the obtained results for a series of cases of current importance beginning with a \bl{$k$-layered toroidal structure dispersion relations}, an isolated solid torus and its complementary toroidal void in II.B.1.
Here, we will  clarify the discrepancies noted above among the results reported by Love~\cite{love:plasma}, and Avaramov \emph{et al.}~\cite{avramov} 
Significant simplification can be achieved by use of perturbation theory to derive the dispersion relations allowing to further calculate the effects of a planar substrate. Thus, using this approach, we investigate  the more realistic case of a nanoring residing on a dielectric substrate, which is generally known to cause a red-shift in the plasmon  energies.
Furthermore, we will compute the dispersion relations for the specific materials  gold, aluminum, silver, and silicon typically used in experiments.  Proceeding with the treatment of the specialized cases, we will then obtain, in II.B.2,  the plasmon dispersion relations for a bimetallic (\emph{i.e.}, singly coated) torus, an isolated solid shell, and a dielectric core-metal shell. Again, here as in the previous subsection, we will consider application of perturbation theory and use of specific materials. 
In subsection II.B.3, we will study  two multilayer cases -- a three layered ring in vacuum, and a four layered system.  As a final example, we introduce two types of  janus nanorings in II.B.4 and discuss how one may formulate the resulting problem.       
Section III is dedicated to the calculation and visualization of the various contributing modes to the response of the nanoring to external fields, where both uniform and nonuniform external fields are treated. Here, employing the Greens function approach, the three-term difference equation is solved for the two source terms corresponding to the external fields. We then extend  the Greens function approach for the general multilayered tori.  The results, which in principle can be used to construct the dyadic Green's function and consequently also the local density of states (LDOS), are verified computationally using the finite element (FEM) and finite difference time domain (FDTD) methods. Here, to illuminate the obtained modes, we present, for the dominating modes, the computationally determined resonances of the single solid ring, \bl{and discuss two cases of janus nanorings.}  
The scattering and absorption cross sections as well as the vibrational eigenmodes that may be excited following the photon scattering and plasmon decay for the same cases are also computed and discussed. Concluding remarks are provided in section IV.   
\section{Plasmon Dispersion Relations}
\begin{figure}[htp]
    \centering
    \includegraphics[width=3.4in]{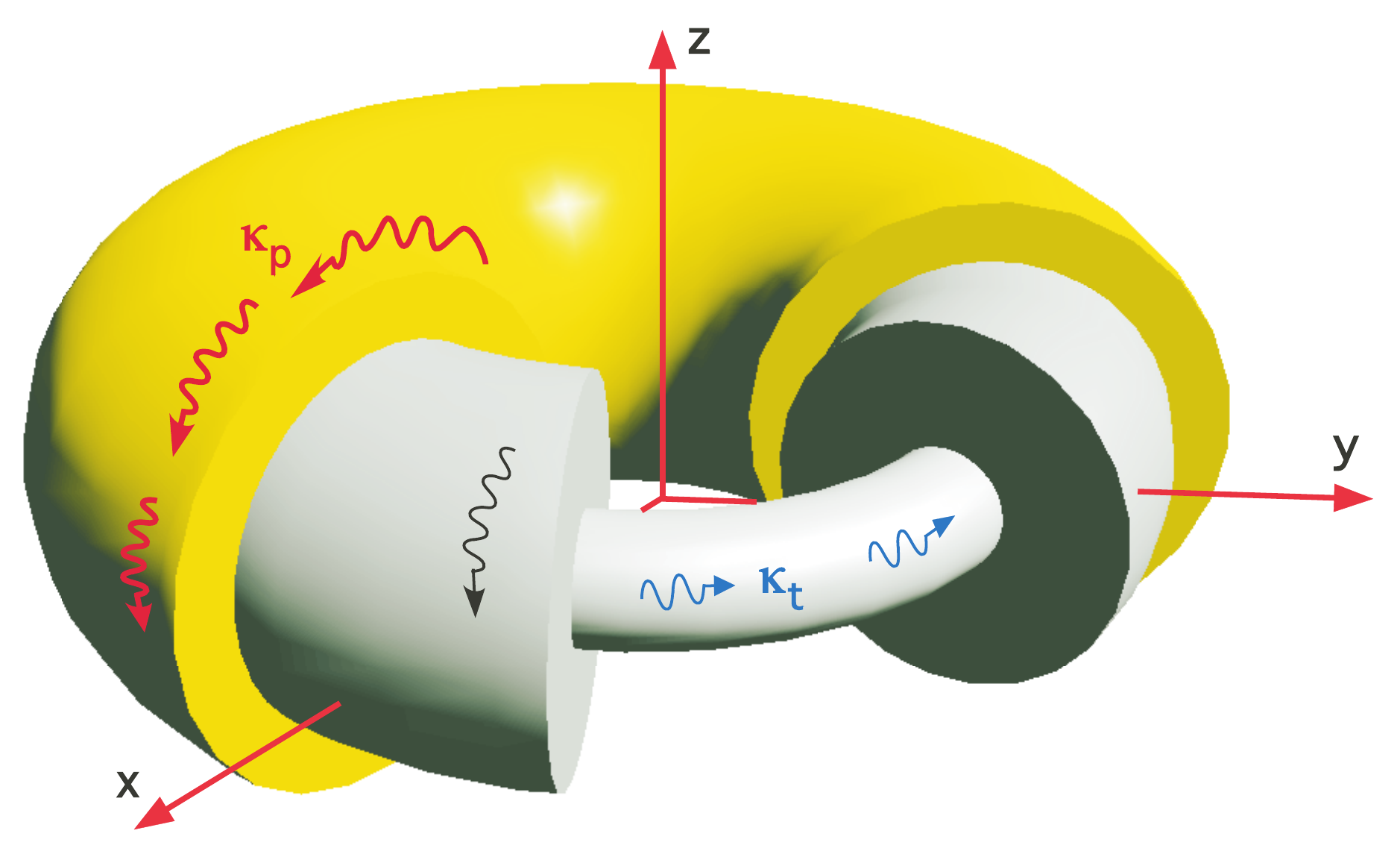}
    \caption[]{\footnotesize Localized plasmons in a multilayered nanoring particle. A three-layered toroidal structure is shown as an example. Schematic depiction of the poloidal and toroidal plasmons of momenta $\kappa_p$ and $\kappa_t$, respectively, excited at the ring interfaces.}
    \label{3torus}
 \end{figure}

The multilayer ring depicted in Fig.~\ref{3torus} possess finite surfaces and therefore the spectrum of the normal modes representing the oscillations of any induced polarization charge will necessarily be discrete.
A domain corresponding to a multilayer metal-dielectric nanoring may  be modeled using the toroidal coordinate system $(\mu,\eta,\varphi)$.
Here, $\mu\in[0,\infty)$, $\eta\in[0,2\pi]$, and $\varphi\in [0,2\pi]$, and the surface of a torus is defined by $\mu=\mu_1$, wherein the circle associated with the toroidal cross section grows from zero radius to infinite radius (with center at infinity) as $\mu$ decreases progressively from $\infty$ to $0$. Constant $\mu_1$ follows a circle of the minor radius $r=a/ \sinh\mu_1$ centered at the major radius $R=a\coth\mu_1$ as the coordinate $\eta$ changes from $0$ to $2\pi$ meaning that for a given $\mu=\mu_1$, $\eta$ traces a circle of fixed radius $r$ as it changes from $0$ to $2\pi$. The coordinate $\varphi$ is the azimuthal angle about the $z$ symmetry axis. As $\varphi$ changes from $0$ to $2\pi$, the circle is revolved around the $z$ axis by $360^{\circ}$ counterclockwise, thus generating a toroidal shape. 

Prior to formulating the normal modes corresponding to charge density oscillations in the multi-domain depicted in Fig.~\ref{3torus}, we note that  the charge density $\rho = - \epsilon_0 \Delta \Phi$ must satisfy
\begin{align}\label{charge_zero}
0 =  \int_{\varphi=0}^{2\pi} \int_{\eta=0}^{2\pi}  \int_{\mu=0}^\infty  
					 h_\mu h_\eta  h_\varphi  \rho (\mu,\eta,\varphi) d\mu d\eta d\varphi,
\end{align}
where $h_\mu$, $h_\eta$, $h_\varphi$ are scale factors given by Eq.~\eqref{scale_fac}, and $\Phi$ \bl{denotes} the electrostatic potential. 
\bl{For a single ring:}
\begin{align}\label{charge_density}
\rho(\mu,\eta,\varphi)   = \dfrac{\epsilon_0\delta(\mu-\mu_1)}{h^2_{\mu}}
									\sum_{m,n}T_n(\eta)V_m(\varphi)W_{mn}(\mu_1),
\end{align}
where $T$ and $V$ are separated solutions and $W_{mn}(\mu_1)$ is the Wronskian of the interior and exterior solutions, in conjunction with Eq.~\eqref{charge_zero}. 
\bl{In many cases the equation of motion for $\rho$, or rather the surface charge density $\sigma$, since here $\rho (\mu, \eta, \varphi) = \delta(\mu - \mu_0)\sigma (\eta,\varphi)/h_{\mu_0}$, can be obtained and solved to calculate the eigenfrequencies~\cite{board}.} 
\bl{While Eq.~\ref{charge_zero} is shown here to generate} a condition for the surface modes (as described in SM~\ref{E}) that is consistent with the structure of the dispersion relations, \bl{ the eigenmode coupling in this case prevents an analytical expression for the eigenfrequencies.}
\bl{Similarly, viewing the plasmons as incompressible irrotational deformations of the conduction electron gas of density $n$, one may study the dynamics of  small deformations of the electron fluid from the following Lagrangian~ \cite{prodan,board}:
$$ L = \frac{nm}{2}\int \eta \dot{\sigma}ds -\frac{1}{2}\int \frac{\sigma(\bf{r})\sigma(\bf{r'})}{|\bf{r}-\bf{r'}|}ds ds',$$ 
where $m$ is the electron mass and the velocity potential satisfies $\Delta \eta =0$. However, as discussed in the next section, the amplitude coupling of the eigenmodes also prevents this formulation to generate analytical expressions for plasmon frequencies of the ring system.
}

\subsection{General derivation for a $k$-layered ring}
\begin{figure}[htp]
    \centering
    \includegraphics[width=3.40in]{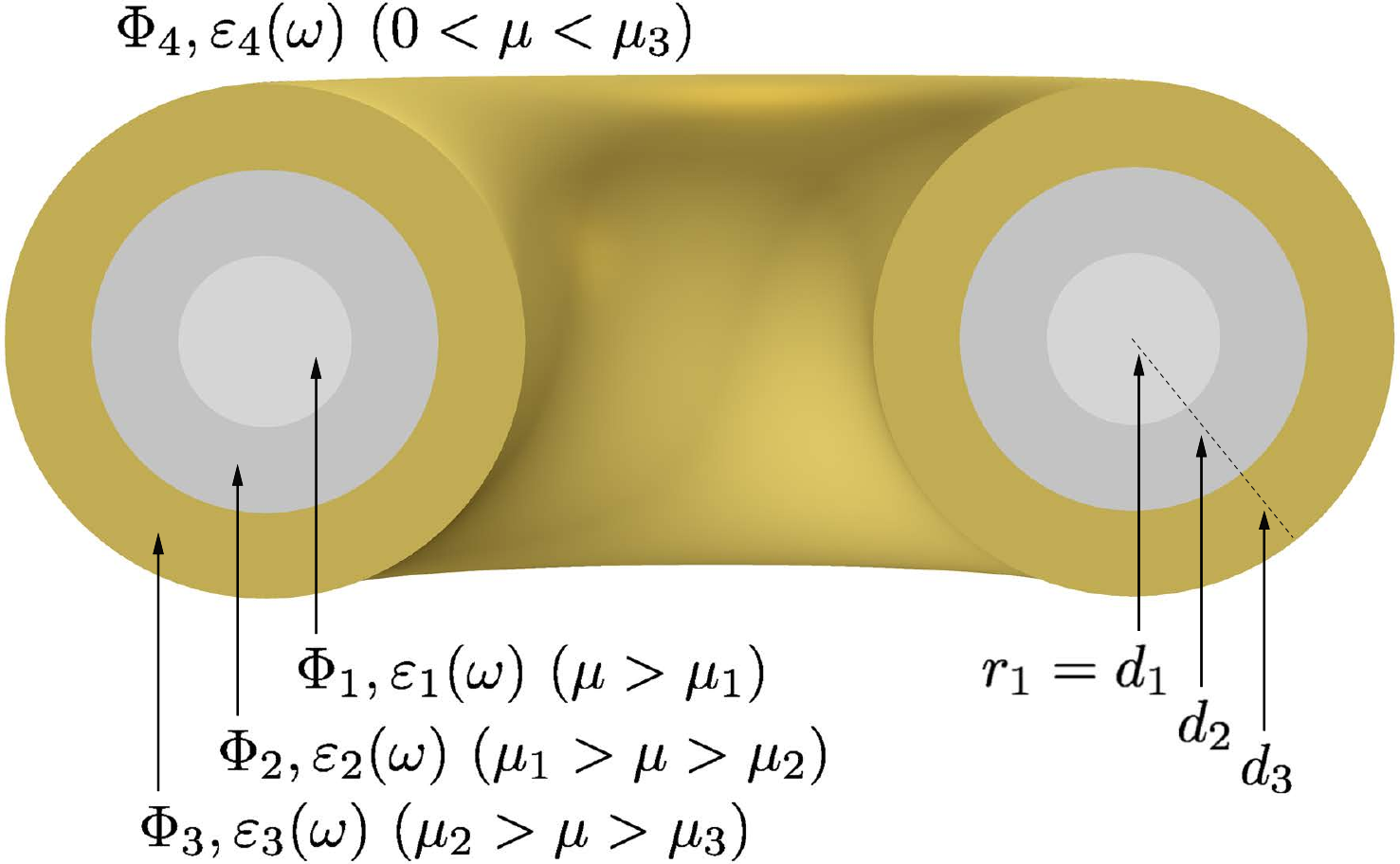}
    \caption[]{\footnotesize Quasi-static scalar electric potential distribution and cross section of a three-layered toroidal structure. Material domains are characterized by their frequency dependent local dielectric function $\epsilon_i$ with boundaries defined by $\mu_i$, with $i=1,2,\cdots, k$. Annotation of the potential $\Phi_i$ in layer with thickness $d_i$ set by the cross sectional radius $r_i$ and boundary $\mu_i$.
    }
    \label{3conc}
 \end{figure}

Nonretarded plasmon dispersion relations describing the resonant surface modes of a material domain may be obtained from a transcendental equation that is generated when imposing the quasi-static boundary conditions at the bounding surfaces of the domain within which the scalar electric field satisfies the Laplace equation~\cite{board}.   
As shown in Fig.~\ref{3torus} and Fig.~\ref{3conc}, a $k$ layered torus can be described as a solid torus with toroidal surface $\mu = \mu_1,$ frequency dependent dielectric function $\eps_1$ and minor radius $r_1=d_1,$  together with $k-1$ sublayers of toroidal shells (representing various material domains), each with corresponding thickness $d_i$ and dielectric function  $\eps_i$, where 
$i=2,\dots, k$.
\bl{The first region corresponds to interior of the solid torus and is given by $\mu \geq \mu_1.$ The remaining $k-1$ concentric toroidal shells share the same major-axis $R$ with the solid torus. Therefore, one can compute the boundary surfaces between  the layers via the relation
	\begin{equation}\label{mui}
		 R=r_1 \cosh \mu_1 = r_i \cosh \mu_i,
	\end{equation}
where $r_1 = d_1= R/\cosh \mu_1,$ $r_i= d_1+ \dots + d_i,$ and $i=2, \dots, k$ (see Fig.~\ref{3conc}). Since 
$r_1 < r_2 < \dots < r_k,$ the relation \eqref{mui} gives $\mu_1 > \mu_2 > \dots > \mu_k,$ where
	\begin{equation}\label{muii}
	\mu_i = \cosh^{-1} \left( \frac{d_1}{d_1+\dots + d_i} \cosh \mu_1 \right),  \quad i=2, \dots, k.
	\end{equation}
We can thus describe the regions enumerated from 2 to $k$ by $ r_{i-1} \leq r \leq r_i$ ($i=2, \dots, k$), respectively. Relation \eqref{mui} transforms the inequalities given in $r$ into the corresponding reversed inequalities in $\mu$ given by $\mu_i \leq \mu \leq \mu_{i-1},$ where  each $\mu_i$  satisfies \eqref{muii}. Finally, the $(k+1)^{th}$ region, which  lies outside the $k$-layered torus, is described by $ 0 \leq \mu \leq \mu_k$ 
(see Fig.~\ref{3conc} for the case $k=3$).}
 The general form of the quasi electrostatic  potential satisfying $\Delta \Phi = 0$ is given by 
\begin{align}\label{eqn_ref}
\Phi(\mu,\eta,\varphi, t) & =    \Theta (\mu - \mu_1)\Phi_1(\mu,\eta,\varphi,t) \notag\\
								& +\sum_{j=1}^{k-1} \Theta (\mu_j-\mu)\Theta (\mu-\mu_{j+1})\Phi_{j+1}(\mu,\eta,\varphi,t) \notag\\
								& + \Theta (\mu_k - \mu) \Phi_{k+1}(\mu,\eta,\varphi,t),
\end{align}
and
\begin{align} \label{k_layer_field1}
  \Phi_j&(\mu,\eta,\varphi,t)  \notag\\  
  			& =f(\mu,\eta) \sum_{m=-\infty}^\infty \sum_{n=-\infty}^\infty 
				 \bigl [  (1-\delta_{j,1})C_{mn}^{\text{j}}(t) P_{n-\frac12}^m(\cosh \mu)\notag\\
					&	+   (1-\delta_{j,k+1}) D_{mn}^{\text{j}}(t) Q_{n-\frac12}^m(\cosh \mu) \bigr ]
						e^{in\eta}  e^{im\varphi},
\end{align}
where $f(\mu,\eta)=\sqrt{\cosh\mu-\cos\eta},$  $j= 1, \dots, k+1$, and $\Theta$ is \bl{the} Heaviside function. \bl{Utilization of} the delta functions assures consistency with the toroidal harmonics $P_{n-\frac12}^m(\cosh \mu)$ and $Q_{n-\frac12}^m(\cosh \mu)$ being unbounded for $\mu \to \infty$ and $\mu \to 0$, respectively, and $C_{mn}^{\text{j}}(t)$ and $D_{mn}^{\text{j}}(t)$ are the non-retarded mode amplitudes at time $t$.
 Prior to invoking the boundary conditions, it is noteworthy to observe that the consequence of the nonlinearity in the  weight function $f$ for the interrelation  of the surface mode amplitudes 
may be regarded as a manifestation of the surface charges affecting each other across the toroidal opening. This conjecture may be supported 
by noting that the dominant modes of the $mn^{\text{th}}$ harmonics may be studied within the first-order approximation:
	$$
		f(\mu,\eta) = \sqrt{\cosh \mu}\ \left[1 - \tfrac12 \frac{\cos \eta}{\cosh \mu}  + 
									\mathcal{O} \left( \left(\tfrac{\cos \eta}{\cosh \mu} \right)^2 \right) \right],
	$$
in  \eqref{k_layer_field1} to obtain
	\begin{equation}
		\Phi_j \approx  
		\left( \sqrt{\cosh \mu} - \frac{\cos \eta}{2 \sqrt{\cosh \mu}}  \right)  
														 \sum_{n,m} L_{mn}^j e^{in\eta}  e^{im\varphi},\notag
	\end{equation}
	where  $L_{mn}^j$ stands for the expression given in the bracket of \eqref{k_layer_field1}. \bl{Setting $L_{mn}^j = -4 \sqrt{\cosh \mu}\, I_{mn}^j$, the above may be rewritten as:}
	\begin{equation}\label{tor_har_1}
		\Phi_j  \approx
		 \sum_{n,m} \Bigl( I_{m,n-1}^j - 4 \cosh \mu \, I_{m,n}^j + I_{m,n+1}^j  \Bigr)  
								e^{in\eta}  e^{im\varphi}. 
 	\end{equation}
This result is also consistent with the lack of distinct $mn$ modes (as described in SM~\ref{B0}), where complete modes are rather seen to be of the form $m\sum_n$ type. \bl{In fact, the electron gas occupies a non simply-connected region of space with genus one, where the potential varies across the central hole. The electron motion in a region of the torus may therefore couple to the electron motions at a different location, leading to the described mode coupling. Such mode coupling is absent for electron motion in simply-connected regions with  genus zero such  as spheres,  ellipsoids, cylinders, etc.      
   }
We now proceed and impose the ($2k$) boundary conditions:
\begin{align}
	\label{k_bc1}
		\left. \Phi_i\right|_{\mu=\mu_i} &= \left. \Phi_{i+1}\right|_{\mu=\mu_i},\\
	\label{k_bc2}
		\eps_i \left.  \frac{\partial \Phi_i}{\partial \ch \mu}\right|_{\mu=\mu_i}
			&= \eps_{i+1} \left.  \frac{\partial \Phi_{i+1}}{\partial \ch \mu}\right|_{\mu=\mu_i},
\end{align}
\bl{where $i=1,\dots, k.$} Since \bl{$\{e^{im\varphi}\}$ is a complete  orthogonal  system,} we can treat Eqs.~\eqref{k_bc1} and  \eqref{k_bc2} separately for each integer $m$. With $m$ maintained fixed, \bl{one} may therefore suppress its notation in upcoming equations. Moreover,  we adopt the simplified notations $P_n^i,$ $Q_n^i,$  $f_i,$   $C^i_n$, $D^i_n$ for $P_{n-\frac{1}{2}}^m(\cosh \mu_i)$, $Q_{n-\frac{1}{2}}^m(\cosh \mu_i)$,   $f(\mu_i,\eta)$, $C^i_{mn}(t)$, $D^i_{mn}(t)$ respectively, and let $P_n^{'i}$, $Q_n^{'i}$, $f'_i$ denote the derivatives of $P_n^i$, $Q_n^i$, $f_i$ with respect to $\cosh \mu$ evaluated at $\mu=\mu_i$. 
 Introducing the $k\times 1$ vectors 
\begin{equation}
\small
\mathcal{C}_n = 
	\begin{bmatrix}\label{C_k}
 	C_n^2\\[0.3em]
	C_n^3\\[0.3em]
 	\vdots\\[0.3em]
	C_n^{k+1}\\[0.3em]
	\end{bmatrix}
	\hspace{2mm}
	\text{ and }
	\hspace{2mm}
\mathcal{D}_n = 
	\begin{bmatrix}
	 D_n^1\\[0.3em]
 	D_n^2\\[0.3em]
 	\vdots\\[0.3em]
	D_n^k\\[0.3em]
	\end{bmatrix},
\end{equation}
and applying the boundary condition \eqref{k_bc1}, one obtains the relation
\begin{equation}\label{PnQn}
		\mathbb{P}_n\mathcal{C}_n=\mathbb{Q}_n\mathcal{D}_n,
	\end{equation}
where the $k\times k$  matrices $\mathbb{P}_n$ and $\mathbb{Q}_n$  are  defined  by  Eqs. \eqref{P_k} and \eqref{Q_k}.
Since both bidiagonal matrices $\mathbb{P}_n$ and $\mathbb{Q}_n$ have non-zero diagonal entries, they are invertible. Thus, one can solve, say,  
$\mathcal{C}_n$ in terms of $\mathcal{D}_n$ to obtain
\begin{equation}\label{CnDn}
	\mathcal{C}_n=\mathbb{P}_n^{-1}\mathbb{Q}_n\mathcal{D}_n.
\end{equation}
Using the substitution \eqref{CnDn} into the second boundary condition \eqref{k_bc2} and after some involved algebra~\cite{pap1}, one can show that 
$\mathcal{D}_n$ satisfies the vector three-term  recurrence relation:
\begin{equation}\label{threeterm}
		\mathcal{W}_{n+1} -\text{R}_n\mathcal{W}_{n}+\mathcal{W}_{n-1}=0,
	\end{equation}
for $n=0,\pm 1, \pm 2,\cdots$, where
\begin{equation}
  \begin{gathered}
   \mathcal{W}_n =\mathrm{J}_n\mathcal{D}_n, \\
   \mathrm{J}_n =
		(\mathbb{P}'_{n} \mathrm{E}_2 \mathbb{P}_{n}^{-1}
			-\mathbb{Q}'_{n}  \mathrm{E}_1  \mathbb{Q}_{n}^{-1} ) \mathbb{Q}_{n},\label{WnJn} \\
  \end{gathered}
\end{equation}
and 
\begin{equation}\label{Rnn}
		\mathrm{R}_n =\mathrm{D}_\mu + \mathbb{A}_n \mathbb{B}_n^{-1},
\end{equation}
where
\begin{equation}
  \begin{gathered}
 	\mathbb{A}_n= \mathbb{P}_n\mathrm{E}_2\mathbb{P}_n^{-1}
				- \mathbb{Q}_n\mathrm{E}_1\mathbb{Q}_n^{-1},\\
	\mathbb{B}_n= \mathbb{P}'_{n} \mathrm{E}_2 \mathbb{P}_{n}^{-1}-
						\mathbb{Q}'_{n} \mathrm{E}_1 \mathbb{Q}_{n}^{-1}.\label{AB}
  \end{gathered}
\end{equation}
For the definition of the $k\times k$ matrices $\mathbb{P}'_{n},$ $\mathbb{Q}'_{n} ,$ $\mathrm{D}_\mu,$ $\mathrm{E}_1,$ and
 $\mathrm{E}_2,$ see Eqs.~\eqref{P_k_1}--\eqref{E1_1}. 
 Note that Eqs.~\eqref{CnDn}--\eqref{AB} contain all \bl{that is needed} to calculate the quasi electrostatic potential $\Phi$ (Eq.~\eqref{eqn_ref}) and normal modes  for a $k$-layered ring in the absence of any applied field. 
\subsection{Surface modes of $k$-layered  ring configurations}
\subsubsection{\bl{Multilayered dispersion relations, an isolated solid torus, and a toroidal void}}
\paragraph{Plasmon dispersion relations for a $k$-layered ring\\\\}
\bl{
Noting~\cite{love:plasma} that the toroidal harmonics $P_{n-\frac12}^m(z)$ and  $Q_{n-\frac12}^m(z),$ and their derivatives with respect to $z$ are symmetric in $n$, it follows from Eqs.~\eqref{Rnn} and~\eqref{AB} that $\text{R}_{n}= \text{R}_{-n}$  for all $n=\pm 1, \pm 2, \cdots$.
Using this together with substitutions 	
	$$
	\mathcal{X}_n = \mathcal{W}_n+\mathcal{W}_{-n} \ 
	\text{ and }  \ 
	\mathcal{Y}_n = \mathcal{W}_n - \mathcal{W}_{-n},
	$$
one can transform the bi-infinite vector three-term recurrence~\eqref{threeterm}  into two  semi-infinite vector three-term recurrences
\begin{align}
	 \mathcal{X}_2  - (\text{R}_1- 2 \text{R}_0^{-1}) \mathcal{X}_1 &=0,  \label{BCTT1}\\
		\mathcal{X}_{n+1} - \text{R}_n  \mathcal{X}_{n}+   \mathcal{X}_{n-1}&=0, 
				\quad   n=2, 3,  \cdots , \label{TTx}
\end{align}
and
\begin{align}
	\mathcal{Y}_2 - \text{R}_1  \mathcal{Y}_1 &= 0, \label{BCTT2} \\
		\mathcal{Y}_{n+1} - \text{R}_n  \mathcal{Y}_{n}+ \mathcal{Y}_{n-1}&=0,
			 \quad n= 2, 3,   \cdots. \label{TTy}
\end{align}
It is easily seen that the above system displays a single   vector three-term recurrence relation together with two initial conditions given by 
Eqs.~\eqref{BCTT1} and~\eqref{BCTT2}. This observation leads us to consider the more general format of the matrix three-term recurrence:
\begin{equation}\label{TTX}
		X_{n+1} - \text{R}_n X_n + X_{n-1}=0, \qquad n=1,2, \cdots,
	\end{equation}
with $\mathcal{X}_n$ and  $\mathcal{Y}_n$ being treated as a single column of the $k\times k$ matrix $X_n.$  It follows that~\eqref{BCTT1} and~\eqref{BCTT2} give the two initial conditions
\begin{align}
\label{in_1}
\frac{X_2}{X_1} &= \text{R}_1- 2\, \dfrac{I}{\text{R}_0},\\
\label{in_2}
\frac{X_2}{X_1} &= \text{R}_1,
\end{align}
where $I$ is the $k \times k$ identity matrix and  the quotient of two $k\times k$ matrices $A$ and $B$ with $B$ non-singular is defined by 
	$
	A/B  :=  A B^{-1},
	$
a convention that will be used throughout the  the paper.
}

\bl{
Let us assume that the sequence of $k\times k$ nonsingular matrices $\{X_n\}$  is a solution to the matrix three-term recurrence\eqref{TTX}.
	 Multiplying Eq.~\eqref{TTX} from the right by $X_n^{-1}$ and defining $V_n$ by $V_n = \dfrac{X_{n+1}}{X_n},$ we arrive at the  first order non-linear matrix recurrence: 
	\begin{equation}\label{Vn}
		V_{n-1} = \frac{I}{\text{R}_n - V_{n}}, \qquad n=2, 3, \cdots.
	\end{equation}
Successive iteration of Eq.~\eqref{Vn} yields the finite matrix continued fraction:
	\begin{equation}\label{Vnn}
	\small
		V_1 =  \frac{I}{\text{R}_2} 
		\cminus \frac{I}{\text{R}_3}
			\cminus \contdots 
				\cminus \frac{I}{\text{R}_n - V_n}\cdot					
	\end{equation}
}
	
\bl{	
It turns out that the limit of the left-hand side of Eq.~\eqref{Vnn} exists as $n \to \infty;$ i.e., one can show~\cite{Perron_Mat,lev}  the  convergence of infinite matrix continued fraction (MCF):
	\begin{equation}\label{MCF1}
	\small
  \frac{I}{\text{R}_2} 
		\cminus \frac{I}{\text{R}_3}
				\cminus \frac{I}{\text{R}_4} \cminus \contdots.
	\end{equation}
The answer as to which unique value the MCF~\eqref{MCF1} converges is addressed in the following result:
}

\bl{
{\em It follows that the MCF~\eqref{MCF1} converges if and only if the matrix three-term recurrence \eqref{TTX} has a minimal solution. Moreover, the minimal solution $X_n$ satisfies}
	\begin{equation}\label{DRG}
	\small
  X_2 X_1^{-1} =\frac{I}{\text{R}_2} 
		\cminus \frac{I}{\text{R}_3}
				\cminus \frac{I}{\text{R}_4} \cminus \contdots.
	\end{equation}
The proof of the above fact is quite involved and is beyond the scope of this paper but is reported elsewhere~\cite{pap1}.
}

\bl{
 	In the scalar case of  \eqref{TTX}, a solution $X_n$ is called {\em minimal}~\cite{Gau} if 				
				$
				\lim_{n\to\infty} \frac{X_n}{Y_n} = 0
				$
holds for any other linearly linearly independent solution $Y_n.$ This notion can be generalized~\cite{Ah1,Ah2} to the matrix case of  Eq.~\eqref{TTX}. As the above result indicates, a matrix three-term recurrence may very well have no minimal solution. These type of non-minimal solutions are called the {\em dominant} solutions in accordance with the terminology for the scalar case.
}
\bl{	
	We turn our attention back to the Eq.~\eqref{Vnn}. Since the MCF~\eqref{MCF1} converges, the matrix three-term recurrence~\eqref{TTX} has the minimal solution, say, $X_n$ and one can let $n\to\infty$ in~\eqref{Vnn}. Assuming that $\mathcal{X}_n,$ defined by Eqns.~\eqref{BCTT1} and~\eqref{TTx}, constitute a single column of $X_n,$ it follows from~\eqref{DRG} and~\eqref{in_1} that 
	\begin{equation}\label{Disp1}
	\small
	 \frac12 \text{R}_0=  \frac{I}{\text{R}_1} 
		\cminus \frac{I}{\text{R}_2}
				\cminus \frac{I}{\text{R}_3} \cminus \contdots.
	\end{equation}	
A similar argument applied to $\mathcal{Y}_n,$ defined by Eqns.~\eqref{BCTT2} and~\eqref{TTy}, together with~\eqref{in_2} implies
	\begin{equation}\label{Disp2}
	\small
	  \text{R}_1=  \frac{I}{\text{R}_2} 
		\cminus \frac{I}{\text{R}_3}
				\cminus \frac{I}{\text{R}_4} \cminus \contdots.
	\end{equation}
Equations~\eqref{Disp1} and~\eqref{Disp2} are the sought dispersion relations for a $k$-layered ring configurations. 	
}
	
\bl{	
	We end this section by making some important remarks regarding the obtained dispersion relations. First of all, the existence of the minimal solution does not guarantee its derivation. In fact, almost all initial conditions generate a dominant solution to the matrix three-term 
	recurrence~\eqref{TTX}.  Simply put, to find the minimal solution of~\eqref{TTX} is equivalent to finding the exact analytic expression for the limit of MCF~\eqref{MCF1}, which is not possible. So the follow up question is how we could assume the minimality of either $\mathcal{X}_n$ or
	$\mathcal{Y}_n$ in obtaining the dispersion relations, where  the minimality of  $\mathcal{X}_n$ or
	$\mathcal{Y}_n$  is referred  to the case in which one of these two vectors constitutes a single column of the minimal solution, say, $X_n$ of~\eqref{TTX}. The answer to this question lies in the fact that the convergence of the MCF~\eqref{MCF1}  is independent of the particular choice of the dielectric values $\eps_1, \eps_2, \dots, \eps_{k+1}.$  These dielectric values are implicitly embedded as  diagonal matrices $\text{E}_1$ and $\text{E}_2$ in the definition of $\text{R}_n$ defined by the Eq.~\eqref{Rnn}.  Thus, by assuming the minimality of either 
	$\mathcal{X}_n$ or $\mathcal{Y}_n,$ one can seek those sets of dielectric values $\eps_1, \eps_2, \dots, \eps_{k+1}$ which satisfy~\eqref{Disp1} and~\eqref{Disp2}, respectively. In doing so, one finds the surface normal modes of a $k$-layered ring. Finally, recall that for two minimal solutions, say, $Y_n$ and $Z_n$ of~\eqref{TTX}, we have $\frac{Y_2}{Y_1}=\frac{Z_2}{Z_1}.$ By considering the initial conditions~\eqref{in_1} 
	and~\eqref{in_2}, it follows that both $\mathcal{X}_n,$ and $\mathcal{Y}_n$ can not be minimal at the same time; that is, no unique set of dielectric values $\eps_1, \eps_2, \dots, \eps_{k+1}$ may satisfy both dispersion relations. In conclusion, we have two independent sets of dispersion relations~\eqref{Disp1} and~\eqref{Disp2} corresponding to the assumptions of $\mathcal{X}_n,$ and $\mathcal{Y}_n$ being the minimal solution, respectively. \\
}
	\paragraph{Clarification of discrepancies in existing models\\\\}
As noted in the introduction, within the context of fusion and Tokamak physics, Love~\cite{love:plasma} and Avramov \emph{et al.}~\cite{avramov} explored the surface modes of a toroidal plasma. More recently, within the context of condensed matter and nearfield optics, Mary \emph{et al.}~\cite{mary}
attempted to obtain the surface plasmon dispersion relations for an isolated solid nanoring. In this section, by considering the $k=1$ case, that is, a solid torus without layers, we will delineate the derivation of the associated 
modes and address the differences among the reported results.  

\bl{
Thus, let us consider a solid ring with the dielectric function $\eps_1$ immersed in a medium  with a  dielectric function $\eps_2.$ Before we proceed, it should be emphasized that all the involved vectors and matrices  introduced in the derivation of the dispersion relations are now simply scalars. Invoking Eqs.~\eqref{CnDn}--\eqref{WnJn}, it  follows from  \eqref{k_bc1} that the inside and outside potentials  $\Phi_1$ and $\Phi_2$ are given by
 \begin{equation} \label{Phi1Wn}
  \Phi_1 =f(\mu,\eta)  \sum_{m,n=-\infty}^\infty 
				 \frac{1}{\text{J}_n} \mathcal{W}_n
				 Q_{n-\frac12}^m(u)  e^{in\eta}  e^{im\varphi},
\end{equation}
and
\begin{equation} \label{Phi2Wn}
  \Phi_2 =f(\mu,\eta)  \sum_{m,n=-\infty}^\infty 
				 \frac{\mathbb{Q}_n}{\mathbb{P}_n \text{J}_n} \mathcal{W}_n
				 P_{n-\frac12}^m(u)  e^{in\eta}  e^{im\varphi},
\end{equation}
respectively, where $u=\cosh\mu$ and  the summation index $m$ has been suppressed in $\mathbb{P}_n,$ 
$\mathbb{Q}_n,$ $\text{J}_n,$ and $ \mathcal{W}_n.$
}

\bl{
Love~\cite{love:plasma} derives the three-term recurrence \eqref{threeterm} for $\mathcal{W}_n,$ but only considers the transformation 
$ \mathcal{X}_n = \mathcal{W}_n + \mathcal{W}_{-n}.$ As a result, only  the dispersion relation \eqref{Disp1} is obtained. We now know that 
$\mathcal{W}_n= \frac12 \mathcal{X}_n + \frac12 \mathcal{Y}_n,$ which clearly indicates that one should also consider  $\mathcal{Y}_n$ as  a minimal solution. This has led us to the second dispersion relation \eqref{Disp2} discussed earlier. Love, considering the more involved transformation $V_n=\dfrac{\mathcal{X}_{n+1}}{\mathcal{X}_n}$ at once, might  have missed the second dispersion relation.
}

\bl{
On the other hand, Avramov \emph{et al.}~\cite{avramov}, aiming to study the high-frequency surface waves on a toroidal isotropic plasma, correctly obtains both dispersion relations \eqref{Disp1} and \eqref{Disp2} for a single toroidal plasma. 
}
It is easily seen that Eqs.~\eqref{BCTT1}--\eqref{TTx} and~\eqref{BCTT2}--\eqref{TTy} can be equivalently expressed as two infinite homogeneous systems of linear equations:
\begin{align}\label{Matrix_1}
\begin{bmatrix}
\dfrac{ \text{R}_0}{2} & -1 & \\[0.3em]
-1 & \text{R}_1 & -1 & \\[0.3em]
&-1&\text{R}_2&-1&\\[0.3em]
 &&\ddots&\ddots&\ddots& \\[0.3em]
\end{bmatrix}
\begin{bmatrix}
\mathcal{X}_0 \\[0.4em] \mathcal{X}_1  \\[0.4em] \mathcal{X}_2 \\[0.4em] \vdots
\end{bmatrix}
=
\begin{bmatrix}
0 \\[0.4em] 0 \\[0.4em] 0 \\[0.4em] \vdots
\end{bmatrix},
\end{align}
\begin{align}\label{Matrix_2}
\begin{bmatrix}
\text{R}_1 & -1 & \\[0.3em]
-1 & \text{R}_2 & -1 & \\[0.3em]
 & -1 & \text{R}_3 & -1 & \\[0.3em]
 &  & \ddots & \ddots & \ddots & \\[0.3em]
\end{bmatrix}
\begin{bmatrix}
\mathcal{Y}_1 \\[0.4em] \mathcal{Y}_2 \\[0.4em] \mathcal{Y}_3 \\[0.4em] \vdots
\end{bmatrix}
=
\begin{bmatrix}
0 \\[0.4em] 0 \\[0.4em] 0 \\[0.4em] \vdots 
\end{bmatrix}.
\end{align}
\bl{
Avramov \emph{et al.}~\cite{avramov} describes that the dispersion relations \eqref{Disp1} and \eqref{Disp2} are obtained  using the statement that the infinite systems \eqref{Matrix_1} and \eqref{Matrix_2} possess nontrivial solutions if  the determinants of the corresponding infinite matrices vanish. Contrary to the finite case, the mentioned statement is incorrect in the infinite case. In general, the determinant of an infinite matrix  may not exist. In fact, there are infinite matrices with non-zero or non-existing determinant for which the corresponding homogenous systems still possess non-trivial solutions. A simple example of such a case is provided  in SM~\ref{D}.  
Moreover, any formal calculation of the determinants of  \eqref{Matrix_1} and \eqref{Matrix_2} does not result in a continued fraction resembling  either  \eqref{Disp1} or \eqref{Disp2}. 
A similar incorrect interpretation of the determinant of an infinite matrix  and non-triviality of the solutions also occurs in Love~\cite{love:plasma}. 
}

\bl{
Our final remark  concerns with the parity of the dispersion modes~\cite{avramov}. Using the standard decomposition  of
$\mathcal{W}_n= \frac12 \mathcal{X}_n + \frac12 \mathcal{Y}_n$   into symmetric ($\mathcal{X}_n$) and antisymmetric
($\mathcal{Y}_n$) terms, together with the symmetry of  $\mathbb{P}_n,$ 
$\mathbb{Q}_n,$ and $\text{J}_n,$ one can split each of the sums indexed by $n$   in \eqref{Phi1Wn} and \eqref{Phi2Wn} into two separate terms with respect to $\sin n\eta$ and $\cos n\eta$ as:
$$\Phi_i= f(\mu,\eta) \sum_{m=-\infty}^\infty ( \Phi_{i,m}^c +  \Phi_{i,m}^s),  \quad i=1,2,$$
where
\begin{align}
 \Phi_{1,m}^c(\mu, \eta, \varphi) &=  f \sum_{n=0}^\infty \frac{1}{2\text{J}_n} \mathcal{X}_n
				 Q_{n-\frac12}^m(u)  \cos n\eta \ e^{im\varphi},\\
  \Phi_{1,m}^s(\mu, \eta, \varphi)  &= f  \sum_{n=1}^\infty  \frac{1}{2\text{J}_n} \mathcal{Y}_n
				 Q_{n-\frac12}^m(u)  \sin n\eta \ e^{im\varphi},
\end{align}
and $f= f(\mu, \eta).$ The expressions for $\Phi_{2,m}^c$ and $ \Phi_{2,m}^s$ can be found in a similar fashion with $Q_{n-\frac12}^m(u)$ and $ \frac{1}{\text{J}_n}$ replaced by 
$P_{n-\frac12}^m(u)$ and $\frac{\mathbb{Q}_n}{\mathbb{P}_n \text{J}_n},$ respectively. The reason behind this decomposition is to identify the even and odd parities of  $\Phi_{i,m}^c$ and $ \Phi_{i,m}^s$ with respect to the  toroidal inversion or point reflection:
$$
(\mu, \eta, \varphi) \mapsto (\mu, 2\pi-\eta, \varphi+\pi).
$$ 
Clearly, for $i=1,2$:
\begin{align}
 \Phi_{i,m}^c &\mapsto  \Phi_{1,m}^c(\mu, 2\pi-\eta, \varphi+\pi)=(-1)^m  \Phi_{i,m}^c,\\
 \Phi_{i,m}^s &\mapsto  \Phi_{1,m}^s(\mu, 2\pi-\eta, \varphi+\pi)=(-1)^{m+1}  \Phi_{i,m}^s.
\end{align}
Under this transformation, $ \Phi_{i,m}^c$ ($i=1,2$) is an even function  for $m= \pm2, \pm4, \cdots$ and an odd function for 
$m= \pm1, \pm3, \cdots.$ A similar statement holds for  $ \Phi_{i,m}^s$ ($i=1,2$) with reversed parity for being even or odd. 
}

\begin{figure}[htp]
\centering
     \includegraphics[width=3.40in]{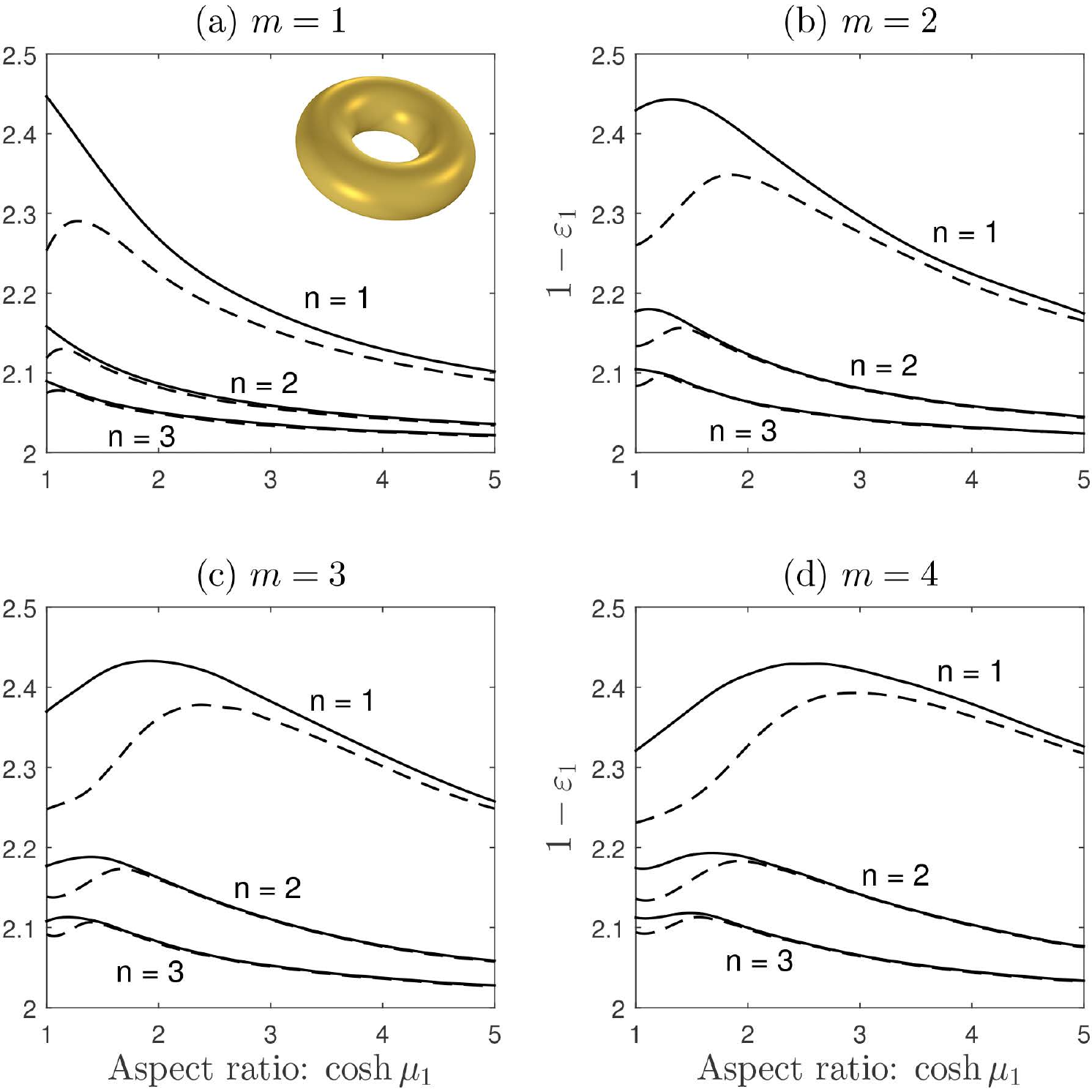} 
   		 \caption[]{\footnotesize Plasmon dispersion relations for a \bl{vacuum bounded} solid ring. \bl{Setting $k=1$, Eqs.~\eqref{Disp1} (dashed) and \eqref{Disp2} (solid) were solved to generate the eigenvalue spectrum as a function of the aspect ratio. The resonance frequency spectra of the dielectric function corresponding to the surface modes    $m=1,2,3,4$ and $n=1,2,3$ were here obtained via the free electron gas description of the metal.}}
   		 \label{SolidTorus}
\end{figure}

\bl{
In conclusion, one can use the parity argument in the following sense. By assuming the minimality of $\mathcal{X}_n,$ the normal modes are calculated from the dispersion relation ~\eqref{Disp1} while $\mathcal{Y}_n$ becomes a dominant solution. Now for a fixed $m,$ if $m$ is even (odd), then  the even  parity corresponds to the solution $\mathcal{X}_n$ ($ \mathcal{Y}_n$) and  the odd parity corresponds to the solution 
$\mathcal{Y}_n$ ($ \mathcal{X}_n$). An exact similar argument applied to the case when $ \mathcal{Y}_n$ is assumed as the minimal solution yields the  dispersion relation ~\eqref{Disp2} with $\mathcal{X}_n$ as a dominant solution, where the odd and even parity can be recognized.  As shown in the next subsection, however, this method works only for small aspect ratio where there is  a significant difference between the two separate sets of the normal modes. For larger aspect ratios, this difference is almost unrecognizable. 
}
\\
\paragraph{Exact numerical solutions\\\\}
The results of the numerical solution of the dispersion relations \eqref{Disp1} and \eqref{Disp2} for a single solid 
torus are shown in Fig.~\ref{SolidTorus}. 
To facilitate comparison~\cite{love:plasma}, the plots are obtained for the mode values of $m=1,2,3,4$ with $n=1,2,3$. The values of the associated Legendre functions $P_{n}^i, \hspace{1mm} Q_{n}^i$ in Eqs.~\eqref{Disp1} and \eqref{Disp2} are computed using the algorithm described in Gil and Segura~\cite{segura:cpc,gil:jcp,gil:cpc}.
Numerical analysis may be performed \bl{on Eq.~\eqref{threeterm}} with $k=1$ in the spherical and cylindrical limits wherein the aspect ratio tends to 1 and $\infty$, respectively. As $Q_n^1$ diverges as $\mu\rightarrow 0$ and $P_n^1$ diverges as $\mu\rightarrow \infty$, one can numerically observe that the coefficient terms associated with $\mathcal{D}_k$ for $k=n-1$, $n$, $n+1$ in Eq.~\eqref{threeterm} diverge in the spherical limit and converge to zero in the cylindrical limit. This observation is consistent with the earlier note on the surface charge influence across the toroidal opening ($n$, $n-1$, and $n+1$ decouple). Also in Fig.~\ref{limit_ring} in SM~\ref{B0} we present various limiting cases of a ring. 

\begin{figure}[htp]
\centering
     \includegraphics[width=3.40in]{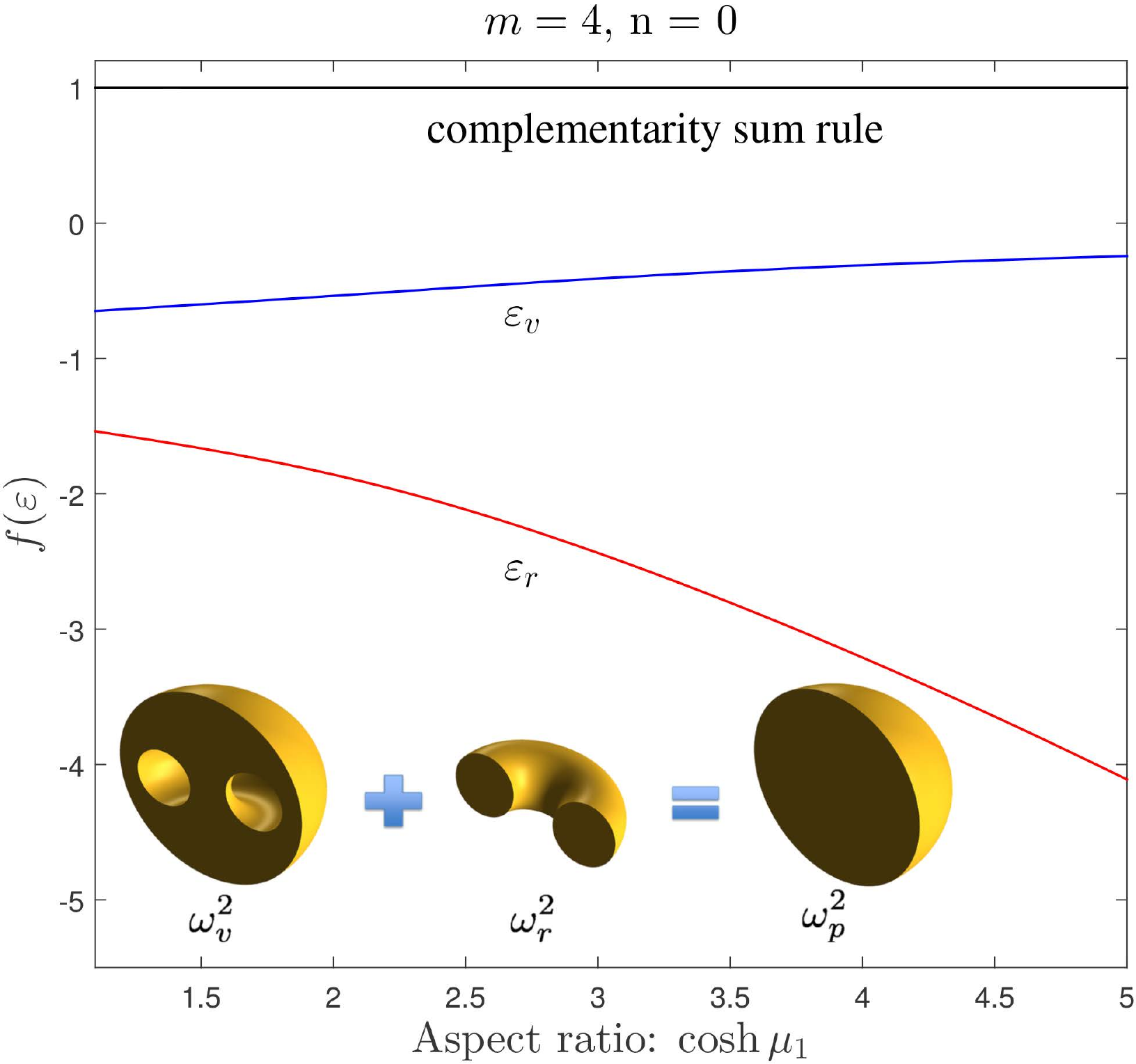} 
   		 \caption[]{\footnotesize 
		\bl{ Surface plasmon sum rule for a cavity-ring pair. 
The inset depicts the geometric complementarity of the surface modes. The sum rule states that the squared eigenmode frequencies of the void $\omega_v^2$ and the ring $\omega_r^2$ add up to those of the bulk $\omega_p^2$. Here, it is understood that the diameter of the spheres representing the domains  $\to \infty$ and the half structures rather than  full structures are used for visualization purposes. Eq.~\eqref{Disp1} was used to seek the respective $\eps_v$ and $\eps_r$. }
} 
   		 \label{SumRule1}
\end{figure}

The presented numerical scheme may be employed directly to also obtain the eigenfrequencies of a single toroidal cavity or alternatively we may consider using the geometric  complementarity and the  sum rule introduced by S. Apell~\emph{et al.}~\cite{sumrule}.
The \bl{surface plasmon} sum rule may therefore be used to check the validity of numerical calculations performed on the \bl{ ring configurations studied}. \bl{Here, to describe the metallic media we have employed the frequency $\omega$ dependent dielectric function:
\begin{equation}\label{epsilon_1}
\eps(\omega)=\eps_\infty - \dfrac{\omega_p^2}{\omega (\omega - i \Gamma \omega)}, 
\end{equation}
where $\omega_p$ is the plasma frequency,  $\Gamma$ is the electron collision rate (damping), and $\eps_\infty$ corrects for the high frequency response due to interband transitions of bound electrons 
(numerically, \emph{e.g.} for gold, $\hbar \Gamma = 0.069$~eV, $\omega_p =  8.95$~eV, and $\eps_\infty = 9.5$)}.
\bl{For the sake of discussions, we occasionally use: $\eps_\infty =1$ and $\Gamma = 0$, that is, free electron gas}. 
Denoting the frequencies of the surface modes of the toroidal void with $\omega_v$ and those of the solid ring with 
$\omega_r$, \bl{the inset in Fig.~\ref{SumRule1} depicts the complementarity of the modes, which
when using Eq.~\eqref{epsilon_1} without damping reads}:
	\begin{equation}\label{sum_rule}
		\dfrac{1}{1-\eps_v}+\dfrac{1}{1-\eps_r}=1.
	\end{equation}
\bl{The satisfaction of the plasmon sum rule  for the  ring-cavity pair is shown in Fig.~\ref{SumRule1}, where  Eq.~\eqref{Disp1} was solved for   
$\eps_1=\eps_r, \eps_2=1$, that is, a vacuum bounded solid ring, and subsequently for 
 $\eps_1=1, \eps_2=\eps_v$, that is, a toroidal vacuum cavity.
In Fig.~\ref{SumRule1}, the dispersion relations Eq.~\eqref{Disp1} were obtained for  $m=4$ and $n=0$.}
Given that the wave equation is not separable in the toroidal system, the results in Fig.~\ref{SolidTorus} are particularly useful. \bl{Further information on the eigenmodes and photonic and plasmonic response may be obtained computationally. 
In addition to computing the eigenfrequencies by solving an eigenvalue equation over an appropriately defined computational domain, we may compute the dominant branches of the retarded dispersion relations from the scattering properties of the nanorings. To begin with,}
employing the FEM method, in Fig.~\ref{ring-eigenmodes} in SM~\ref{E1} we present the  computationally determined \bl{transient} response of a gold nanoring to linearly polarized incident photons, \bl{and for comparison in Fig.~\ref{ring-dc}, the quasi-static response to different polarizations.
\begin{figure}[htp]
  \centering
  \begin{tabular}{cc}
    \includegraphics[scale =.265]{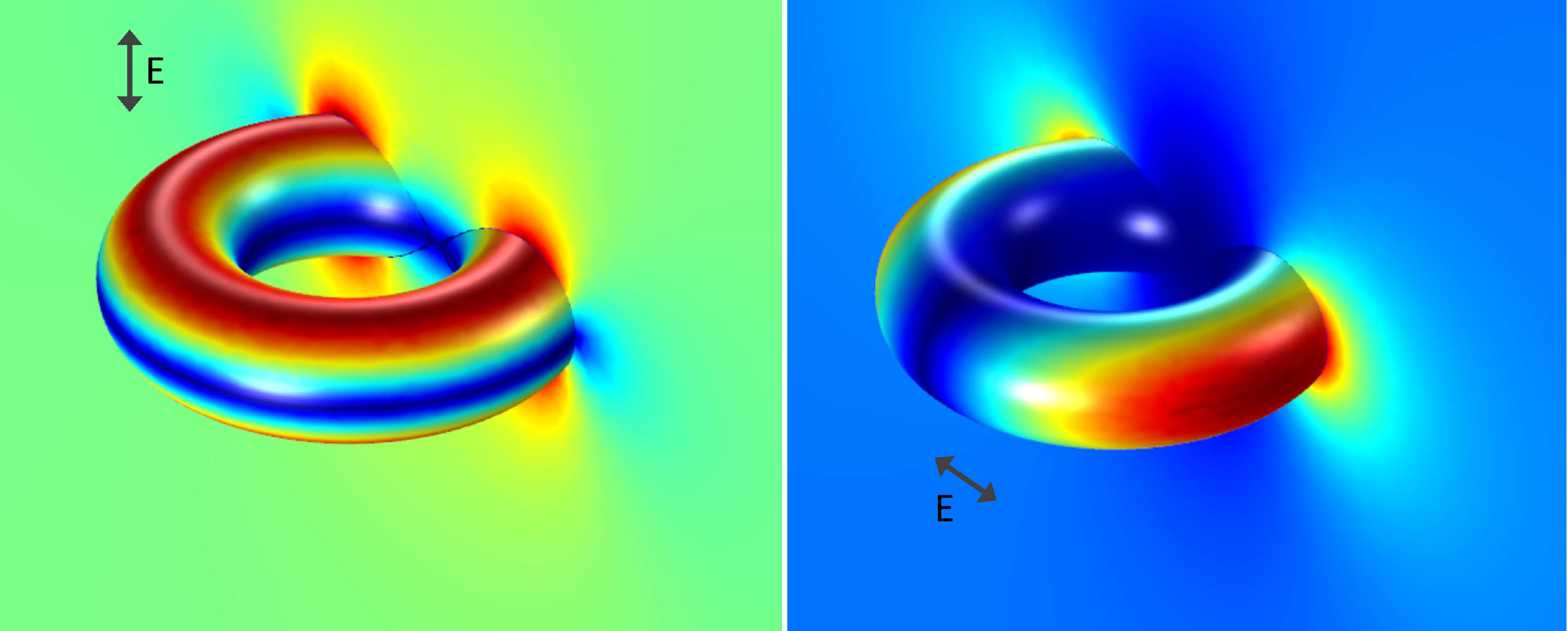} 
  \end{tabular}
  \caption{\bl{Quasi-static response of a gold nanoring to a static field parallel (left) and perpendicular (right) to the torus symmetry axis.}}
  \label{ring-dc}
\end{figure} 
\begin{figure}[htp]
\centering
     \includegraphics[width=3.40in]{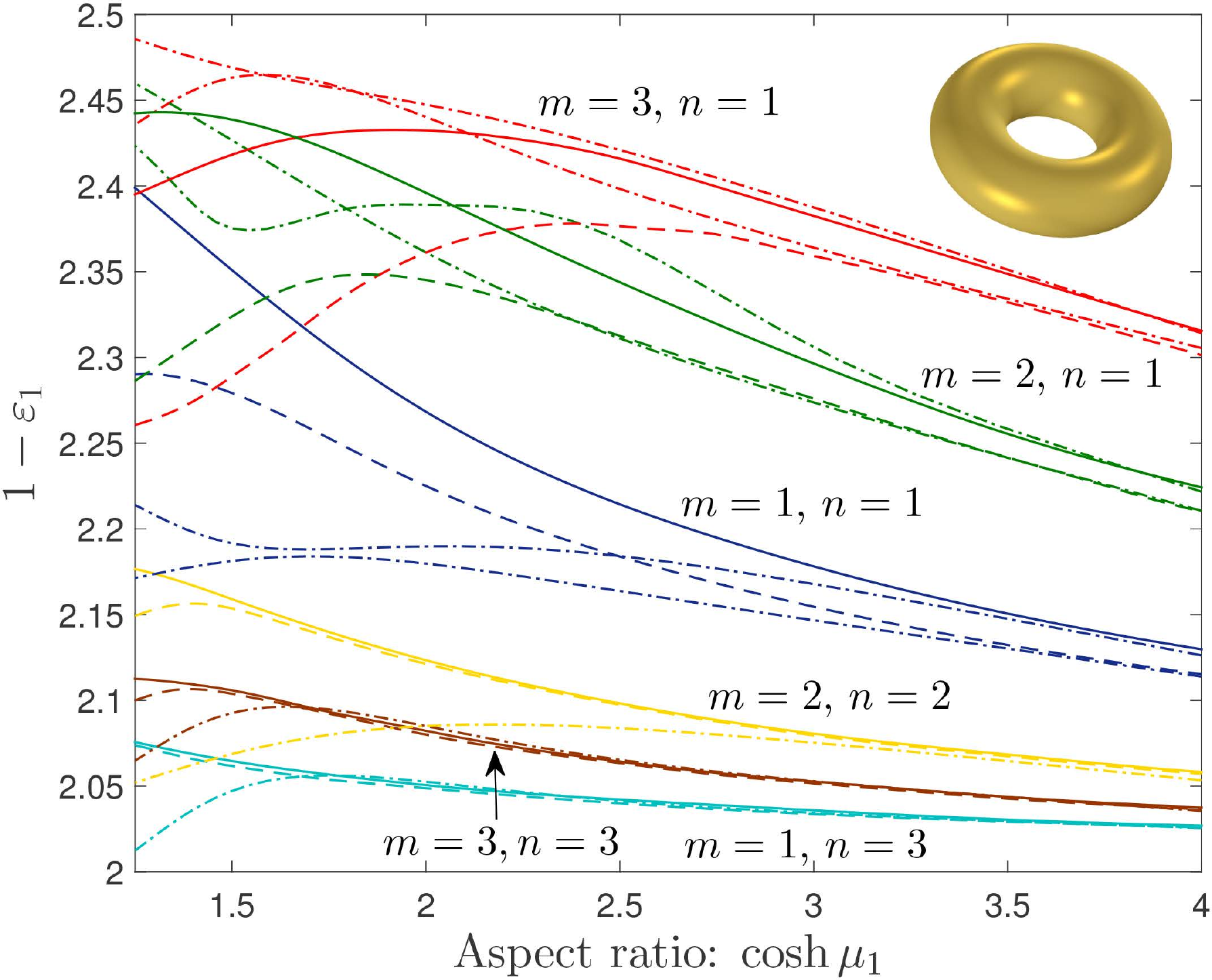} 
   		 \caption[]{\footnotesize \bl{ Retarded and nonretarded plasmon dispersion relations for a vacuum bounded nanoring. The quasi-static branches, the  dashed and solid curves, are obtained from Eq.~\eqref{Disp1} and Eq.~\eqref{Disp2}, respectively. The retarded branches, the dot-dashed curves, are obtained from the scattering properties of the nanorings exemplified in 
	Fig.~\ref{ring-eigenmodes} in SM~\ref{E1}. The mode indices are annotated.}}
		 \label{fdtd_gold}
\end{figure}
Parametric studies based on such results allow the determination of the retarded dispersion relations. In Fig.~\ref{fdtd_gold} we present a comparison of the nonretarded dispersion relations to some of the dominant retarded dispersion relations of a gold nanoring obtained computationally using FDTD. 
The FDTD converts the computational domain occupied by the ring and the surrounding medium into a lattice with a basic cell (the Yee cell) over which the field components are staggered in space and time. The fields are then evolved (time marching via leapfrog scheme) while satisfying Maxwell's equations. This procedure obtains retarded fields and quantities. 
As can be seen, the agreement is quite good for larger aspect ratios, which is reasonable in light of Fig.~\ref{SolidTorus}. We also observe that, for the same toroidal mode number, the agreement improves with higher poloidal mode numbers.}
\\
\paragraph{Perturbation technique\\\\}
\bl{Since the dispersion relations cannot be obtained analytically from the  continued fractions, approximate analytical solutions may be sought employing perturbation theory. Here we show that a decoupling of the poloidal surface modes as presented by the three-term recursion can be achieved using the perturbation approach. To demonstrate the mode decoupling, \bl{after specializing Eq.~\eqref{eqn_ref} for $k=1$, and decomposing the parity of the poloidal solutions,} we consider the even surface modes \emph{i.e.},} $\cos n\eta$ modes with fixed $m$ (the \bl{odd modes $\sin n\eta$ may be treated similarly}). 
\bl{Setting $\varepsilon=\varepsilon_1/\varepsilon_2$,} $u=\cosh \mu$, $u_1 = \cosh \mu_1$, for convenience,
\bl{ the conditions Eq.~\eqref{k_bc1} and Eq.~\eqref{k_bc2}, yield the Fourier modes:}
\begin{equation} 
c_n = b_n,
\end{equation}
\begin{align}\label{eq7}
C_n b_n & = \frac{1}{2u_0}\sum_{\sigma = \pm 1}  \left( C_{n+\sigma} + \frac{1-\varepsilon}{2} \right) b_{n+\sigma}, \notag\\
 C_n & = \left. \varepsilon \frac{d\log \left[ \sqrt{u} Q_{n- \frac{1}{2}}^m (u) \right] }{d\log{u}}   -\frac{d\log \left[ \sqrt{u} P_{n- \frac{1}{2}}^m (u) \right] }{d\log{u}} \right|_{u=u_1},
\end{align}
where we defined $b_{-1} =0$. This equation can be solved using perturbation theory. 
We shall work in the scaling limit $u_1,m\to\infty$, with $m/u_1$ held fixed.
In this limit,
\begin{align}
&\frac{d\log \left[ \sqrt{u} P_{n- \frac{1}{2}}^m (u) \right] }{d\log{u}} = 
 \frac{d}{d\log{u}}  \biggl[ \log K_n\left( \dfrac{m}{u} \right)  \notag \\ \notag
& + \frac{ \frac{3-24n^2+64( \frac{m}{u} )^2 + 48 n^4}{16} - \left( 1-4n^2+8( \frac{m}{u} )^2 \right) \frac{d\log K_n\left( \frac{m}{u} \right)}{d\log{u}}}{24m^2} \notag \\ 
& + \dots \biggr],
\end{align}
\begin{align}
&\frac{d\log \left[ \sqrt{u} Q_{n- \frac{1}{2}}^m (u) \right] }{d\log{u}} = 
\frac{d}{d\log{u}}  \biggl[ \log I_n\left( \frac{m}{u} \right) \notag \\ \notag
& +\frac{ \frac{3-24n^2+64( \frac{m}{u} )^2 + 48 n^4}{16} - \left( 1-4n^2+8( \frac{m}{u} )^2 \right) \frac{d\log I_n\left( \frac{m}{u} \right)}{d\log{u}}}{24m^2} \notag \\ 
& + \dots \biggr],
\end{align}
and the right-hand side of Eq.~\eqref{eq7} vanishes. Therefore, Eq.~\eqref{eq7} reduces to
\begin{align}\label{eq7s}  
C_nb_n &=  \left[ \varepsilon \frac{d\log I_n\left( \frac{m}{u} \right)}{d\log{u}} -\frac{d\log K_n\left( \frac{m}{u} \right)}{d\log{u}} + \mathcal{O} \left( \frac{1}{u_1} \right) \right] b_n \notag \\
&= 0 ,
\end{align}
at $u=u_1$, and the modes decouple. For the $n=n_0$ mode, we obtain
\begin{equation}
\varepsilon = \varepsilon_{mn_0}^0 = \frac{I_{n_0}( \frac{m}{u_1} )K_{n_0}'( \frac{m}{u_1} )}{I_{n_0}'( \frac{m}{u_1} )K_{n_0}( \frac{m}{u_1} )},\end{equation}
\bl{where $I$ and $K$ are the modified Bessel functions of the first and second kind, respectively.}
For first order in perturbation theory, set $b_{n_0}=1$, and expand
\begin{align} 
C_n &= C_n^0 + \frac{1}{u_1} C_n^1 + \frac{1}{u_1^2} C_n^2 + \dots\ ,\notag\\\notag
\varepsilon &= \varepsilon_{mn_0}^0 + \frac{1}{u_1} \varepsilon_{mn_0}^1 + \frac{1}{u_1^2} \varepsilon_{mn_0}^2 + \dots \ ,\\\notag
b_{n} &= \frac{1}{u_1} b_{n}^1 +\frac{1}{u_1^2} b_{n}^2 + \dots, \ \ (n\ne n_0),\\
\end{align}
where
\begin{equation}\label{eq13} 
\begin{split}
C_n^0 &= (\varepsilon_{mn_0}^0 -\varepsilon_{mn}^0 )\frac{d\log I_n\left( \frac{m}{u} \right)}{d\log{u}} ,   \nonumber\\
 C_n^1 &= \varepsilon_{mn_0}^1\frac{d\log I_n\left( \frac{m}{u} \right)}{d\log{u}}  ,\nonumber\\
C_n^2 &=  \varepsilon_{mn_0}^2 \frac{d\log I_n\left( \frac{m}{u} \right)}{d\log{u}}  +\frac{1- \varepsilon_{mn_0}^0}{3} \\
&   + \frac{1}{24(\frac{m}{u})^2} \frac{d}{d\log{u}} \biggl\{ \biggl( 1-4n^2+8\left( \frac{m}{u} \right)^2 \biggr)\\
& \times \biggl[ \frac{d\log K_n\left( \frac{m}{u} \right)}{d\log{u}}  - \varepsilon_{mn_0}^0 \frac{d\log I_n\left( \frac{m}{u} \right)}{d\log{u}}\biggr]   \biggr\}. \nonumber\\ 
\end{split}
\end{equation}
Notice that $C_{n_0}^0 =0$.
At $\mathcal{O} (1/u_1)$, Eq.~\eqref{eq7} becomes
\begin{equation} 
C_n^1 \delta_{nn_0} + C_n^0 b_n^1 = \frac{1-\varepsilon_{mn_0}^0}{4}\sum_{\sigma = \pm 1}   \delta_{n+\sigma n_0},
\end{equation}
therefore
\begin{equation}
\varepsilon_{mn_0}^1 = 0, \ C_{n}^1 =0 ,\ b_{n_0\pm 1}^1 =  \frac{1-\varepsilon_{mn_0}^0}{4C_{n_0\pm 1}^0},\ b_n^1 = 0,
\end{equation}
where $(n\ne n_0\pm 1) $.
At $\mathcal{O} (1/u_1^2)$, Eq.~\eqref{eq7} becomes
\begin{equation}  
C_n^2 \delta_{nn_0} + C_n^0 b_n^2 = \frac{1}{2}\sum_{\sigma = \pm 1}  \left( C_{n+\sigma}^0 + \frac{1-\varepsilon_{mn_0}^0}{2} \right) b_{n+\sigma}^1. \end{equation}
For $n=n_0$, this gives
\begin{align}
C_{n_0}^2  & =  \frac{1}{2}\sum_{\sigma = \pm 1}  \left( C_{n_0+\sigma}^0 + \frac{1-\varepsilon_{mn_0}^0}{2} \right) b_{n_0+\sigma}^1  \notag\\ 
& =  \frac{1-\varepsilon_{mn_0}^0}{8}\sum_{\sigma = \pm 1}  \left( 1 + \frac{1-\varepsilon_{mn_0}^0}{2C_{n_0+\sigma}^0} \right).
\end{align}
In the special case $n_0=0$, the above reduces to
\begin{align} C_{0}^2  = \frac{1}{2}  \left( C_{1}^0 + \frac{1-\varepsilon_{m0}^0}{2} \right) b_{1}^1  =   \frac{1-\varepsilon_{m0}^0}{8}  \left( 1 + \frac{1-\varepsilon_{m0}^0}{2C_{1}^0} \right).
\end{align}
Having obtained $C_{n_0}^2$, we then extract $\varepsilon_{mn}^2$ from Eq.~\eqref{eq13}.
\begin{figure}[htp]
\includegraphics[width=3.25in]{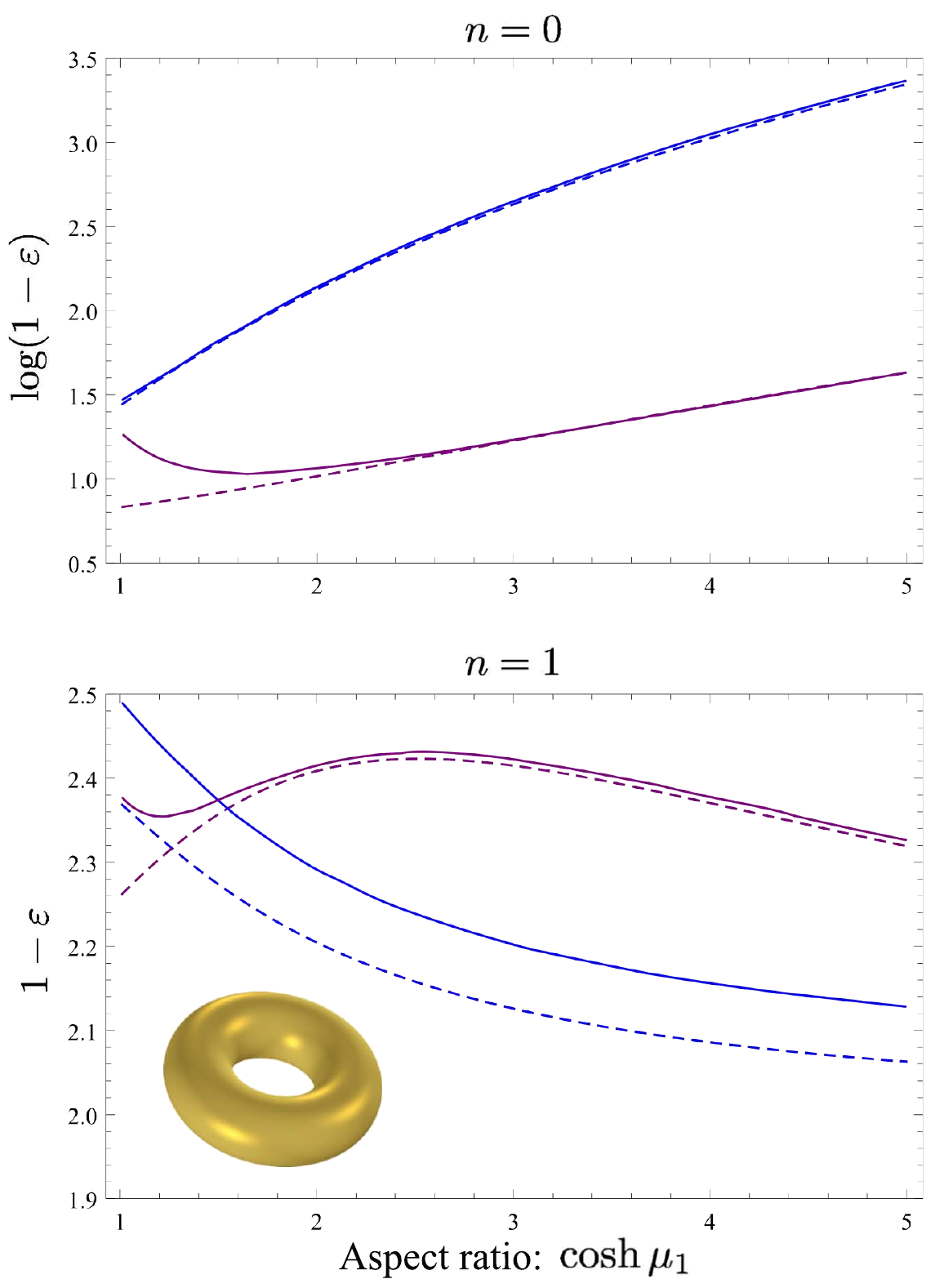}
\caption{Zeroth and first order perturbation calculation of the resonant dielectric values for a single  ring in vacuum corresponding to \bl{modes} $n=0$, $n=1$, and $m=1$ (blue), $m=4$ (purple). Zeroth (first) order are dashed (solid) lines.}
\label{fig2} 
\end{figure}
A comparison between zeroth order and next-to-leading order is shown in Fig.~\ref{fig2}. When corrections are included, the analytic formulas agree well with numerical results down to small aspect ratios. They disagree near the minimum aspect ratio $u_1=1$, indicating the effect of mode amplitude coupling in this regime, and thus need for inclusion of higher orders.
\\
\paragraph{\bl{Plasmon redshift due to substrate effect}\\\\}
From an experimental point of view, most measurement efforts will consider use of a supporting substrate. It is known that, as in the case of thin solid films, generally surface plasmon energy redshifts when a dielectric substrate is considered. Similar modifications have been reported for a hyperboloidal nanostructures near a dielectric surface~\cite{pass}.  \bl{Employing the perturbation approach, here we calculate the redshift of the plasmon energies by obtaining the first order correction of the vacuum bounded plasmon dispersion due to the substrate.}

Suppose that the ring is placed at \bl{ $z=0$ plane on} a dielectric substrate (occupying the $z<0$ region) characterized by $\varepsilon'$. Let $\Phi_0 (\vec{r})$, $\vec{r} = (x, y, z)$ in Cartesian coordinates be the potential without the substrate. 
Then the potential with the substrate is found using the method of images to be
\begin{equation}
\Phi (\vec{r}) = \left\{ \begin{array}{ccc} \Phi_0 (x,y,z) + \dfrac{1-\varepsilon'}{1+\varepsilon'} \Phi_0 (x,y,-z) & , & z>0 \\ \\
\dfrac{2}{1+\varepsilon'} \Phi_0 (x,y,z) & , & z <0 \end{array} \right. 
\end{equation}
\bl{We define the} toroidal coordinates $u_+ = \cosh\mu_+$, $\eta_+$ related to the Cartesian coordinates $\rho = \sqrt{x^2+y^2}$, $z$ by
\begin{align} 
\rho &= \frac{\sqrt{u_+^2-1}}{u_+ - \cos\eta_+}, \notag\\\notag
 z &= z_0 + \frac{\sin\eta_+}{u_+- \cos\eta_+} ,\\\notag
 z_0 &= \frac{1}{\sinh\mu_1} = \frac{1}{\sqrt{u_1^2 -1}}.\\
\end{align}
For the image ring, we similarly define  toroidal coordinates $u_- = \cosh\mu_-$, $\eta_-$ by 
\begin{equation}
\rho = \frac{\sqrt{u_-^2-1}}{u_- - \cos\eta_-},\ \, \ \ \ \ z = -z_0 + \frac{\sin\eta_-}{u_-- \cos\eta_-}.
\end{equation}
Concentrating on the $\cos n\eta$ modes and fixed $m$, we have for $z>0$,
\begin{eqnarray}
\notag & \Phi_0(x,y,z) = e^{im\varphi} \dfrac{\sqrt{u_+ - \cos\eta_+}}{\sqrt{u_1}}  \sum_{n=1}^\infty \cos n\eta_+\\
&\times \left\{ \begin{array}{ccl}  b_n  \dfrac{Q_{n- \frac{1}{2}}^m (u_+)}{Q_n^1},  \hspace{0.5in}  u_+ \ge u_1\\ \\
  c_n  \dfrac{P_{n- \frac{1}{2}}^m (u_+) }{P_n^1},  \hspace{0.5in}  u_+ \le u_1 \end{array} \right. 
\end{eqnarray}
respectively for inside and outside the ring. Now for the image we have:
\begin{align}
 \Phi_0 (x,y,-z) &= e^{im\varphi}\frac{\sqrt{u_- - \cos\eta_-}}{\sqrt{u_1}} \notag \\
 		&	\times \sum_{n=1}^\infty c_n \sin n\eta_- 
 						\frac{P_{n- \frac{1}{2}}^m (u_-) }{P_n^1}.
 \end{align}
Continuity at $u_+=u_1$ implies
\begin{equation}
b_n = c_n,
\end{equation}
as before, because the potential of the image is already continuous.
Continuity of the $u_+$-component implies
\begin{align}\label{eq16}
 &\varepsilon\sum_{n=1}^\infty   b_n \cos n\eta_+ \frac{d}{du}_{+} \log \left[ \sqrt{u_+} Q_{n- \frac{1}{2}}^m (u_+) \right]  \notag\\
& - \sum_{n=1}^\infty b_n \cos n\eta_+ \frac{d}{du}_{+} \log \left[ \sqrt{u_+} P_{n- \frac{1}{2}}^m (u_+) \right] 
 \notag\\
& =(1-\varepsilon ) \biggl[ \sum_{n=1}^\infty b_n \cos n\eta_+ \frac{d}{du}_+ \log \sqrt{1 - \frac{\cos\eta_+}{u_+}}  \notag\\
& +\frac{1-\varepsilon'}{1+\varepsilon'}\frac{d}{du}_+ \frac{\sqrt{u_- - \cos\eta_-}}{\sqrt{\mu_1}} \sum_{n=1}^\infty b_n \cos n\eta_- \dfrac{P_{n- \frac{1}{2}}^m (u_-) }{P_n^1} \biggr],
\end{align}
at $u_+ = u_1$. Notice that both $u_-$ and $\eta_-$ depend on $u_+$. To solve this, we shall treat the right-hand side as a perturbation. At zeroth order, the modes decouple. After fixing $n=n_0$, we obtain the zeroth order estimate
\begin{equation}
\varepsilon = \varepsilon_{mn_0}^0 = \left. \frac{\frac{d}{du} \log \left[ \sqrt{u} P_{n- \frac{1}{2}}^m (u) \right] }{\frac{d}{du} \log \left[ \sqrt{u} Q_{n- \frac{1}{2}}^m (u) \right]} \right|_{u = u_1}.
\end{equation}
\bl{Thus, in the zeroth order, the substrate effects on plasmon energies are absent. To calculate the redshift, we proceed to the first-order calculation.}
Introducing the expansion (bookkeeping) parameter $\xi$, we re-write Eq.~\eqref{eq16} as
\begin{align}\label{eq16a} 
&\varepsilon\sum_{n=1}^\infty b_n \cos n\eta_+ \frac{d}{du}_+ \log \left[ \sqrt{u_+} Q_{n- \frac{1}{2}}^m (u_+) \right]   \notag\\
& -\sum_{n=1}^\infty b_n \cos n\eta_+ \frac{d}{du}_+ \log \left[ \sqrt{u_+} P_{n- \frac{1}{2}}^m (u_+) \right] 
\notag\\ 
& =\xi
(1-\varepsilon ) \biggl[ \sum_{n=1}^\infty b_n \cos n\eta_+ \frac{d}{du}_+ \log \sqrt{1 - \frac{\cos\eta_+}{u_+}} \notag\\
& +  \frac{1-\varepsilon'}{1+\varepsilon'}\frac{d}{du}_+ \frac{\sqrt{u_- - \cos\eta_-}}{\sqrt{u_1}} \sum_{n=1}^\infty b_n \cos n\eta_- \frac{P_{n- \frac{1}{2}}^m (u_-) }{P_n^1} \biggr].
\end{align}
We fix $b_{n_0} =1$, and expand
\begin{align}
\varepsilon &= \varepsilon_{mn_0}^0 + \xi (\varepsilon_{mn_0}^1 + \overline\varepsilon_{mn_0}^1)+ \mathcal{O} (\xi^2), \notag\\\notag
b_n &= \xi b_{n}^{1} + \mathcal{O} (\xi^2), \\
\end{align}
where $ ( n\ne n_0)$ and $\overline\varepsilon_{mn_0}^1$ is the first-order correction due to the substrate.\\

At $\mathcal{O}(\xi)$, Eq.~\eqref{eq16a} becomes
\begin{align}\label{eq16b}
&\sum_{n=1}^\infty (\varepsilon_{mn_0}^0 - \varepsilon_{mn}^0)b_{n}^{1} \cos n\eta_+ \frac{d}{du}_+ \log \left[ \sqrt{u_+} Q_{n- \frac{1}{2}}^m (u_+) \right] \notag\\
& +(\varepsilon_{mn_0}^1 + \overline\varepsilon_{mn_0}^1) \cos n_0\eta_+ \frac{d}{du}_+ \log \left[ \sqrt{u_+} Q_{n_0-\frac{1}{2}}^m (u_+) \right]  
\notag\\
&=
(1-\varepsilon_{mn_0}^0 ) \biggl(  \cos n_0\eta_+ \frac{d}{du}_+ \log \sqrt{1 - \frac{\cos\eta_+}{u_+}}
\notag\\
& + \frac{1-\varepsilon'}{1+\varepsilon'}\frac{d}{du}_+ \frac{\sqrt{u_- - \cos\eta_-}}{\sqrt{u_1}} \cos n_0 \eta_- 
			\frac{P_{n_0-\frac{1}{2}}^m (u_-) }{P_{n_0}^1} \biggr).
\end{align}
Multiplying by $\cos n_0\eta_+$ and integrating over $[0,\pi]$, we deduce the \bl{sought} correction to the $n=n_0$ mode due to the substrate,
\begin{equation}\label{eq16c}
\begin{split}
\overline\varepsilon_{mn_0}^1 &= \frac{1-\varepsilon'}{1+\varepsilon'}
\frac{2(1-\varepsilon_{mn_0}^0)}{\pi\frac{d}{du_+} \bl{\log \left[ \sqrt{u_+} Q_{n_0-\frac{1}{2}}^m (u_+) \right]} }  
\\ & \times \int_0^\pi d\eta_ +
\frac{d}{du}_+ \frac{\sqrt{u_- - \cos\eta_-}}{\sqrt{u_1}} \cos n_0 \eta_- \frac{P_{n_0-\frac{1}{2}}^m (u_-) }{P_{n_0}^1},
\end{split}
\end{equation}
which may be simulated to obtain the amount of the shift induced in the plasmon energies. \bl{While the correction vanishes  for $\eps' \to 1$, that is, for a vacuum bounded isolated ring, it diverges for $\eps' \to -1$. Interestingly, for the surface plasmons on a plane-bounded semi-infinite metal, the quasi-static resonance occurs when the real part of the dielectric function is -1. } 
\\
\paragraph{\bl{Plasmon redshift  for a} ring on a substrate with finite thickness\\\\}
Thin solid films are ubiquitous within plasmonics and nano-optics. We therefore suppose that the substrate has finite thickness $d$, \bl{that is, it possesses two planar interfaces and therefore two dispersion branches, which in the metallic case correspond to symmetric and anti-symmetric electron density oscillations~\cite{pass}. For thick films (large $d$), the two branches degenerate into a single branch, which in the nonretarded limit approaches -1.}  It is perhaps best to solve this problem iteratively. We start with the solution for the semi-infinite slab, \emph{i.e.}, for $z>0$
\begin{equation} 
\Phi_I^{(0)} = \Phi_0 (x,y,z) + \frac{1-\varepsilon'}{1+\varepsilon'} \Phi_0 (x,y,-z),
\end{equation}
for $-d<z<0$
\begin{equation} 
\Phi_{II}^{(0)} = \frac{2}{1+\varepsilon'} \Phi_0 (x,y,z).
\end{equation}
By design, this potential satisfies the correct boundary conditions at the interface $z=0$.
For correct boundary conditions at $z=-d$, we take the image of $\Phi_{II}^{(0)}$ with respect to $z=-d$ and obtain the modified potential
 \begin{align}
 \Phi_{II}^{(1)} &= \Phi_{II}^{(0)} (x,y,z) - \frac{1-\varepsilon'}{1+\varepsilon'} \Phi_{II}^{(0)} (x,y,-z-2d) \notag\\
 		&=\frac{2}{1+\varepsilon'}\Phi_{0} (x,y,z) - \frac{2(1-\varepsilon')}{(1+\varepsilon')^2} \Phi_{0} (x,y,-z-2d),
\\
\Phi_{III}^{(1)} &= \frac{2\varepsilon'}{1+\varepsilon'} \Phi_{II}^{(0)} (x,y,z) = \frac{4\varepsilon'}{(1+\varepsilon')^2} \Phi_{0} (x,y,z),
\end{align}
where $\Phi_{III}$ is defined in the region $z<-d$, satisfying $\Phi_{II}^{(1)} = \Phi_{III}^{(1)}$, and $\varepsilon' \frac{\partial\Phi_{II}^{(1)}}{\partial z} = \frac{\partial\Phi_{III}^{(1)}}{\partial z}$ at $z=-d$.
Next, we correct $\Phi_{II}$ and $\Phi_{I}$ so that they satisfy the correct boundary conditions at $z=0$. $\Phi_{I}^{(0)}$ and $\Phi_{II}^{(0)}$ already do, so we only need to consider the second image and its image with respect to $z=0$.
We obtain
\begin{align}
\Phi_{I}^{(2)} & = \Phi_{I}^{(0)} - \frac{2\varepsilon'(1-\varepsilon')}{(1+\varepsilon')^2} \Phi_{II}^{(0)} (x,y,-z-2d) \notag\\
 & = \Phi_0 (x,y,z) + \frac{1-\varepsilon'}{1+\varepsilon'} \Phi_0 (x,y,-z) \notag\\
 & - \frac{4\varepsilon'(1-\varepsilon')}{(1+\varepsilon')^3} \Phi_{0} (x,y,-z-2d),
 \\
\Phi_{II}^{(2)} &= \Phi_{II}^{(1)} + \left( \frac{1-\varepsilon'}{1+\varepsilon'} \right)^2 \Phi_{II}^{(0)} (x,y,z-2d) \notag\\ 
 & = \frac{2}{1+\varepsilon'}\Phi_{0} (x,y,z) - \frac{2(1-\varepsilon')}{(1+\varepsilon')^2} \Phi_{0} (x,y,-z-2d) \notag\\ 
 &+  \frac{2(1-\varepsilon')^2}{(1+\varepsilon')^3} \Phi_{0} (x,y,z-2d),
\end{align}
satisfying  $\Phi_{I}^{(2)} = \Phi_{II}^{(2)}$, and $ \frac{\partial\Phi_{I}^{(2)}}{\partial z} = \varepsilon'\frac{\partial\Phi_{II}^{(2)}}{\partial z}$ at $z=0$.
Alternating between $z=0$ and $z=-d$, we thus build the potential as a series of images, \bl{which can be subsequently used to obtain a similar correction as in Eq.~\eqref{eq16c}.}
\\
\paragraph{Plasmon dispersion in metallic and semiconductor nanorings\\\\}
Experiments involving surface enhanced spectroscopies, nanophotonics and plasmonics typically employ 
metallic nanorings made of aluminum, gold, and silver as well as semiconductors such as silicon and germanium. 
For the case of vacuum bounded modes (\emph{i.e.}, $\eps_2=1$),  the plasmon resonance frequencies as a function of aspect ratio can provide useful information regarding the characteristics of the excitation photons impinging on the nanorings. 
\bl{To obtain the dispersion relations for specific materials, we employ an interpolations scheme between the solutions of Eq.~\eqref{Disp1} with $k=1$, and  the experimental data composed by Palik~\cite{palik}.}
\begin{figure}[htp]
      \includegraphics[width=3.15in,keepaspectratio]{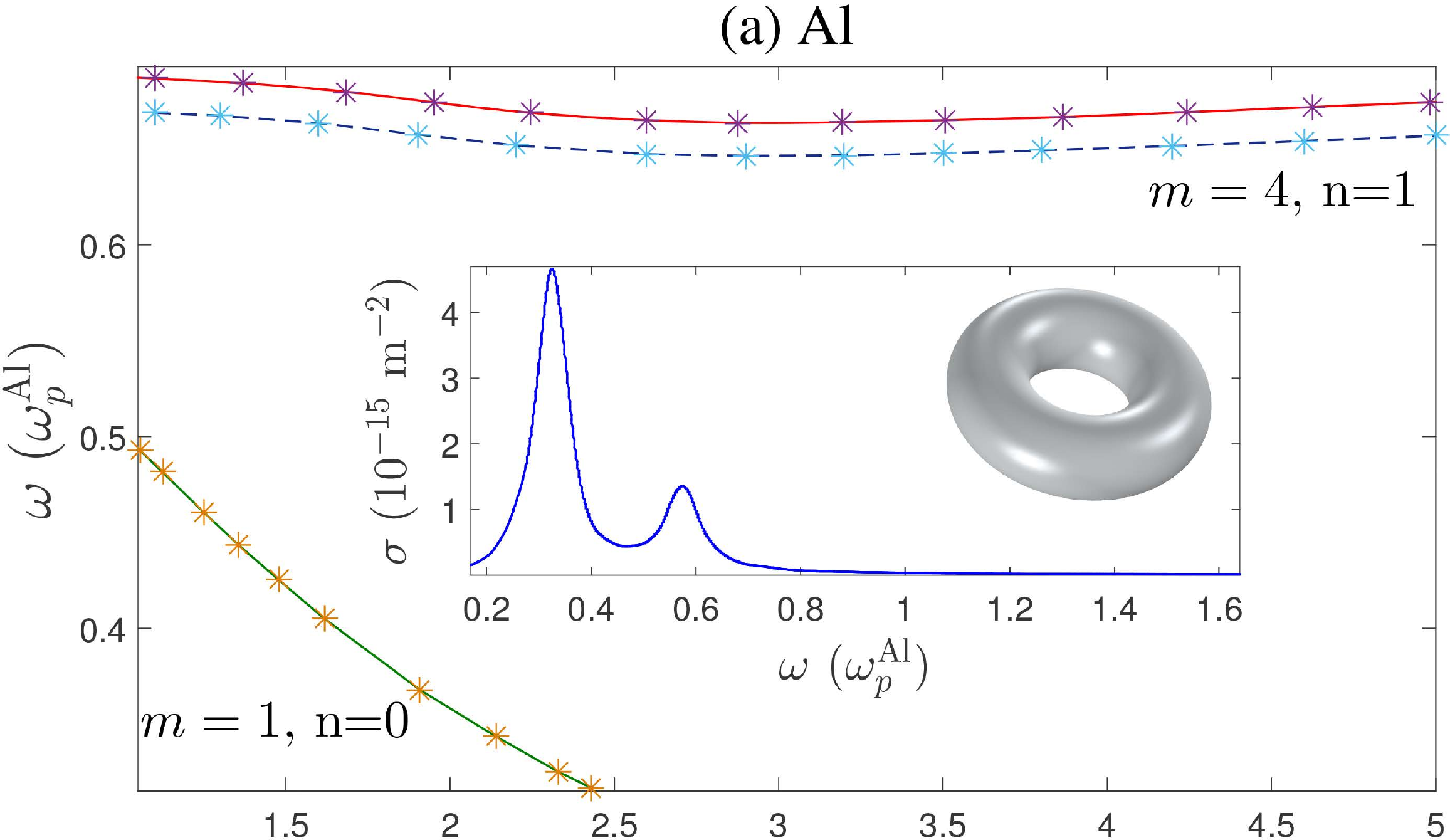} 
     
     \vspace{0.25cm}
     
     \includegraphics[width=3.15in,keepaspectratio]{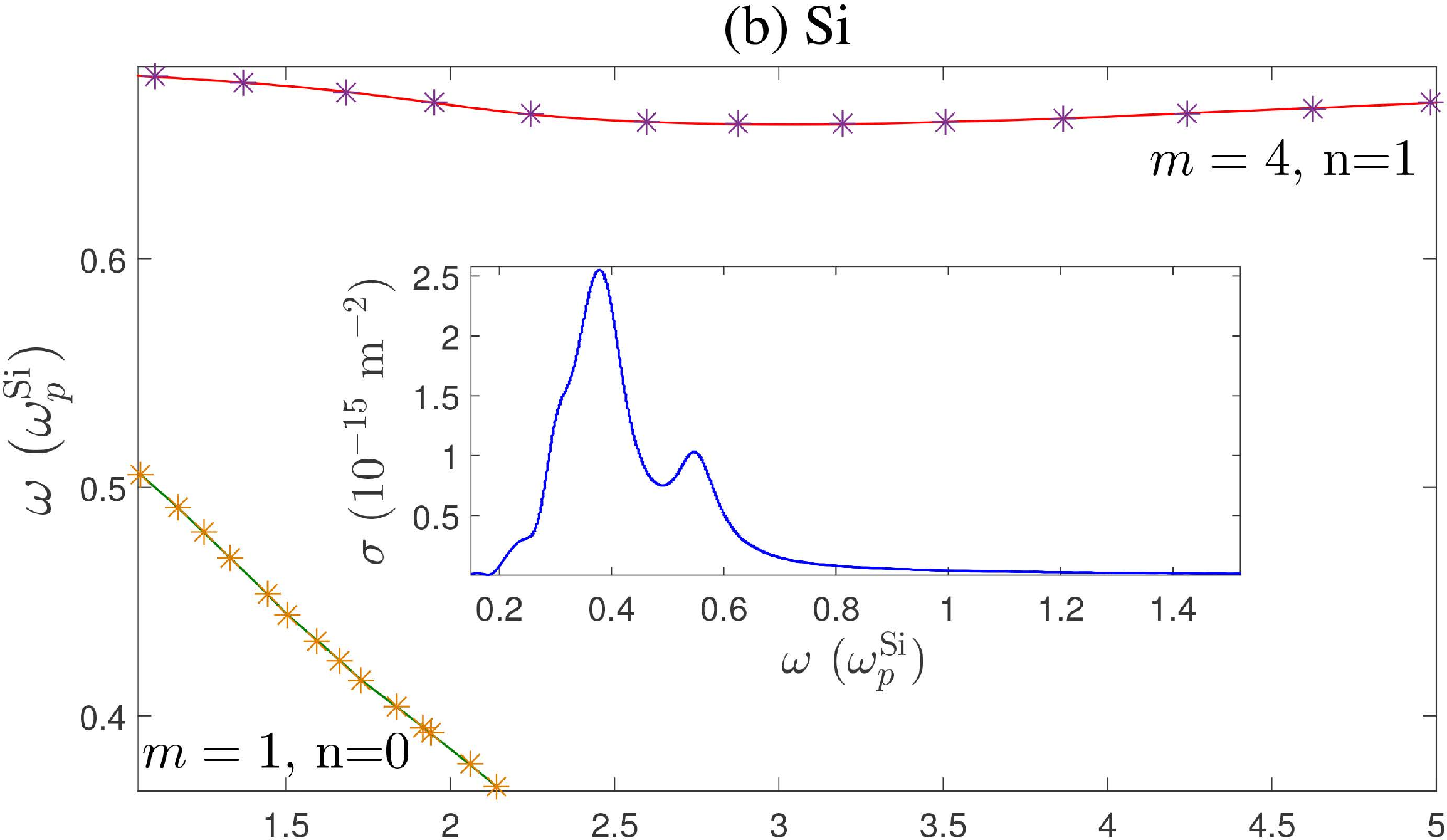} 
     
     \vspace{0.25cm}
     
     \includegraphics[width=3.15in,keepaspectratio]{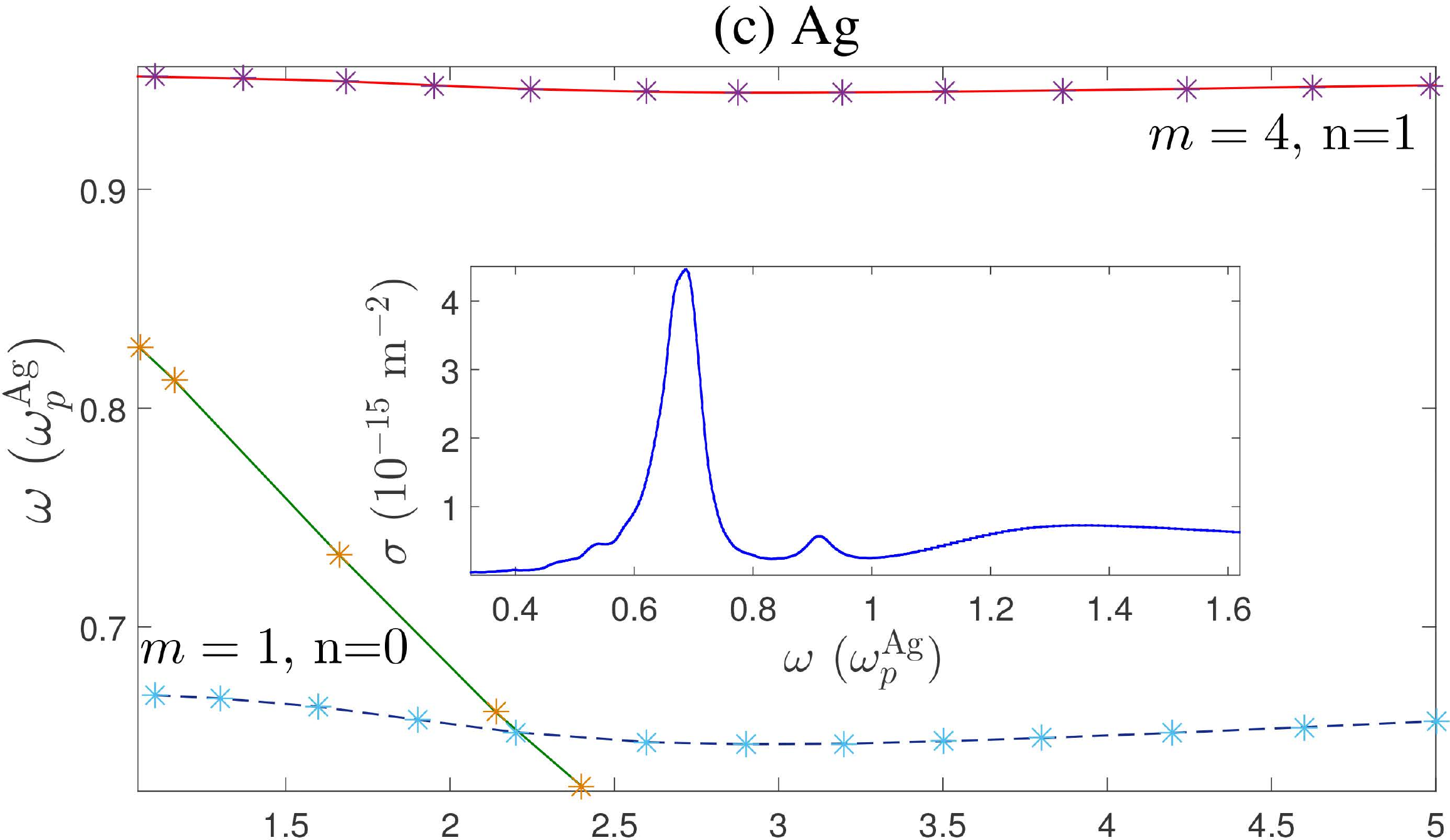}
     
     \vspace{0.25cm}
     
     \includegraphics[width=3.15in,keepaspectratio]{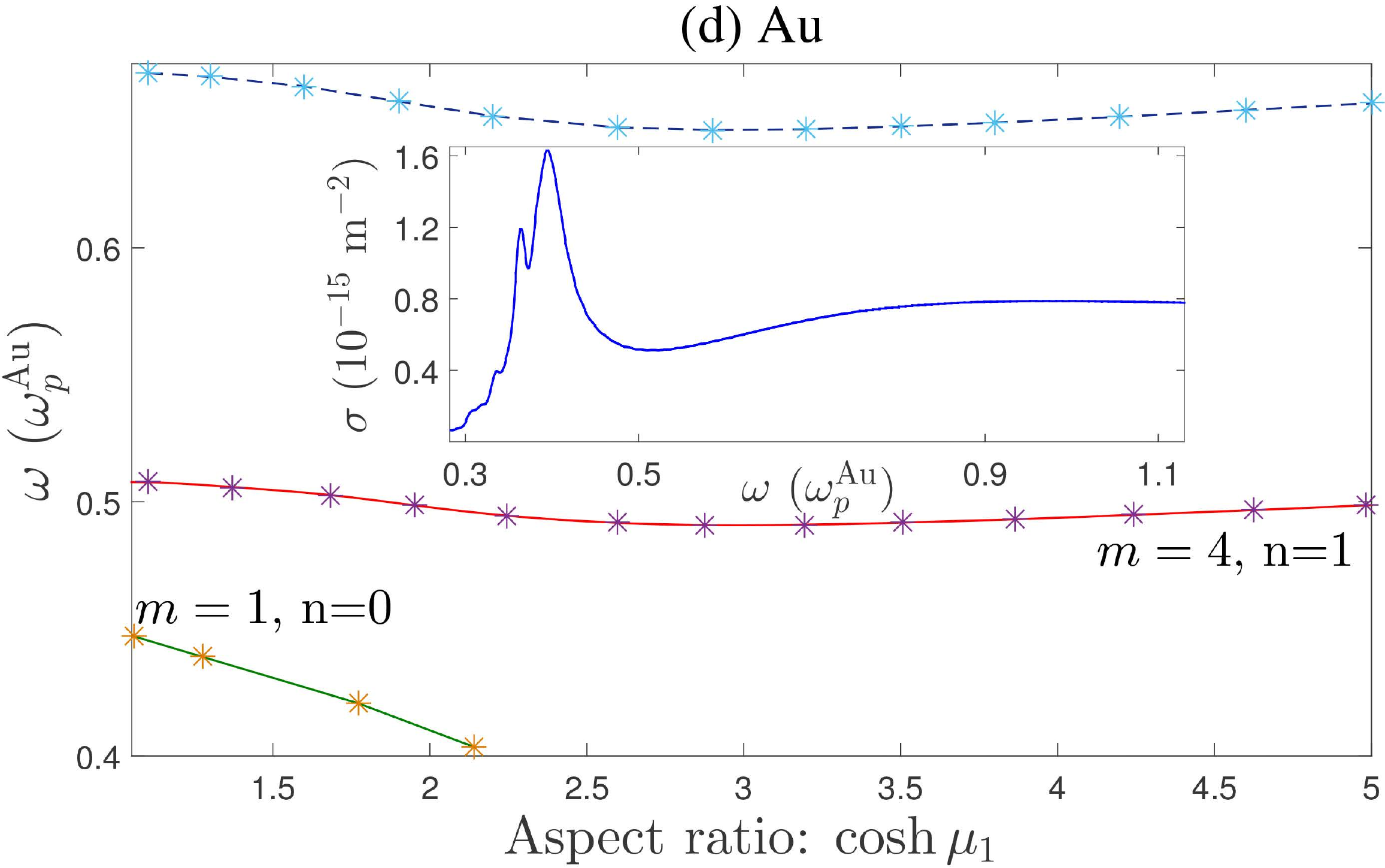}
   		 \caption[]{\footnotesize \bl{
		 Resonance frequencies of metallic and semiconductor  solid rings as a function of the aspect ratio. 
		 The modes $m=1,4$, and $n=0,1$ are obtained from Eq.~\eqref{Disp1} with $k=1$.
		 The  $(m,n)=(4,1)$  plasmon  dispersion (dashed) in a toroidal free electron gas employing $\omega_p^{\text{Al}}=3.643\times10^{15}$, $\omega_p^{\text{Ag}}=9.264\times10^{14}$, and $\omega_p^{\text{Au}}=1.324\times10^{15}$ are shown for aluminum, silver, and gold.  The insets present the computed spectral dependence of the scattering cross section displaying the dominant resonance modes $m=1$, $n=0$ for aluminum, silicon, gold and silver (note that, for Si, $\omega_p^{\text{Si}}=3.948\times10^{15}$ was used simply for scaling purposes.)}}
   		 \label{metalring_1}
\end{figure}
This results in the dispersion relations \bl{shown in Fig.~\ref{metalring_1} for the modes $m=1$, $n=0$ and $m=4$, $n=1$.}
\bl{The usefulness of these results can be appreciated when seeking the response of the system.}
For example, for an aspect ratio $\cosh\mu=2$, following the nonretarded resonance values for the dominant mode $m=1$, $n=0$, presented in \bl{Fig.~\ref{metalring_1}}, the corresponding absorption cross sections are displayed \bl{in the insets of Fig.~\ref{metalring_1}}. The fully retarded cross sections \bl{in the insets of Fig.~\ref{metalring_1}} were here computed by numerically solving Maxwell's equations over a \bl{spatial} domain containing the solid gold ring in vacuum using the \bl{FDTD} method~\cite{fdtd}.  
 As can be seen the agreement is quite good between the two cases. \bl{Also the FEG dispersion curves presented in Fig.~\ref{metalring_1} exhibit qualitatively similar patterns but quantitatively are different due to the effect of $\omega_p$ values.}

\begin{figure}[htp]
     \includegraphics[width=3.40in]{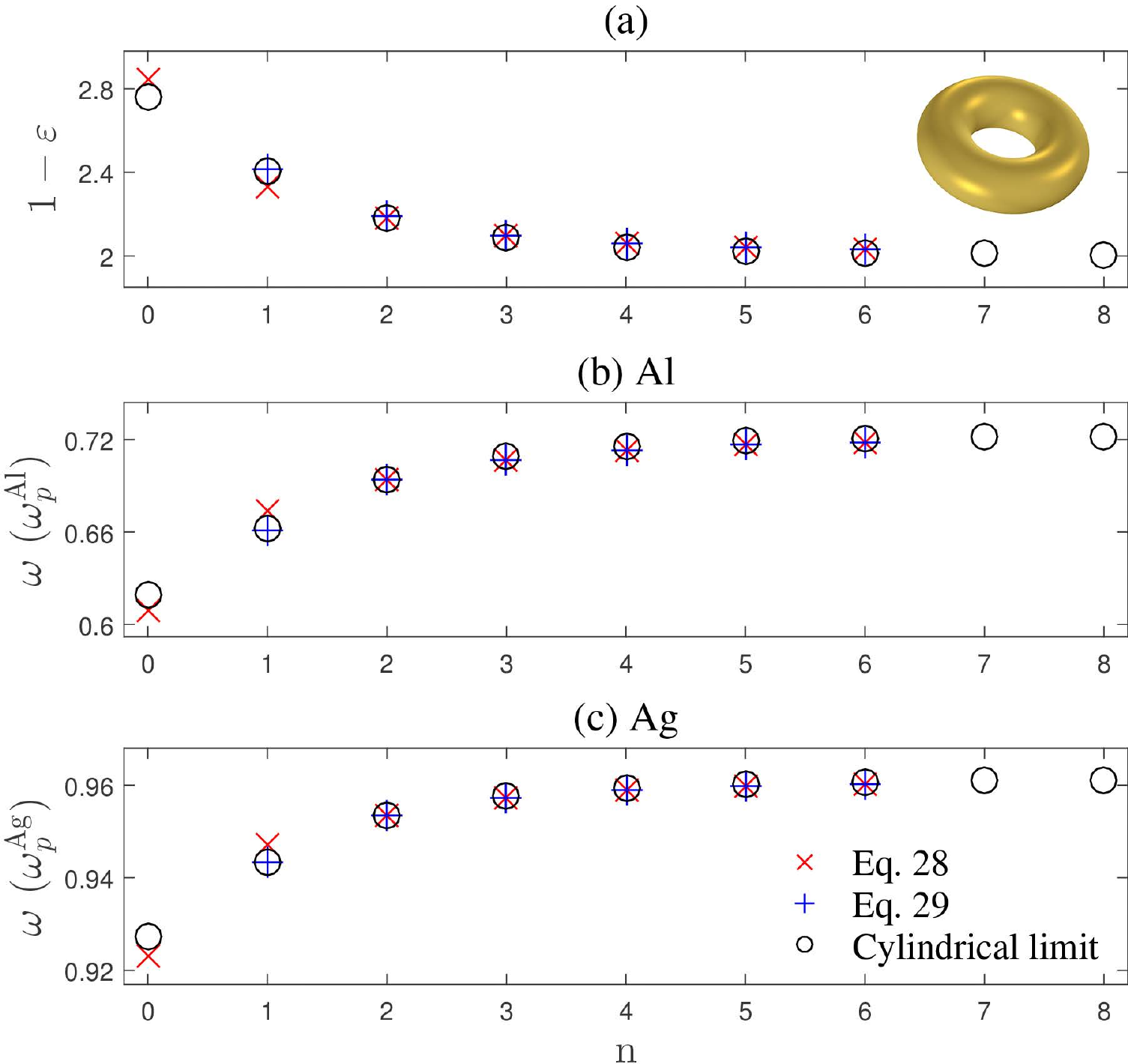} 
   		\caption[]{\footnotesize 
		\bl{Dependence of the resonance on the poloidal mode number for an isolated solid ring. 
   		 (a) Comparison using Eqs.~\eqref{Disp1}, \eqref{Disp2} with $k=1$ and $m=4$, versus the cylindrical limit.  The corresponding resonance frequencies for aluminum (b) and silver (c).
		 }}
   		 \label{levelsfreq}
\end{figure}

It is further useful to study the behavior of the modes with respect to poloidal variations. \bl{Therefore, we fix $\cosh\mu=2$ and consider a solid ring  in vacuum ($\varepsilon_2=1$) to  study the corresponding resonance values of $\varepsilon_1$ satisfying the dispersion relation  Eqs.~\eqref{Disp1}, \eqref{Disp2} with $k=1$ or, for comparison, the cylindrical limit solution~\cite{love:plasma}}. The results are shown as $1-\varepsilon$
in Fig.~\ref{levelsfreq}(a)  for various $n$ at the toroidal mode $m=4$. 
The obtained  solutions are then matched with the experimental data~\cite{palik} to determine the resonance frequency spectrum of specific materials. This is as shown in Figs.~\ref{levelsfreq}(b), \ref{levelsfreq}(c) for aluminum and silver.
Working in the cylindrical limit, it is straightforward to obtain solutions  for comparison with the exact solutions using the continued fraction Eq.~\eqref{Disp1}. Thus,  for the sake of sensing the asymptotic behavior $ n=0,1,\cdots,8$ in the cylindrical limit suffice. 
Comparison with  the exact solutions for $n =0,1,\cdots,6$  clearly reveals the convergence to the asymptotic limit.  
Clearly, while $1-\varepsilon$,  for higher values of $n$,  decreases asymptotically to  2 (and the decrease is much more rapid as $\cosh\mu \to\infty$), 
the metallic resonance frequencies grow larger for both aluminum and silver nanorings, as one may expect (implying higher energy photons for excitation of higher $n$ modes). 
\subsubsection{A singly coated toroid and its complementary void}
\paragraph{\bl{Surface modes of coated rings} in vacuum\\\\}
\bl{The specific case of a coated ring is of particular interest due to its potential use in modeling an oxide layer or a thin film or a monolayer of a nonlinear material in strong exciton-plasmon coupling studies, or for generating thin walled (shells) representing carbon based molecules or a hollow cage of atoms such as fullerenes, etc.} 

\begin{figure}[htp]
     \includegraphics[width=3.40in]{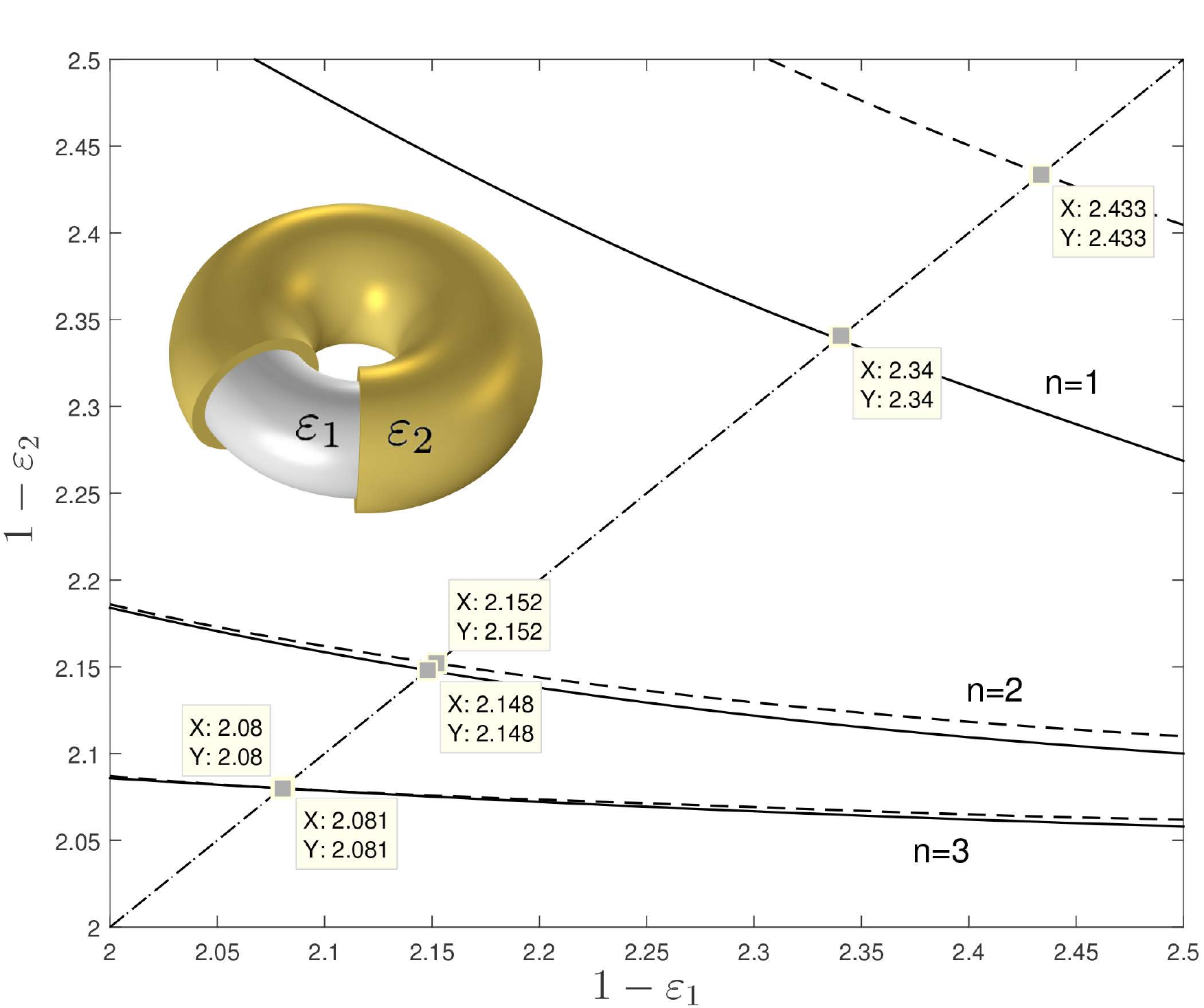} 
   		 \caption[]{\footnotesize \bl{
		 Controlling the plasmon dispersion for a singly coated ring. 		 
		 Resonance dielectric response in the $\varepsilon_1 \varepsilon_2$ plane displaying the relative variation in the dielectric functions of a singly coated ring for fixed aspect ratios. The modes are obtained from Eq.~\eqref{Disp1} (solid), Eq.~\eqref{Disp2} (dashed)  with $k=2$,  $m=2$, and $n=1,2,3$. 
The intersection of $\varepsilon_1=\varepsilon_2$ line with the displayed modes correspond to the resonance values of single ring with appropriate aspect ratio as shown in  Fig.~\ref{SolidTorus}.
}}
		 \label{EpsVer1}
\end{figure}

Plasmon excitation in a  singly coated torus embedded in an external  medium or vacuum may be studied through analysis of the dispersion relations for the surface modes in a layered torus with fixed $\mu_1$, $\mu_2$. We begin by considering variations in $\eps_1$, $\eps_2$ for 
the system in vaccum, \emph{i.e.},  $\eps_3=1$. 
The dispersion relations associated with this model are given by setting  $k=2$ in Eqs.~\eqref{Disp1} and \eqref{Disp2}. To check the correctness of the  numerical simulation  in case of a multilayered torus, we first perform a limiting study of dielectric function to compare  the dispersion relations for the  $k=2$ case  to those of the single solid torus  (with $k=1$ in Eqs.~\eqref{Disp1} and \eqref{Disp2}, respectively).
The results are shown 
\bl{in Fig.~\ref{EpsVer1} for the modes $m=2$ and $n=1,2,3$ considered} for fixed aspect ratios $\cosh\mu_1, \cosh\mu_2$. 
These solutions in the $\eps_1\eps_2$ plane are obtained by using the corresponding dispersion relations Eqs.~\eqref{Disp1} and \eqref{Disp2} with $k=2$. We then consider the values when $\eps_1=\eps_2=\eps$. This corresponds to the case of a single solid torus since the similar $\eps_1,\eps_2$ values in a $2$-layered torus corresponds to  a single solid torus with dielectric constant $\eps$ and fixed aspect ratio $\cosh\mu_2$. The common epsilon values $1-\eps_1=1-\eps_2=1-\eps$ are marked on the $\eps_1=\eps_2$ line.
The values marked on the $\eps_1=\eps_2$ line in Fig.~\ref{EpsVer1} agree well with the values generated from the solutions for a single solid ring as shown in Fig.~\ref{SolidTorus}. \bl{For example in Fig.~\ref{EpsVer1} for the mode $m=2$ and $n=1$, the common $\eps$ values are marked at $2.34$, $2.433$ for Eqs.~\eqref{Disp1},~\eqref{Disp2}, respectively. The fixed outer aspect ratio in this case is $\cosh\mu_2=1.6$. The agreement is seen in Fig.~\ref{SolidTorus}, at the fixed aspect ratio $\cosh\mu_2=1.6$ for the mode $m=2$ and $n=1$. 
It can be seen that these marked values are in agreement with the values corresponding to the single solid torus solutions selected based on the ($m,n$) at the desired aspect ratio. This is true for all other values marked in Fig.~\ref{EpsVer1}.}

\begin{figure}[htp]
     \includegraphics[width=3.3in]{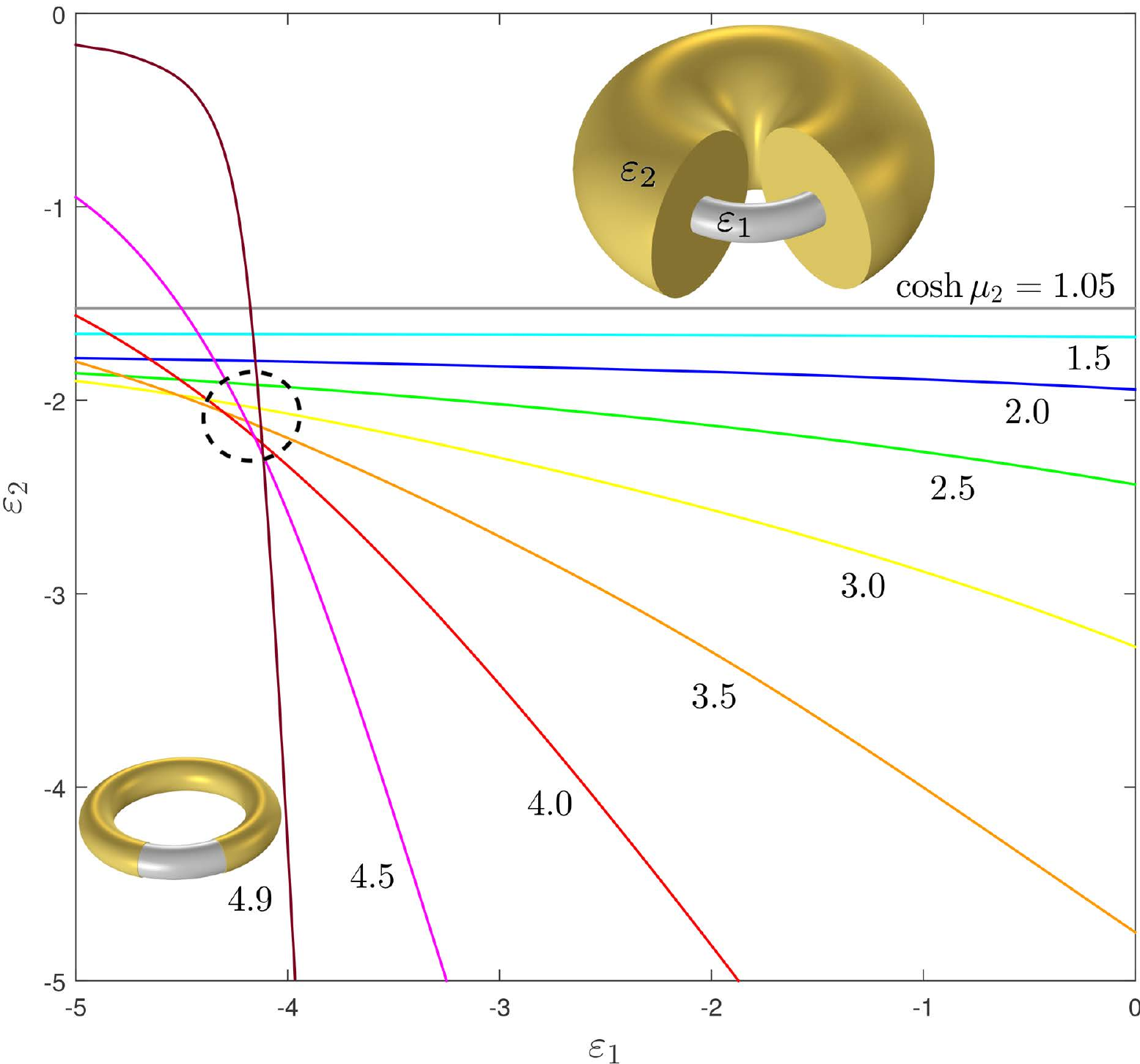} 
   		 \caption[]{\footnotesize  \bl{
		Redistribution of the eigenvalues in the   $\eps_1 \eps_2$ plane for a coated ring in vacuum. For  a fixed inner aspect ratio, varying the outer aspect ratio (1.05 - 4.9) redistributes the modes (Eq.~\eqref{Disp1} with $k=2$ for $m=4$ and $n=0$.
		For dielectric values within the domain demarcated by the dashed circle, sensitivity to aspect ratio is reduced.}}
     \label{CarbonTube1}
\end{figure}

Another interesting numerical experiment regards a singly coated torus with a fixed $\mu_1$ but varying $\eps_1$, $\eps_2$, and $\mu_2$. 
As a specific example of a physical arrangement modeled by this case, one may note a closed form of coated carbon nanotube, for which plasmon excitation may be studied in a similar fashion as that for a nanotube~\cite{stokli}.   
Thus, considering a coated ring in vacuum  ($\eps_3=1$), Fig.~\ref{CarbonTube1} shows 
 the dispersion relation (Eq.~\eqref{Disp1} with $k=2$) for mode $m=4$ and $n=0$ in the $\eps_1\eps_2$ plane for a fixed aspect ratio $\cosh\mu_1=5$. As  can be seen a redistribution of the dielectric values occur as the outer aspect ratio $\cosh\mu_2$ approaches the inner 
 $\cosh\mu_1$, that is, the $\eps_1$ range narrows down while the $\eps_2$ range broadens up within the considered dielectric constant value ranges. 

\paragraph{\bl{Surface modes of coated rings by} perturbation technique\\\\}
As in the case of a single ring,  significant simplification can be obtained when treating the eigenmodes of the  coated ring using the perturbation technique.
 \bl{Thus, writing the function $f$ in 
 Eq.~\eqref{k_layer_field1} as $f=\sqrt{u - \xi\cos\eta}$, and
 specializing Eq.~\eqref{eqn_ref} for $k=2$, and decomposing the parity of the poloidal solutions, we consider the odd surface modes, \emph{i.e.}, $\sin n\eta$ modes with fixed $m$ (the even modes $\sin n\eta$ may be treated similarly). 
At zeroth order, $\xi =0$ and the modes decouple. Let us concentrate on the $n$th mode.
Using the  conditions in Eq.~\eqref{k_bc1} and Eq.~\eqref{k_bc2} for $i=1,2$, and assuming
$\varepsilon_v=\varepsilon_1/\varepsilon_3$ and $\varepsilon_c=\varepsilon_2/\varepsilon_3$, we have:}  
\begin{align}
b_n&=d_n \frac{\sqrt{u_1}P_n^1}{\sqrt{u_2}P_n^2}
				+ e_n \frac{\sqrt{u_1}Q_n^1}{\sqrt{u_2}Q_n^2}, \\
c_n &= d_n + e_n,
\end{align}
and
\begin{align}
 &\frac{d}{du} \bigg[ \bigg( \varepsilon_c \dfrac{e_n}{\sqrt{u_2}Q_n^2}
 		 -\varepsilon_v \dfrac{b_n}{\sqrt{u_1} Q_n^1}\bigg)   
				\sqrt{u} Q_{n- \frac{1}{2}}^m (u)\notag\\
				&	+ \varepsilon_c d_n  \frac{ \sqrt{u} P_{n- \frac{1}{2}}^m (u) }{ \sqrt{u_2} 
							P_n^2} \bigg] \bigg|_{u=u_1}
  = 0,
\end{align}
\begin{align} 
&\frac{d}{du} \bigg\{ \varepsilon_c e_n   \log \left[ \sqrt{u} Q_{n- \frac{1}{2}}^m (u) \right]   \notag\\
	&	+ (\varepsilon_c d_n -c_n)  \log \left[ \sqrt{u} P_{n- \frac{1}{2}}^m (u) \right] \bigg\} \bigg|_{u=u_2}
 	 = 0.
\end{align}
This system has a solution if the determinant vanishes,
\begin{equation}
\begin{vmatrix}\label{pert_det}
0&1&-1&-1 \\
-1 & 0 &  h_{u_1}/h_{u_2} & f_{u_1}/f_{u_2}\\
0 & -Dh_{u_2} & \varepsilon_c Dh_{u_2} & \varepsilon_c  Df_{u_2}\\
-\varepsilon_v Df_{u_1}/f_{u_1} & 0 & \varepsilon_c Dh_{u_1}/h_{u_2} & \varepsilon_c Df_{u_1}/f_{u_2}
\end{vmatrix}
 = 0,
\end{equation}
where
\begin{align*}
f_u&=\sqrt{u} Q_{n- \frac{1}{2}}^m (u),\hspace{2mm} h_u=\sqrt{u} P_{n- \frac{1}{2}}^m (u),\\
Df_{u_1}&=\dfrac{d f_u}{du}\bigg|_{u=u_1}, \hspace{2mm} Dh_{u_1}=\dfrac{d h_u}{du}\bigg|_{u=u_1} , \\
Df_{u_2}&=\dfrac{d\log{f_u}}{du}\bigg|_{u=u_2}, \hspace{2mm} Dh_{u_2}=\dfrac{d\log{h_u}}{du}\bigg|_{u=u_2}.
\end{align*}
The solutions can thus be obtained following these results, \bl{as discussed for the special case of a thin toroidal membrane in the next section}.
\\
\paragraph{\bl{Plasmon hybridization and  sum rules for single toroidal shell}\\\\}
The utility of the calculated expressions describing the generalized dispersion relations can be appreciated when considering more elaborate toroidal configurations. Here, we consider a single shell cavity structure, for which we can obtain the eigenvalues corresponding to a variety of parameter configurations. 
\begin{figure}[htp]
\centering
     \includegraphics[width=3.3in]{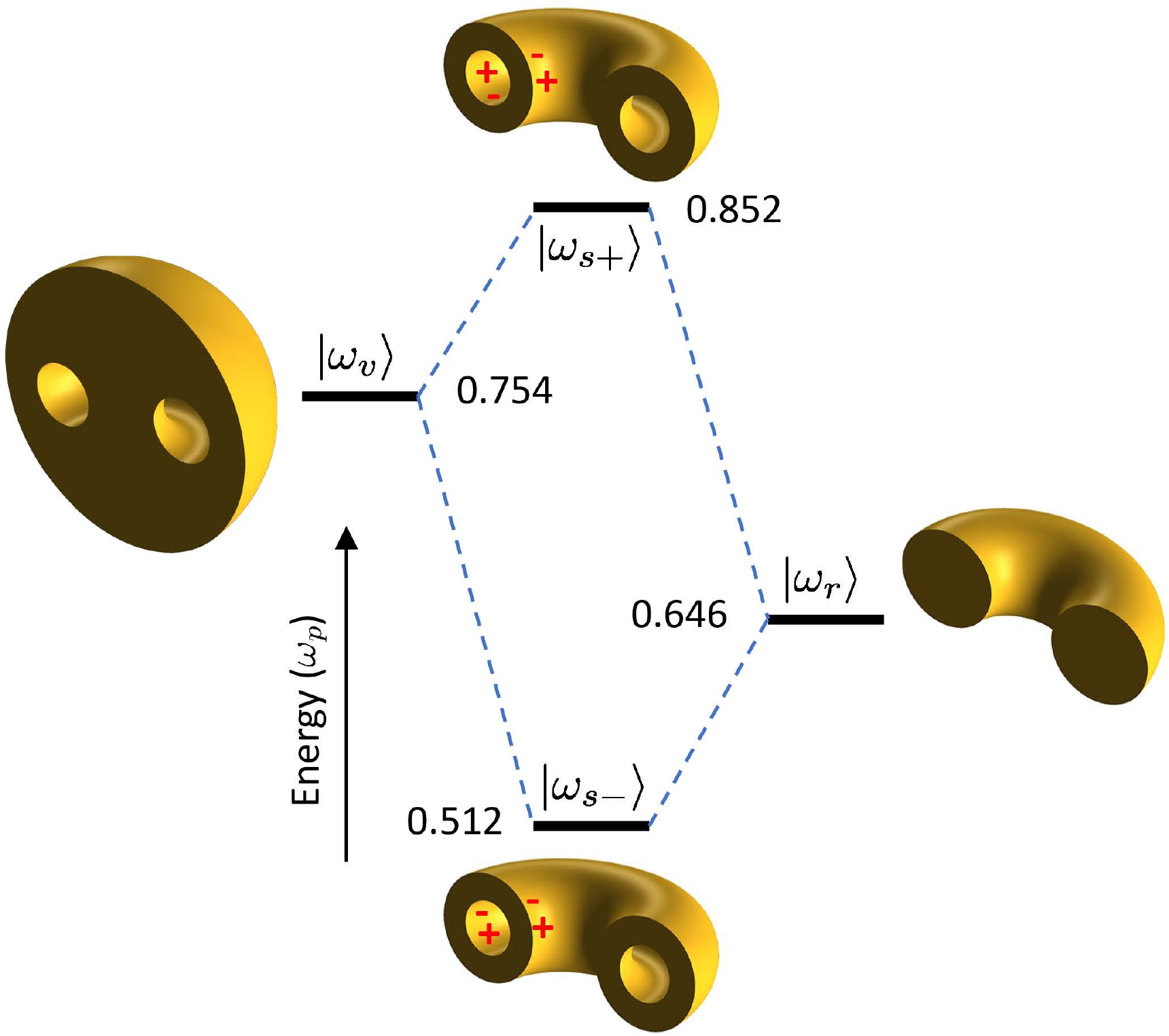} 
   		 \caption[]{\footnotesize 
		 		 \bl{Hybridized collective electronic states of a toroidal shell. The sum rule for surface plasmon frequencies in this case relate the frequencies of symmetric ($\omega_{s-}$) and antisymmetric ($\omega_{s+}$) charge oscillations on the bounding surfaces of the shell to the surface mode frequencies of the original solid ($\omega_r$) and cavity domains ($\omega_v$) making up the shell. The thinner the shell, the stronger the coupling and thus the larger the energy splitting $\Delta \omega_{s\pm}$.}}
				 \label{hybridiz}
\end{figure}
\bl{These modes describe the out-of-phase and in-phase linear combinations of the induced screening charges
 when the shell is subjected to an electromagnetic disturbance such as an incident photon, electron, or ion. Therefore, polarization effects prevent the direct application of the previously considered sum rule, that is, the sum of the square frequencies of the complementary cavity, the shell, and the solid ring does not add up to those of the square bulk $\omega^2_p$.
Instead, the frequencies of the lower energy symmetric $\omega_{s-}$ and the higher energy antisymmetric $\omega_{s+}$ modes form the new sum rule:
 \begin{equation}\label{sr2}
 \omega^2_{s+} + \omega^2_{s-} =  \omega^2_r +  \omega^2_v = \omega^2_p,
  \end{equation}
  where $ \omega_r$ and  $ \omega_v$ now correspond to the plasmon frequencies of the original surfaces making up the shell, that is, the frequencies of a solid ring with a radius $r$ equal to the radius of the outer shell surface, and a cavity with a radius equal to the radius of the shell's inner surface.
  The continued fractions in Eqs.~\eqref{Disp1} and \eqref{Disp2} for $k=2$, generate all the needed frequencies to verify the sum rule. Furthermore, within the zeroth order, one readily extracts, from Eq.~\eqref{pert_det}, a second order polynomial behavior for $\varepsilon$, from which we obtain the frequencies of the symmetric and antisymmetric charge density oscillations:   
\begin{equation}\label{pert_omega}
\omega_{s\pm} = \frac{\omega_p}{\sqrt{1+\mathcal{A}_n^m \mp \Delta_n^m}},
\end{equation}
where the energy splitting for a vacuum bounded shell  is given by:
\begin{equation}\label{pert_delta}
\Delta_n^m = \sqrt{(\mathcal{A}_n^m)^2 - \mathcal{B}_n^m},
\end{equation}
where $\mathcal{A}$ and $\mathcal{B}$ are as given in the SM\ref{F}.
}
\bl{Since the shell possesses two distinct surfaces, any excited modes will couple unless for large thicknesses of the shell so that any engendered screen charges on the inner and outer surfaces would decouple. In the latter case, the structure of the modes would be expected to approach those of either $\omega_r$ or $\omega_v$. 
These energy considerations can be captured within the plasmon hybridization picture~\cite{prodan}, as shown in Fig.~\ref{hybridiz}.  
Characterizing the plasmon mode of the bulk with the energy state $|\omega_p\rangle$, and forming a single interface, the plasmon sum rule was shown to be satisfied by the energy of the new surface modes $|\omega^{mn}_r\rangle$ and $|\omega^{mn}_v\rangle$. Forming a second interface to create the shell, the modes $|\omega^{mn}_{r,v}\rangle$ hybridize to form the states $|\omega^{mn}_{s\pm}\rangle$, as described in Fig.~\ref{hybridiz}.  For $(m,n)=(4,1)$ as an example, the numerical annotation in Fig.~\ref{hybridiz} shows, the higher mode energy of the void $\omega_v /\omega_p = 0.754 $, the lower energy of the solid ring  $\omega_r /\omega_p = 0.646$, and an energy splitting of $(\omega_{s+} -\omega_{s-})/\omega_p = 0.34$ between the hybridized states.  
Specializing the exact Eqs.~\eqref{Disp1} and \eqref{Disp2} for $k=2$, $m=3$, and $n=1,2,3$, as well as the 0-order perturbative solutions Eq.~\eqref{pert_omega}, Fig.~\ref{sr2plot} demonstrates that the plasmon sum rule is satisfied in all cases.}
\begin{figure}[htp]
     \includegraphics[width=3.40in]{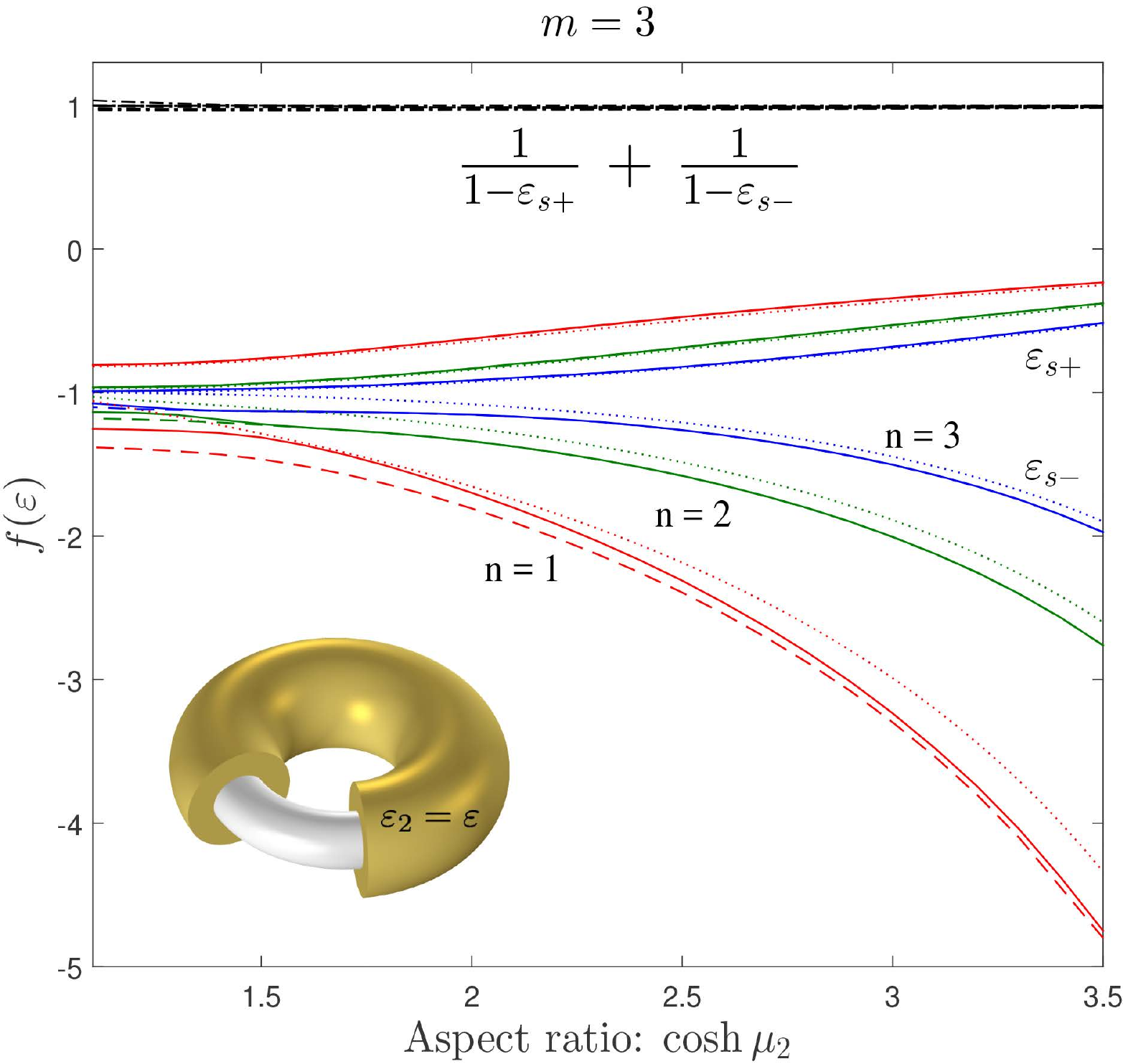} 
      \caption[]{\footnotesize 
		 		 \bl{Validation of the surface plasmon sum rule for the toroidal shell with inner aspect ratio $\cosh\mu_1=5$, employing both the exact and perturbative solutions. Solid, dashed and dotted curves are solutions to  Eq.~\eqref{Disp1},  Eq.~\eqref{Disp2}, and perturbative Eq.~\eqref{pert_omega}, respectively. The antisymmetric $\varepsilon_{s+}$, symmetric $\varepsilon_{s-}$ solutions are computed by considering $\varepsilon_1=\varepsilon_3=1$. For each $n$ and $m$, the sum rule Eq.~\eqref{sr2} is readily seen to be satisfied (see the $f(\epsilon) \approx 1$ overlapping lines) for both exact and perturbed solutions.} }
     \label{sr2plot}
\end{figure}
For shells made of specific materials,  the plasmon dispersion relations  may be obtained following the same numerical procedure as before.  For example,  for aluminum and silver shells  in vacuum, the corresponding resonance frequencies are shown for $\cosh\mu_1=5$ in Fig.~\ref{metalshell}.  \bl{Employing the presented approach, other useful shell configurations may be studied, as discussed further in SM~\ref{H}. Notably,  the case of two shells ($k=5$) is amenable to the hybridization picture, in which the plasmon modes $|\omega^{mn}_{s\pm}\rangle$ are seen to further hybridize and form states $|\omega^{mn}_{s\pm\pm}\rangle$ creating highly blue-shifted antibonding-antibonding $|\omega^{mn}_{s++}\rangle$ and red-shifted bonding-bonding $|\omega^{mn}_{s--}\rangle$ plasmon modes. 
}
\begin{figure}[htp]
     \includegraphics[width=3.40in]{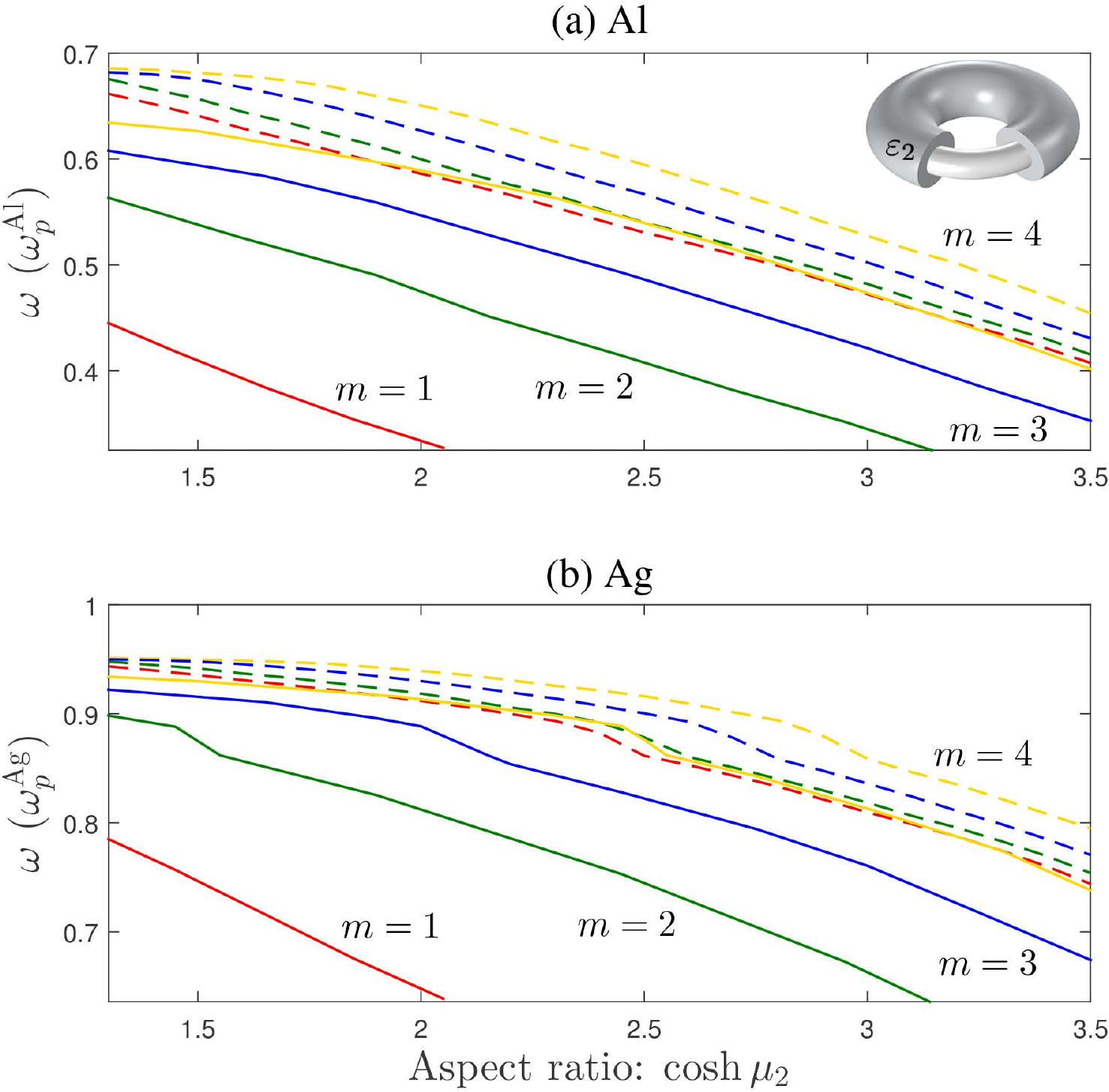} 
   		 \caption[]{\footnotesize 
   		 \bl{Comparison of plasmon dispersion relations for metallic shells. 
		  The resonance values $\eps_2$ satisfying Eq.~\eqref{Disp1} with $k=2$ for $m=1,2,3,4$ and $n=0$ (solid curves), $n=1$ (dashed curves) are displayed for a ring shell with $\eps_1=1=\eps_3$, $\cosh\mu_1=5$.
		  }}
   		 \label{metalshell}
\end{figure} 
\subsubsection{Metal-dielectric \bl{stratified toridal medium}}
\begin{figure}[htp]
     \includegraphics[width=3.40in]{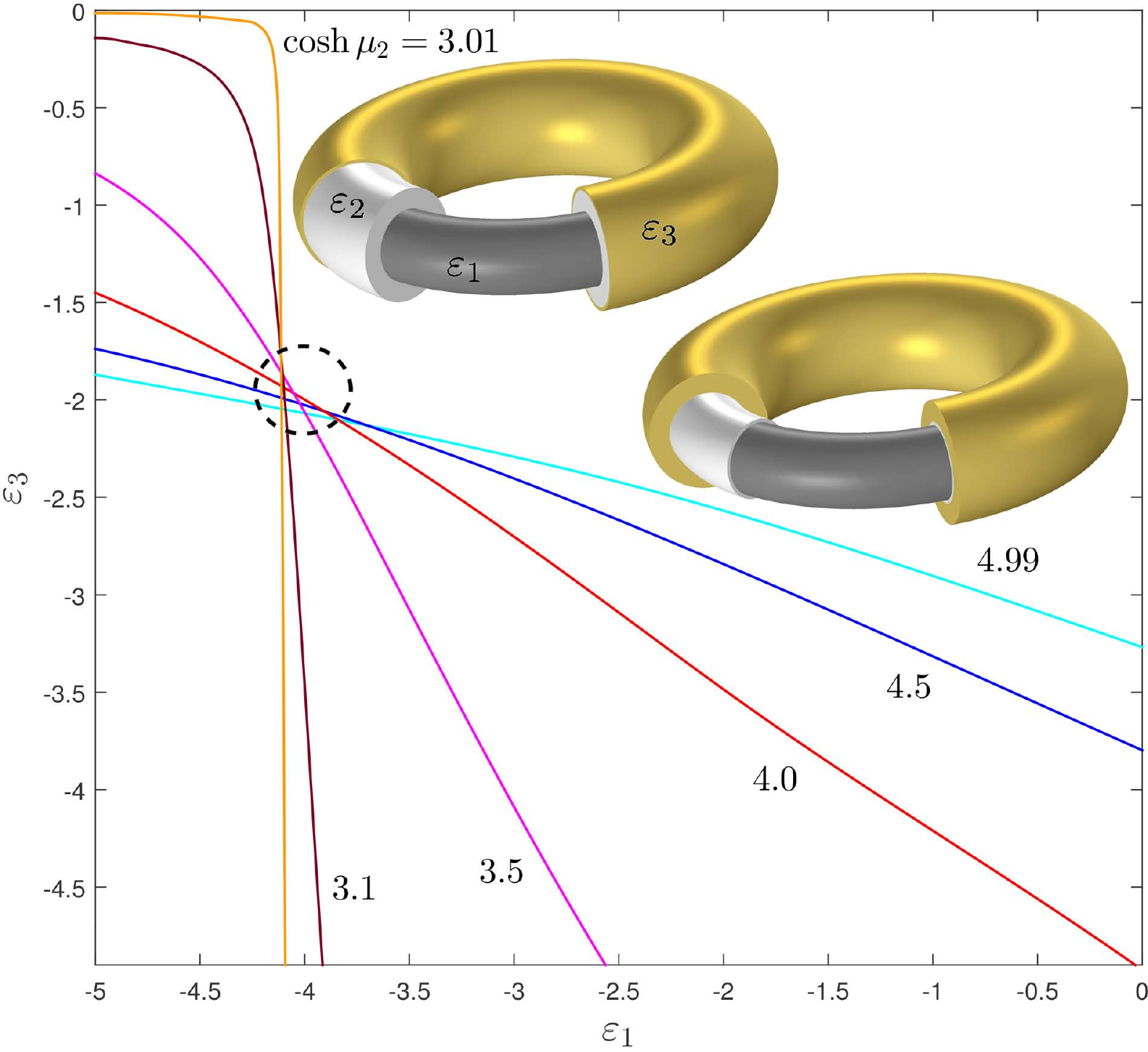} 
   		 \caption[]{\footnotesize 
   		 \bl{Thickness dependence of the resonant dielectric response of a three layered torus with fixed aspect ratios  $\cosh\mu_1=5$, $\cosh\mu_3=3$ and dielectric constants $\eps_2=\eps_4=1$.   For dielectric values within the domain demarcated by the dashed circle, sensitivity to aspect ratio is reduced. The dispersion relation is obtained from Eq.~\eqref{Disp1} with $k=3$, $m=4$, and $n=0$.}}
   		 \label{threelayer}
\end{figure}

Similar to applications capitalizing on the photon scattering and propagation properties in layered media (\emph{e.g.}, interference filters and field enhancement in dielectric or metalo-dielectric stacks~\cite{lereu}), insight into the mode structures and resonance spectra of multilayered tori can be useful. 
Using the general approach presented here, preliminary information can indeed be obtained for specific cases that are sufficiently practical to be of relevance in applications, such as, a three layered and a four layered torus. For example, with a fixed $\eps_2=1$, fixed aspect ratios $\cosh\mu_1, \cosh\mu_3$, and variable aspect ratio $\cosh\mu_2$, the dispersion relations (Eq.~\eqref{Disp1} with $k=3$) describing the toroidal mode $m=4$ and $n=0$ may be obtained for the three layered torus  in vacuum ($\eps_4=1$). 
Fig.~\ref{threelayer} displays the results in the $\eps_1\eps_3$ plane for fixed aspect ratios of $\cosh\mu_1=5, \cosh\mu_3=3$. 
It is seen that as $\cosh\mu_2 \to \cosh\mu_3$, the eigenvalues redistribute, that is, the $\eps_1$ range narrows down while that of $\eps_3$ broadens up. 
\bl{In Fig.~\ref{threelayer} one can make an additional observation that there seems to be a crossing point or small region in the neighborhood of $\eps_1 = -4$, $\eps_3 = -2$ where there is always a solution irrespective of (or weakly dependent on) the aspect ratio. This can be potentially interesting for fabrication and design consideration.}
\begin{figure}[htp]
     \includegraphics[width=3.40in]{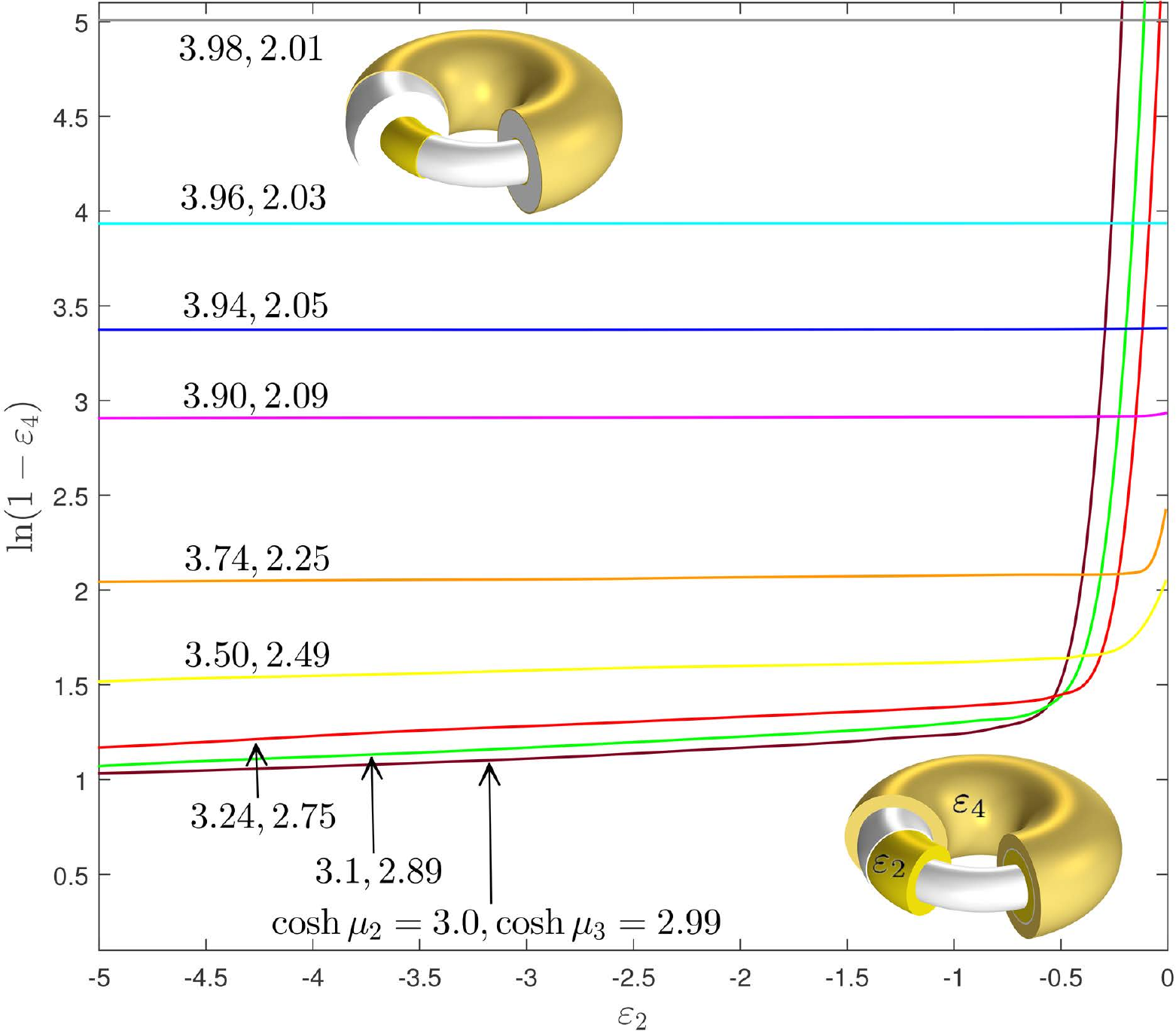} 
   		 \caption[]{\footnotesize  \bl{Interfacial control of the dielectric response of a four layered torus with fixed aspect ratios  $\cosh\mu_1=3.99$, $\cosh\mu_4=2$ and dielectric constants $\eps_1=\eps_3=\eps_5=1$.  The dispersion relation is obtained from  Eq.~\eqref{Disp1} with $k=4$, $m=4$, and $n=0$.}}
   	 \label{fourlayer1}
\end{figure}
Similarly, one may consider a four layered torus in which two concentric shells are separated by vacuum so that $\mu_1$, $\mu_4$ are fixed and  $\eps_1=\eps_3=1$. The results are shown in the $\eps_2 \eps_4$ plane for various $\cosh\mu_2$ and  $\cosh\mu_3$ values as shown in the Fig.~\ref{fourlayer1}. 
The dispersion relations were obtained from Eq.~\eqref{Disp1} with $k=4$ for the toroidal mode $m=4$ and $n=0$. 
Other effects such as  the dependence on the distance or layer thickness, \bl{as shown in Fig.~\ref{fourlayer2}(a) in SM~\ref{H}},
or obtaining positive values satisfying Eq.~\eqref{Disp1}, \bl{as shown in Fig.~\ref{fourlayer2}(b) in SM~\ref{H}} may be studied.
 As can be seen, these results suggest that the geometric parameters $\cosh\mu_i$, as well as the material parameters $\eps_i$ may be used to engineer the eigenvalue distribution that best serves a given application. 
\subsubsection{Janus rings}
\begin{figure}[htp]
  \centering
    \includegraphics[width=8.0cm]{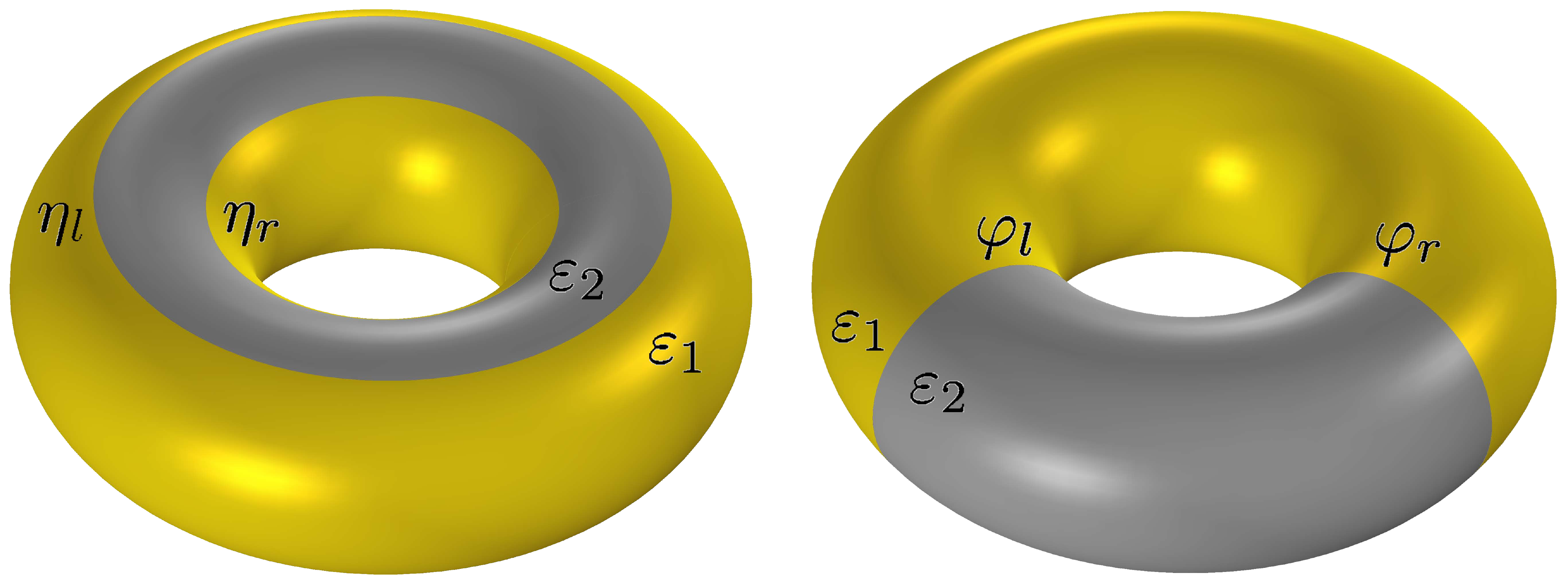} 
  \caption{\footnotesize \bl{Inhomogenous toroidal particles are formed when dissimilar material domains are interfaced. Two different filling configurations may be discerned:  the $z$-type (left) with at least one material filling the space $\eta_l \le \eta \le \eta_r$ and the $\varphi$-type (right) with at least one material filling the space $\varphi_l \le \varphi \le \varphi_r$.}}
  \label{janus0}
\end{figure}

Janus nanoparticles~\cite{honegger,zentgraf} possess  unique \bl{optical and electronic} properties that are finding important applications.  
\bl{We here introduce and analytically formulate two new janus systems with different asymmetric properties,} which have not appeared previously. \bl{The introduced janus particles may be considered special cases of the inhomogeneous toroidal domain shown in 
Fig.~\ref{janus0}.}
 \bl{For the sake of illuminating their electromagnetic response, we here suffice with a computational determination} of their fields as a full treatment of their surface modes are beyond the scope of the present article, \bl{for reasons noted below}.   

\begin{figure}[htp]
  \centering
    \includegraphics[width=8.0cm]{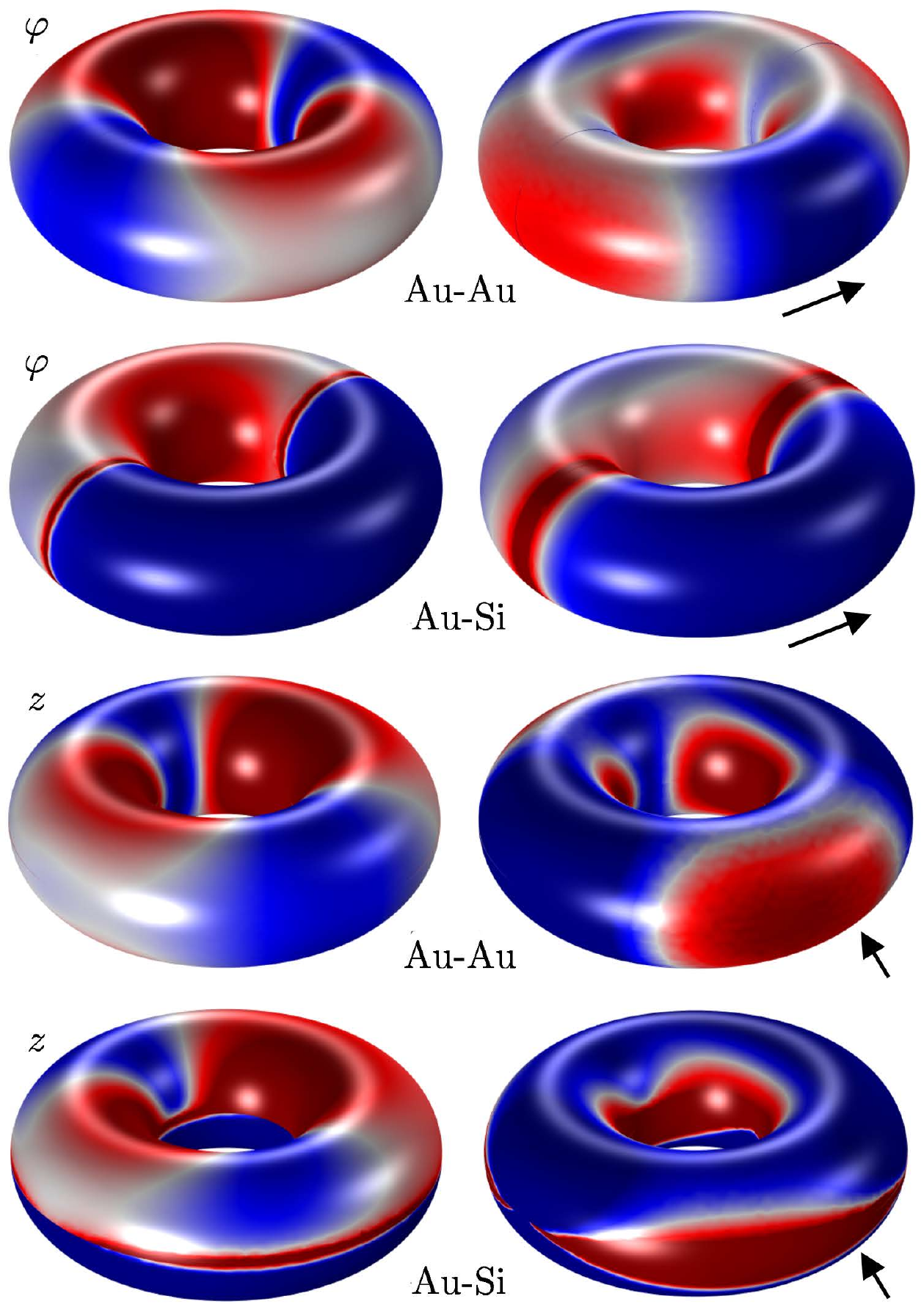} 
      \caption{\footnotesize \bl{Janus nanorings are formed when two dissimilar material domains are interfaced at $z=z_0$ ($z$-type), or at $\varphi = \varphi_0$ and  $\varphi = \pi + \varphi_0$ ($\varphi$-type). Here,  the retarded responses of Au-Si and Au-Au janus rings to linearly polarized photons of wavelength $\lambda =$652.6~nm and polarization indicated by the arrows are visualized by the norms of the induced surface  current density (left) and field distribution (right). The localization of the fields and currents can readily be observed around the transition regions.}}
  \label{janus}
\end{figure}

The $z$-type janus ring is \bl{formed when $\eta_r - \eta_l = \pi$, that is, composition of two half rings  described with dielectric functions $\eps_1(\omega)$ and $\eps_2 (\omega)$,} respectively, as depicted in Fig.~\ref{janus0} and Fig.~\ref{janus}. 
To formally define the problem, we consider the potentials inside and outside such a janus torus:  
\begin{align}\label{superposed}
\Phi_{\text{in}}^1 & =  f(\mu,\eta) \  \sum_{m=-\infty}^\infty \sum_{n=-\infty}^\infty
							D_{mn}^1Q_{n-\frac12}^m(\cosh \mu) \notag\\
													& \times e^{i(n\eta+m\varphi)}, 
											\	\mu \ge \mu_{1},\ 0\le\eta<\pi, \notag\\
\Phi_{\text{in}}^2 & =  f(\mu,\eta) \  \sum_{m=-\infty}^\infty \sum_{n=-\infty}^\infty
							D_{mn}^2Q_{n-\frac12}^m(\cosh \mu) \notag\\
													& \times e^{i(n\eta+m\varphi)}, 
											\	\mu \ge \mu_{1},\ \pi\le\eta<2\pi, \notag\\
\Phi_{\text{out}} & =   f(\mu,\eta) \  \sum_{m=-\infty}^\infty \sum_{n=-\infty}^\infty
									C_{mn}P_{n-\frac12}^m(\cosh \mu)\notag \\
									& \times e^{i(n\eta+m\varphi)},
											\	0 \le \mu \le \mu_{1}.\notag\\
\end{align}
The boundary conditions which $\Phi_{\text{in}}^1$, $\Phi_{\text{out}}^2$ must satisfy on $\mu_{1} \le \mu < \infty$ reduce to
\begin{align}\label{bc_janus}
\left. \Phi_{\text{in}}^1\right|_{\eta=0/\varphi=0} &= \left. \Phi_{\text{in}}^2\right|_{\eta=0/\varphi=0},\\
\label{bcc_janus}
	\eps_1 \left.  \frac{\partial \Phi_{\text{in}}^1}{\partial n}\right|_{n=z=0/n=y=0}
			&= \eps_2 \left.  \frac{\partial \Phi_{\text{in}}^2}{\partial n}\right|_{n=z=0/n=y=0}.
\end{align}
While on $\mu=\mu_1$ the potential equations $\Phi_{\text{in}}^1$, $\Phi_{\text{in}}^2$ and $\Phi_ {\text{out}}$ must satisfy
\begin{align}
	\label{bc1_janus}
		\left. \Phi_{\text{in}}^i\right|_{\mu=\mu_1} &= \left. \Phi_{\text{out}}\right|_{\mu=\mu_1}, \,\, i=1, 2,\\
	\label{bc2_janus}
		\eps_i \left.  \frac{\partial \Phi_{\text{in}}^i}{\partial \ch \mu}\right|_{\mu=\mu_1}
			&= \eps_{3} \left.  \frac{\partial \Phi_{\text{out}}}{\partial \ch \mu}\right|_{\mu=\mu_1}, \,\, i=1, 2.
\end{align}
The $y$ and $z$ are as given in Eq.~\eqref{yz_coeff} in SM~\ref{B}. In case of $z$-type janus rings we choose $\eta=0$, $z=0$.
\bl{To obtain the normal modes and resonance frequencies associated with the charge density oscillations,} it would be prudent to first establish whether for each $m=0,1,2,\cdots,$ the functions $\big\{Q_{n-\frac{1}{2}}^m\big\}_{n=0}^\infty$ are pairwise orthogonal in $L^2(\sqrt{\cosh\mu-1} d\mu)$, the space of square integrable functions with respect to the measure $\sqrt{\cosh\mu-1} d\mu$. To the best of our knowledge, such an orthogonality relation is not known. 
Also care must be taken with singularities that may arise from \bl{the boundary condition Eq.~\eqref{bcc_janus}, see also Eq.~\eqref{diff_yz}-\eqref{diff_tor} in SM~\ref{C}.}
However, bypassing the analytical complexity, we may obtain the eigenmodes and field distributions  computationally employing FEM or the FDTD. 
For example, for \bl{janus rings} made up of gold~\cite{palik} and silicon~\cite{palik}, employing FEM \bl{to solve the frequency domain form of Maxwell's wave equation}, we  computationally visualize the induced field and current distributions for an incident \bl{various} linearly polarized fields of wavelength \bl{652.6}~nm, as shown in \bl{Fig.~\ref{janus}. The field and current density confinement and localization properties of the janus effect are clearly seen from the comparison to the uniform Au ring for two polarization.} 

\bl{The dissimilarity in the material domains in the $\varphi$-type janus ring takes the form of two half rings with transition regions satisfying
 $\varphi_r - \varphi_l = \pi$, and described by dielectric functions $\eps_1(\omega)$ and $\eps_2(\omega)$,} respectively, as depicted in \bl{Fig.~\ref{janus0}, and Fig.~\ref{janus}}.
In Eq.~\eqref{superposed} we now have $0\le\varphi<\pi$ and $\pi\le\varphi<2\pi$. To satisfy the boundary conditions, we consider $\varphi=0$, $y=0$. 
 The same arguments as in the case of $z$-type janus rings  would be true for $\varphi=0$ case as well, but now one needs to establish the orthogonality condition for functions $\big\{Q_{n-\frac{1}{2}}^m\big\}_{m=0}^\infty$ with $n$ fixed. 
 Altogether Eqs.~\eqref{bc_janus}--\eqref{bcc_janus} and \eqref{bc1_janus}--\eqref{bc2_janus} shall result in three difference equations which can be transformed into a matrix vector product form as in Eq.~\eqref{WnJn}. One may now follow the same procedure described in section II to obtain proper dispersion relations with needed symmetry checks performed.
For \bl{the $\varphi$-type janus rings, field modifications under the same conditions as the $z$-type, are shown in Fig.~\ref{janus}. Noting that the contrasts have been adjusted individually in Fig.~\ref{janus} for visualization purposes, the silicon domain of the Au-Si janus rings (both types) is marked with minimum field and current when compared with the uniform material (Au-Au) cases. Following the above formulation, the plasmon dispersion relations may be obtained.}  
\section{Eigenmodes and field distribution}
Nanofocusing and field gradient engineering in multilayer rings are of particular interest due to toroidal and poloidal geometric features of the metallic, dielectric, and metallo-dielectric transitions. In the quasi-static limit, photon scattering by the multilayered torus can be formulated as an inhomogeneous Neumann problem. Here, considering  a multilayer nanoring, we obtain a three-term difference amplitude equation with source terms characterizing the external radiation. We will then employ a Green function approach to solve for the  amplitudes and thus the potential and fields.
We will now  proceed to obtain the appropriate Green's function for the general case of a multilayer ring.

\subsection{Green function for \bl{the response of a stratified toroidal medium to nonuniform fields}}
In order to study the scattering properties of the studied system, we begin by treating a $k$-layered ring in presence of electric monopoles and dipoles of charge $q_s$  at $(\mu_{0,s},\eta_{0,s},\varphi_{0,s})$ outside the toroidal boundary $\mu=\mu_k > \mu_{0,s}$. Expanding the electrostatic Green function for free space and superimpose the potential of $q_s$ to the exterior potential of the source-free ring, we can write: 
\bl{ \begin{align}\label{k_layer2}
		 \Phi_j &=  f(\mu,\eta) \sum_{m=0}^\infty \sum_{n=-\infty}^\infty \sum_s \notag\\ &
				\times \Bigl [ (1-\delta_{j,1})    C^{'\text{j}}_{mn,s} P_{n-\frac12}^m(\cosh \mu)
					 e^{in(\eta-\eta^j_{_{mn,s}})} \notag\\
					&	+ (1-\delta_{j,k+1}) D^{'\text{j}}_{mn,s} Q_{n-\frac12}^m(\cosh \mu)e^{in(\eta-\eta'^{j}_{_{mn,s}})}\notag\\
						&	+\delta_{j,k+1}      Q_{n-\frac12}^m(\cosh \mu)   
								 K^{'}_{mn,s} e^{in(\eta-\eta_{0,s})}\Bigr ]  \notag\\
									& \times   \cos{m(\varphi-\varphi_{0,s})},
\end{align}}
where $s$ counts the number of charges, $\delta$ denotes Kronecker delta, $j=1,\cdots,k+1$ and \bl{$\eta^j_{_{mn,s}}$, $\eta'^{j}_{_{mn,s}}$} are phase factors that are not necessarily coherent with $\eta_{0,s}$ and
\begin{align}\label{charge}
		K^{'}_{mn,s}&=\dfrac{1}{4\pi\eps_0\eps_{k+1}}\dfrac{q_s}{\pi a}f(\mu_{0,s},\eta_{0,s})(2-\delta_{m0})	\notag\\
			&	\times \dfrac{\Gamma(n-m+1/2)}{\Gamma(n+m+1/2)} P_{n-\frac12}^m(\cosh \mu_{0,s}),
\end{align}
and where $\eps_0$ is the permittivity of free space.
As in the case of multiple charges~\cite{kuk:cpc}, Eq.~\eqref{k_layer2} together with the Green's function discussed in this section, allows us to compute inside and outside potentials in case of a multilayer nanoring structures, when multiple charges are placed outside the toroidal boundary $\mu=\mu_k$. 

One can evaluate the potential represented in Eq.~\eqref{k_layer2} by computing $C^{\text{j}}_{mn}$, $D^{\text{j}}_{mn}$ for individual charges. Thus in the following equations we adopt the notation $C^\text{j}_{mn}=C^{'\text{j}}_{mn,s}e^{-in\eta^j_{_{mn,s}}},D^\text{j}_{mn}=D^{'\text{j}}_{mn,s}e^{-in\eta'^j_{_{mn,s}}},K_{mn}=K^{'}_{mn,s}e^{-in\eta_{0,s}}$. We suppress index $m$ in the same way as described in Section II.A. Introducing  the $k \times k$ matrix 
\begin{equation}\label{K_k}
\small
\mathbb{K}_n=
	\begin{bmatrix}
	0&  & &0 \\[0.3em]
	&  \ddots & &\\[0.3em]
	&   & 0& \\[0.3em]
	0	&  & & Q_n^k	\\[0.3em]
	\end{bmatrix} 
	,
\end{equation}
and the $k\times 1$ vector 
\begin{equation}\label{M_k}
\small
\mathcal{K}_n=
	\begin{bmatrix}
	0 \\[0.3em]
	\vdots \\[0.3em]
	0 \\[0.3em]
	K_n
	\end{bmatrix}
	,
\end{equation}
and applying the boundary condition Eq.~\eqref{k_bc1}, one obtains the relation
\begin{equation}\label{GC_n}
		\mathcal{C}_n= \mathbb{P}_n^{-1} [\mathbb{Q}_n\mathcal{D}_n+\mathbb{K}_n\mathcal{K}_n].
	\end{equation}
\bl{
Using the substitution \eqref{GC_n} into the second boundary condition \eqref{k_bc2} and after some algebra~\cite{pap1}, one can show that 
$\mathcal{D}_n$ satisfies the vector three-term  recurrence relation:
\begin{equation}\label{threeterm_2}
\begin{split}
		\mathcal{W}_{n+1}-\text{R}_n\mathcal{W}_{n}+\mathcal{W}_{n-1}&=
			\mathcal{V}_{n+1}-\text{D}_{\mu}\mathcal{V}_{n}+\mathcal{V}_{n-1},\\
\end{split}
\end{equation}
for $n=0,\pm1, \pm2, \dots,$ where
\begin{align}
	\mathcal{V}_n&=( \mathbb{K}'_n \mathrm{E}_2-
				\mathbb{P}'_{n}\mathrm{E}_2\mathbb{P}_{n}^{-1}\mathbb{K}_{n})\mathcal{K}_n,
\end{align}
and all other used notations, except for $\mathcal{C}_n$ given by Eq.~\eqref{GC_n}, are in accordance with the ones already introduced in Section II.A. We note that the three-term  recurrence \eqref{threeterm_2} holds for each fixed $m=0,\pm1, \pm2, \cdots$. 
}

\bl{
Generalizing Green's function method~\cite{love:jmp,kuk}, we obtain $\mathcal{W}_{n}$ in Eq.~\eqref{threeterm_2} by noting that  
	\begin{equation}\label{Green_Sol}
\mathcal{W}_n=\sum\limits_{N=-\infty}^{\infty}G_{n,N}\mathcal{V}_N,
	\end{equation}
forms a solution to Eq.~\eqref{threeterm_2} provided that for each $N$, the $k \times k$ matrices $G_{n,N}$ satisfy
\begin{align}\label{Green}
		G_{n+1,N}-&\mathrm{R}_n G_{n,N}+ G_{n-1,N} \notag\\
		&	=\delta_{n,N+1} I- \delta_{n,N} \text{D}_{\mu}
				+\delta_{n,N-1}I,
\end{align}
for $n=0,\pm 1, \pm 2,\dots,$ where $I$ denotes the $k \times k$ identity matrix. Using Kronecker delta's definition at $n=N-1,\ N, \ N+1$, Eq.~\eqref{Green} can be written as:
\begin{align}
G_{N,N}-\mathrm{R}_{N-1} G_{N-1,N}+ G_{N-2,N} &=I, \label{GN-1} \\
G_{N+1,N}-\mathrm{R}_N G_{N,N}+ G_{N-1,N} &= -\text{D}_{\mu}, \label{GN} \\
G_{N+2,N}-\mathrm{R}_{N+1} G_{N+1,N}+ G_{N,N} &=I. \label{GN+1} 
\end{align}
For other $n$ values, we have
\begin{align}
	G_{n+1,N}-\mathrm{R}_n G_{n,N}+ G_{n-1,N} &=0 \  \text{   for  }  \  n \geq N+2, \label{GN+2}\\
	G_{n+1,N}-\mathrm{R}_n G_{n,N}+ G_{n-1,N} &=0 \  \text{   for  }  \  n \leq N-2. \label{GN-2}
\end{align} 
}

\bl{
Suppressing the index $N,$  the three term matrix recurrences \eqref{GN+2} and \eqref{GN-2} display a single recurrence
\begin{equation}\label{XN}
	X_{n+1}-\mathrm{R}_{n} X_{n}+  X_{n-1} =0, 
\end{equation}
where either $n\geq N+2$ or $n\leq N-2$.  Since the MCF
\begin{equation}\label{MCFN+2}
	\frac{I}{\text{R}_{N+2}} 
		\cminus \frac{I}{\text{R}_{N+3}}
			\cminus \frac{I}{\text{R}_{N+4}} \cminus \contdots 
\end{equation}
is the tail of the converging MCF~\eqref{MCF1}, it follows from the italic statement preceding Eq.~\eqref{DRG} that the three-term recurrence \eqref{GN+2} has a minimal solution. Letting $G_{n,N}$ denote this minimal solution, Eq.~\eqref{DRG} yields
	\begin{equation}\label{G_1}
G_{n+1,N}G_{n,N}^{-1}=
	\frac{I}{\text{R}_{n+1}} 
		\cminus \frac{I}{\text{R}_{n+2}}
			 \cminus \contdots,	
				\quad n \geq N+1. \\
\end{equation}
Denoting the right-hand side of \eqref{G_1} by $\boldsymbol{\gamma}_{n+1},$ we get
\begin{equation}\label{GNN+2}
	G_{n+1,N} =  \boldsymbol{\gamma}_{n+1} G_{n,N} \quad \text{ for } \quad  n \geq N+1.
\end{equation}
}
\bl{
For the backward recurrence \eqref{XN} with $n\leq N-2,$ a similar argument presented in section II.B.1.a together with the substitution 
$V_{n} = X_{n-1} X_{n}^{-1}$ gives the following first order non-linear matrix recurrence: 
	\begin{equation}\label{Vt}
		V_{n+1} = \frac{I}{\text{R}_{n} - V_{n}}.
	\end{equation}
Using the fact $\text{R}_n=\text{R}_{-n},$  the backward MCF:
$$\frac{I}{\text{R}_{N-2}} 
		\cminus \frac{I}{\text{R}_{N-3}}
			\cminus \frac{I}{\text{R}_{N-3}} \cminus \contdots
$$
equals
$$\frac{I}{\text{R}_{-N+2}} 
		\cminus \frac{I}{\text{R}_{-N+3}}
			\cminus \frac{I}{\text{R}_{-N+4}} \cminus \contdots.
$$
The last MCF converges using the same argument  discussed above  for the MCF \eqref{MCFN+2}. Thus, the backward recurrence has a minimal solution. Denoting this minimal solution by  $G_{n,N},$ it follows from Eq.~\eqref{Vt} that
\begin{equation}\label{G_2}
G_{n-1,N}G_{n,N}^{-1}=
	\frac{I}{\text{R}_{n-1}} 
		\cminus \frac{I}{\text{R}_{n-2}}
			 \cminus \contdots,
				\quad n \leq N-1. \\
\end{equation}
Similarly, denoting the right-hand side of \eqref{G_2} by $\boldsymbol{\zeta}_{n-1},$ we get
 \begin{equation}\label{GNN-2}
	G_{n-1,N} =  \boldsymbol{\zeta}_{n-1} G_{n,N} \quad \text{ for } \quad  n \leq N-1.
\end{equation}
}
\bl{
Setting $n=N+1$ and $n=N-1$ in \eqref{GNN+2} and \eqref{GNN-2}, respectively, one may now find  $G_{N+1,N},$ $G_{N,N},$ and $G_{N-1,N}$ via 
Eqs.~\eqref{GN-1}--\eqref{GN+1}  as~\cite{kuk}:
\begin{align}
G_{N-1,N} &=\frac{\boldsymbol{\zeta}_{N-1}}
					{\text{R}_N-\boldsymbol{\gamma}_{N+1}-\boldsymbol{\zeta}_{N-1}}
						(\text{D}_{\mu}-\text{R}_N),\label{N-1N}\\
G_{N,N} &=\frac{I}{\text{R}_N-\boldsymbol{\gamma}_{N+1}-\boldsymbol{\zeta}_{N-1}}
			(\text{D}_{\mu}-\boldsymbol{\gamma}_{N+1}-\boldsymbol{\zeta}_{N-1}), \label{NN}\\ 
G_{N+1,N} &=\frac{\boldsymbol{\gamma}_{N+1}}
						{\text{R}_N-\boldsymbol{\gamma}_{N+1}-\boldsymbol{\zeta}_{N-1}}
							(\text{D}_{\mu}-\text{R}_N).\label{N+1N}
\end{align}
}
\bl{
Finally, for each fixed $n,$ one can use Eqs. \eqref{GNN+2} and \eqref{GNN-2} together with Eqs.~\eqref{N-1N}--\eqref{N+1N} to obtain  
$\mathcal{W}_n$ via \eqref{Green_Sol}. 
Since $\mathrm{J}_n$ in Eq.~\eqref{WnJn} is invertible, one can rewrite Eq.~\eqref{WnJn} in terms of $\mathcal{D}_n$ as
     \begin{equation}\label{Dn_1}
	 \mathcal{D}_n=\mathrm{J}_n^{-1}\mathcal{W}_n.
	 \end{equation}
Thus one can obtain the $k\times 1$ vector $\mathcal{D}_n$ as considered in Eq.~\eqref{C_k}. On the same lines, since $\mathbb{P}_n$ is invertible, one can use Eq.~\eqref{GC_n} to obtain another $k\times 1$ vector $\mathcal{C}_n$ as presented in Eq.~\eqref{C_k}. The obtained vector values can now be substituted into the potential Eq.~\eqref{k_layer2}.
}
\subsection{Green function for the difference equation of the multilayer nanoring in a uniform field}
\subsubsection{Polarization parallel to the symmetry axis of the ring}
The general form of the potential for a uniform field polarized parallel to $z$ axis outside the toroidal boundary $\mu=\mu_k$ is given by 
\begin{align} \label{k_layer_field2}
		 \Phi_j &=  f(\mu,\eta) \displaystyle\sum_{n=-\infty}^\infty 
				 \Bigl[(1-\delta_{j,1})C_{n}^{\text{j}} P_{n-\frac12}(\cosh \mu)\notag\\
					&	+ (1-\delta_{j,k+1})D_{n}^{\text{j}} Q_{n-\frac12}(\cosh \mu) \notag\\
					&		+\delta_{j,k+1}K_{n} Q_{n-\frac12}(\cosh \mu)\Bigr] 
								e^{in\eta},
\end{align}
where $j=1,\cdots,k+1$ and $\delta$ denotes Kronecker delta and
\begin{equation}\label{K2}
     K_{n}=\dfrac{\sqrt{8}aE_0}{i\pi}n,
\end{equation}
similar to the uniform electrostatic field defined in Love~\cite{love:jmp}.
Because of the axial symmetry in case of a uniform field along $z$ axis, the potential is independent of the coordinate $\varphi$. Hence we do not have $m$-sums in Eq.~\eqref{k_layer_field2}. Also there are no phase differences in the $\eta$ solutions and hence they are just given by $e^{in\eta}$ everywhere.
One could now follow the same procedure defined through Eqs.~\eqref{K_k}--\eqref{N+1N} to obtain Eqs.~\eqref{N-1N}--\eqref{N+1N} but with $K_{n}$ 
as defined in Eq.~\eqref{K2}. The rest of the procedure to obtain $\mathcal{D}_n$ and $\mathcal{C}_n$ is discussed following Eq.~\eqref{N+1N}.
\subsubsection{Polarization not parallel to the symmetry axis of the ring}
The general form of the potential corresponding to a field with a   $x-z$ polarization  axis outside the toroidal boundary $\mu=\mu_k$ is given by 
\begin{align} \label{k_layer_field3}
			 \Phi_j &=  f(\mu,\eta) \displaystyle\sum_{n=-\infty}^\infty \sum_{m=0}^1 
				 \Bigl[(1-\delta_{j,1})C_{mn}^{\text{j}} P_{n-\frac12}^m(\cosh \mu)\notag\\
					&	+ (1-\delta_{j,k+1})D_{mn}^{\text{j}} Q_{n-\frac12}^m(\cosh \mu)\notag\\
						&	+\delta_{j,k+1}K_{mn} Q_{n-\frac12}^m(\cosh \mu)\Bigr]
						e^{in\eta}  \cos{m\varphi},
\end{align}
where $j=1,\cdots,k+1$ and $\delta_{j,k+1}$ denotes Kronecker delta and
\begin{align}\label{K3}
	 K_{mn} &=  \frac{\sqrt{2}a}{\pi}\mathcal{E}_{m+1} (-i)^{m-1} 
		 				(1+\delta_{m0})(1+m\delta_{n0})\notag\\
							&\times \left(n^{m+1} - \frac14 m\right)
								\frac{\Gamma(n - m + \frac12)}{\Gamma(n + m + \frac12)}.
\end{align}
The derivation of the applied potential 
\begin{align*}
\Phi_{ap}=f(\mu,\eta)\sum_{n=-\infty}^\infty \sum_{m=0}^1 K_{mn} Q_{n-\frac12}^m(\cosh \mu)
						e^{in\eta}  \cos{m\varphi},
\end{align*}
in Eq.~\eqref{k_layer_field3} is discussed in SM~\ref{B}. The method to obtain Green's function and coefficients in Eq.~\eqref{k_layer_field3} follows the same lines to the procedure discussed in previous subsections. 
\subsection{Poloidal and toroidal \bl{modes} in response to specific excitations}\label{LDOS}
\subsubsection{\bl{Response of a vacuum bounded solid ring to nonuniform fields}}
\begin{figure}[H]
\begin{center}
     \includegraphics[width=3.4in]{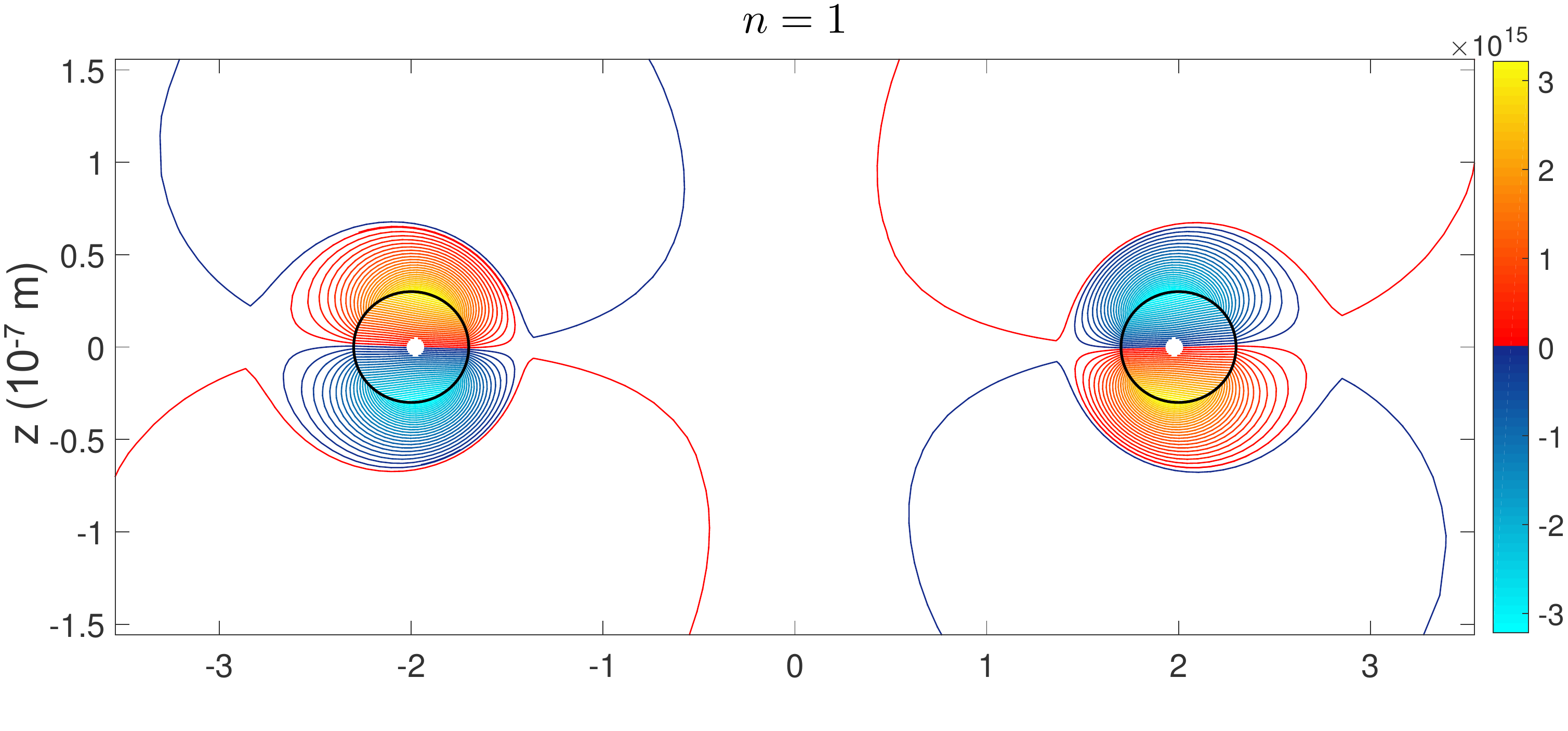}\\
     \includegraphics[width=3.4in]{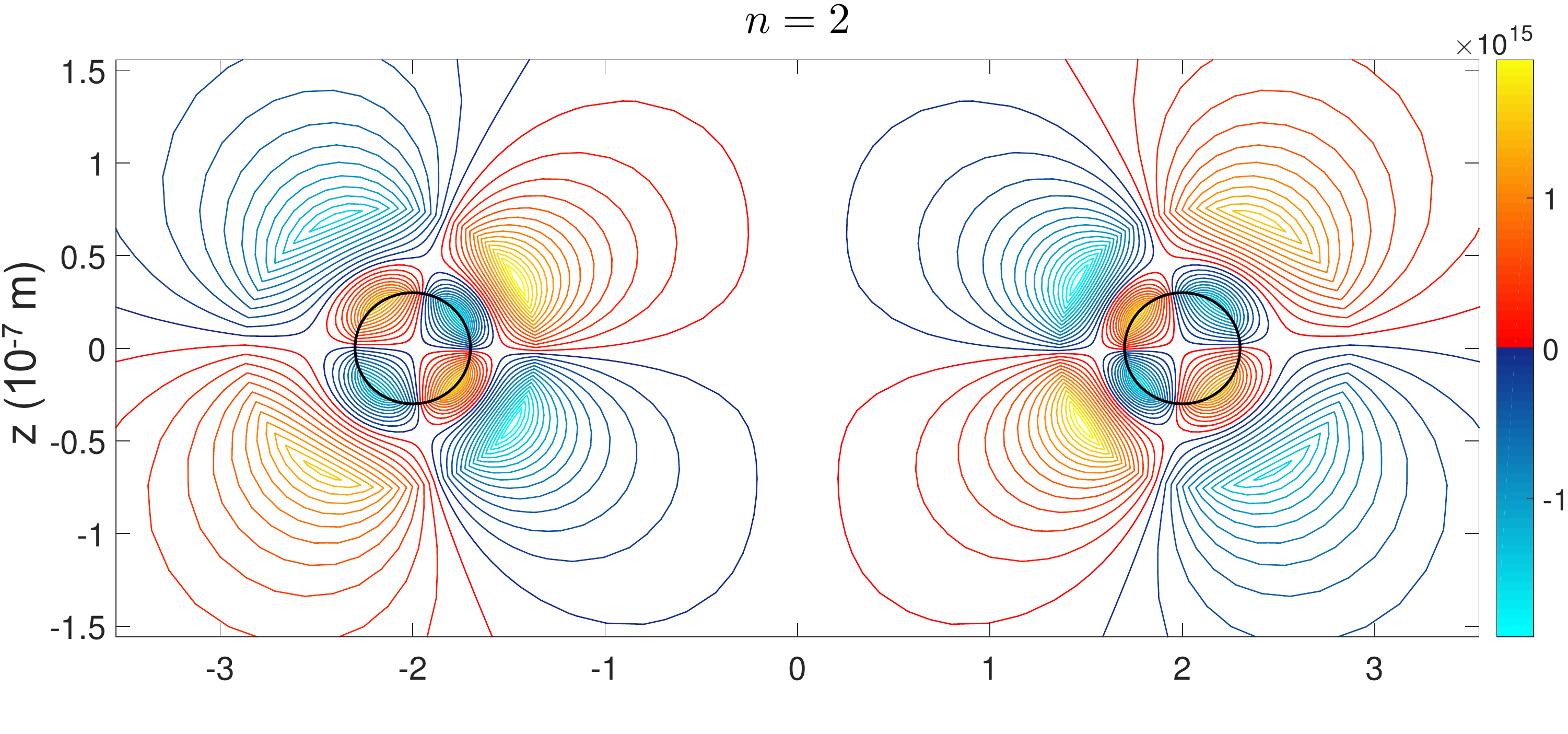}\\
      \includegraphics[width=3.4in]{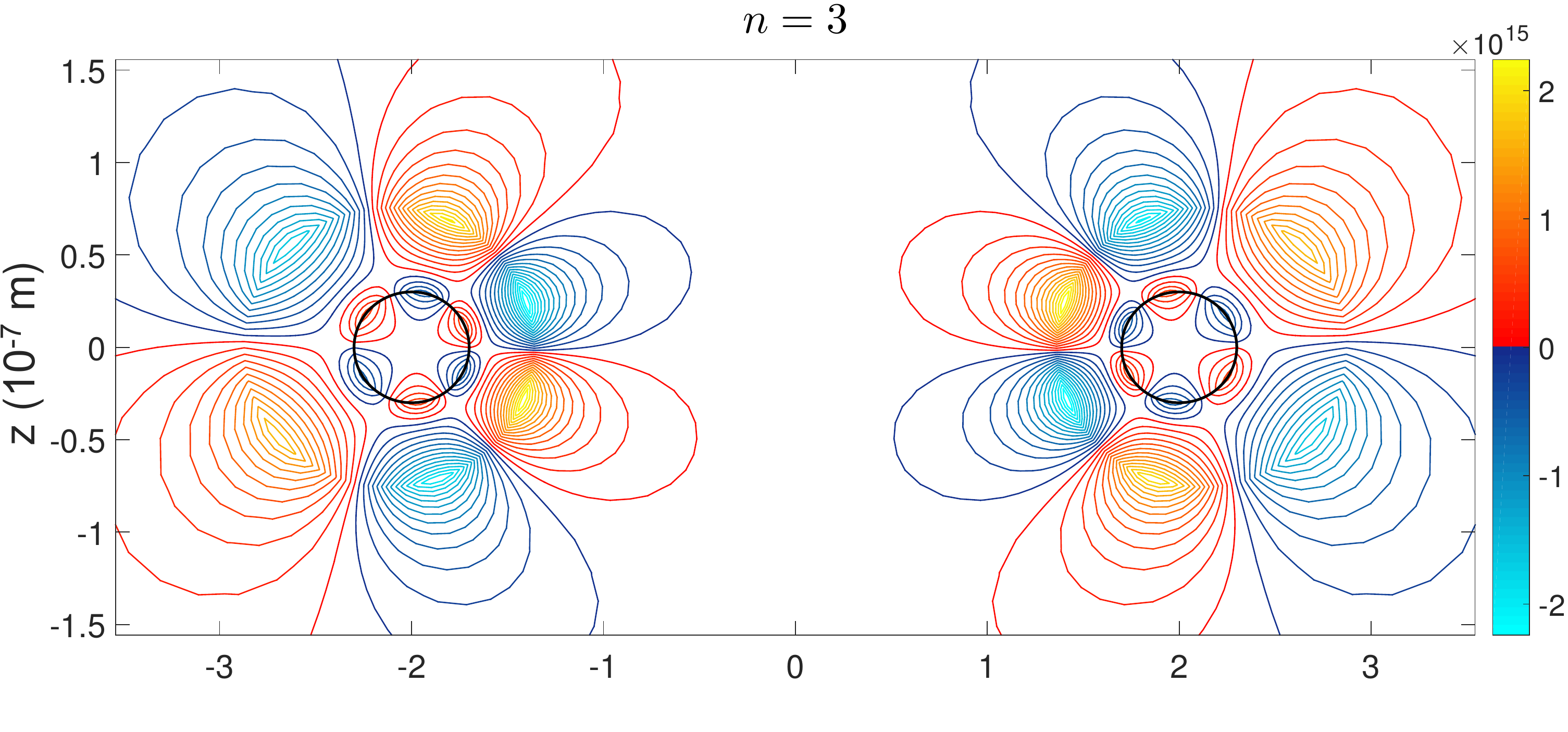}\\
       \includegraphics[width=3.4in]{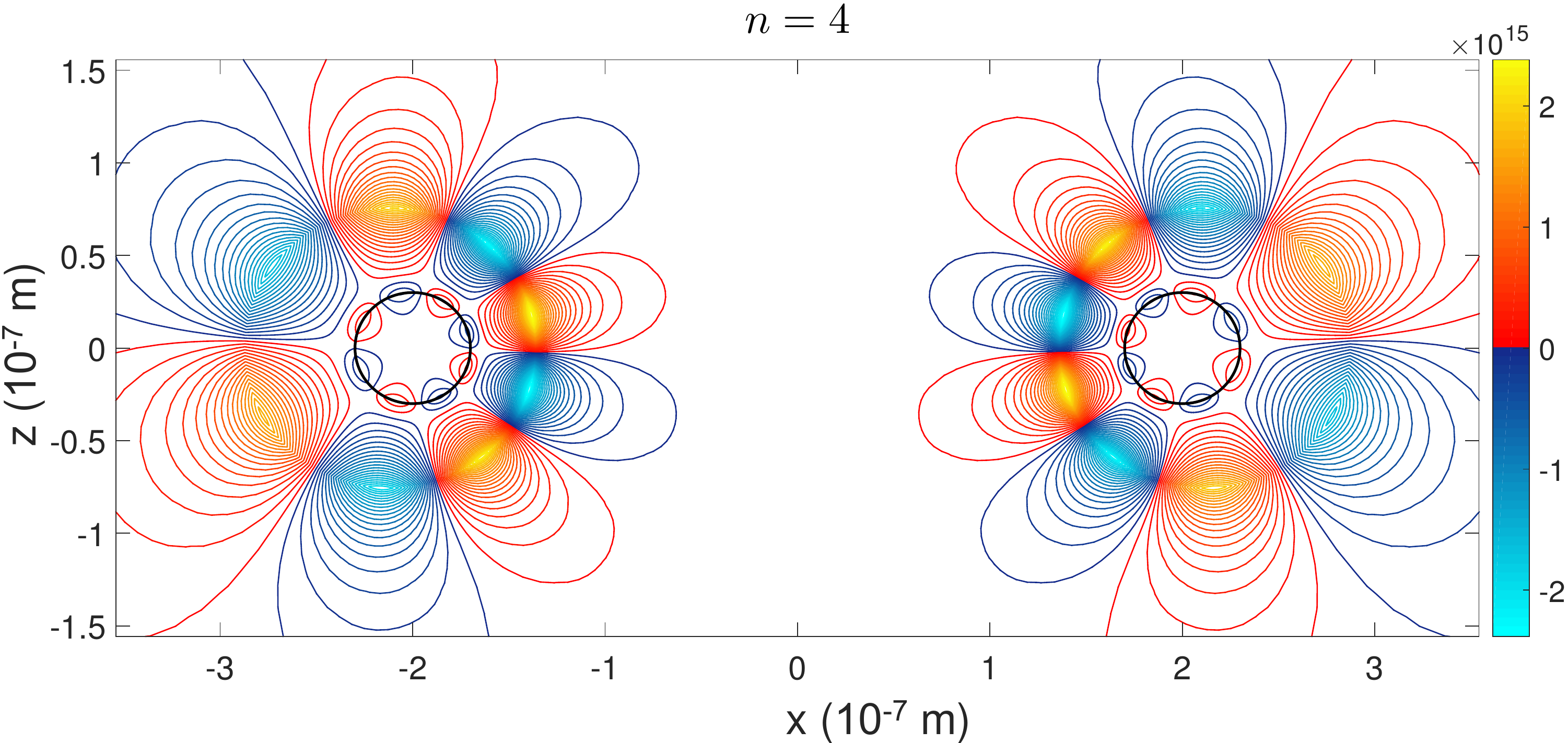}\\
     \caption[]{\footnotesize \bl{Poloidal eigenmodes of a vacuum bounded solid ring. Simulation was made for a  ring with $a=\SI{0.1977}{\micro\meter}$,  $\cosh{\mu_1}=6.6667$, $\eps_1=-1.5$, and $\eps_2=1$.   
  The visualization was made for angles  $\varphi=0,\pi$ and $0 < \eta < 2\pi$.
The column displays the $m=1$ and $n=1,2,3,4$ modal potential distributions developed in response to a nearby dipolar emitter at $\eta_0=0, 5$ with $\cosh{\mu_0}=2.8$, $\varphi_0=0$}.}
     \label{dipole_m1_n1}
   	\end{center}
\end{figure}

\bl{The torus develops induced surface charges in  response to the field of an external charge such as an electron at $(\mu_0,\eta_0,\varphi_0)$ outside the toroidal boundary $\mu=\mu_1>\mu_0$, which can be calculated 
after solving the second-order difference equation \eqref{threeterm_2} using the Green function with $k=1$. The combined potentials now read:}
\begin{align}
 \label{point_charge_in}
\Phi_{1} & =  f(\mu,\eta) \  \sum_{m=0}^\infty \sum_{n=-\infty}^\infty
				\sum_{N=-\infty}^\infty
							d{_{m,n,N}}Q_{n-\frac12}^m(\cosh \mu) \notag\\	
											& \times	\cos{m(\varphi-\varphi_0)}
													 e^{in\eta}, 
											\	\mu \ge \mu_{1},\\
\label{point_charge_out}
\Phi_{2} & =   f(\mu,\eta) \  \sum_{m=0}^\infty \sum_{n=-\infty}^\infty
									\bigl[\bigl(	\sum_{N=-\infty}^\infty d{_{m,n,N}}
							- K_{mn}\bigr)\notag\\
							& \times	P_{n-\frac12}^m(\cosh \mu)
								\dfrac{Q_n^{1}}{P_n^{1}}
									+ K_{mn}
										Q_{n-\frac12}^m(\cosh \mu)\bigr] \notag\\
										& \times \cos{m(\varphi-\varphi_0)}
													 e^{in\eta} ,
														\	0 \le \mu \le \mu_{1},
\end{align}
where
\begin{align}\label{dnN}
 d{_{m,n,N}} &= \dfrac{\lambda_{_{N}}^m}{\Pi_{n}^{m}(\mathrm{R}_N^m-\gamma_{_{N+1}}^m-\zeta_{_{N-1}}^m)}\notag\\
	& \times \bigg[\delta_{_{n,N}}(2\cosh{\mu_1}-\gamma_{_{N+1}}^m-\zeta_{_{N-1}}^m) \notag\\
 			&	+(2\cosh{\mu_1}-\mathrm{R}_N^m)
 					\bigg(\theta(n-N)\prod_{k=N+1}^n\gamma_{k}^m \notag\\
 					&	+\theta(N-n)\prod_{k=n}^{N-1}\zeta_{k}^m\bigg)\bigg],\notag\\
\Pi_{n}^m&=\eps_1Q_n^{'1}-\eps_2P_n^{'1}
					Q_n^{1}/
						P_n^{1},\notag\\
\lambda_{n}^{m}&=\eps_2(Q_n^{'1}
						 -P_n^{'1}
								Q_n^{1}/P_n^{1})
										K_{mn},\notag\\
\end{align}
and \bl{$K_{mn}=K^{'}_{mn}e^{-in\eta_0}$ with $K^{'}_{mn}$ as defined in Eq.~\eqref{charge}} and $\theta(x)=1$ if $x>0$, and 0 otherwise. 
\bl{The total potential obtained by summation of $m$ and $n$ modes as in Eqs.~\eqref{point_charge_in} and \eqref{point_charge_out} is shown in Fig.~\ref{point_charge} in SM~\ref{G}. The poloidal and toroidal mode potential distribution induced by a nearby dipole using Eq.~\eqref{k_layer2} by considering only relevant modes is shown in Fig.~\ref{dipole_m1_n1}. 
The potential (encompassing contributions from toroidal and poloidal modes  in Eq.~\eqref{k_layer2}) is shown in Fig.~\ref{arb_charge} and in the inset of Fig.~\ref{ldos}. 
The plotting procedure for these figures is briefly discussed in SM~\ref{G}.
}
\bl{The utility of the results may be appreciated if we consider forming a dipole of moment $p$ and from its particular placements at ${\bf r}_0 = (\mu_0,\eta_0,\varphi_0)$ and orientation build the dyadic Green's function:
$$\mathcal{G} (r,r_0) = \nabla \Phi (r) c^2\eps/p\omega^2,$$ 
where $c$ is the speed of light and $\omega$ is the frequency at which $\eps$ is specified. 
\begin{figure}[H]
   \begin{center}
       \includegraphics[width=3.4in]{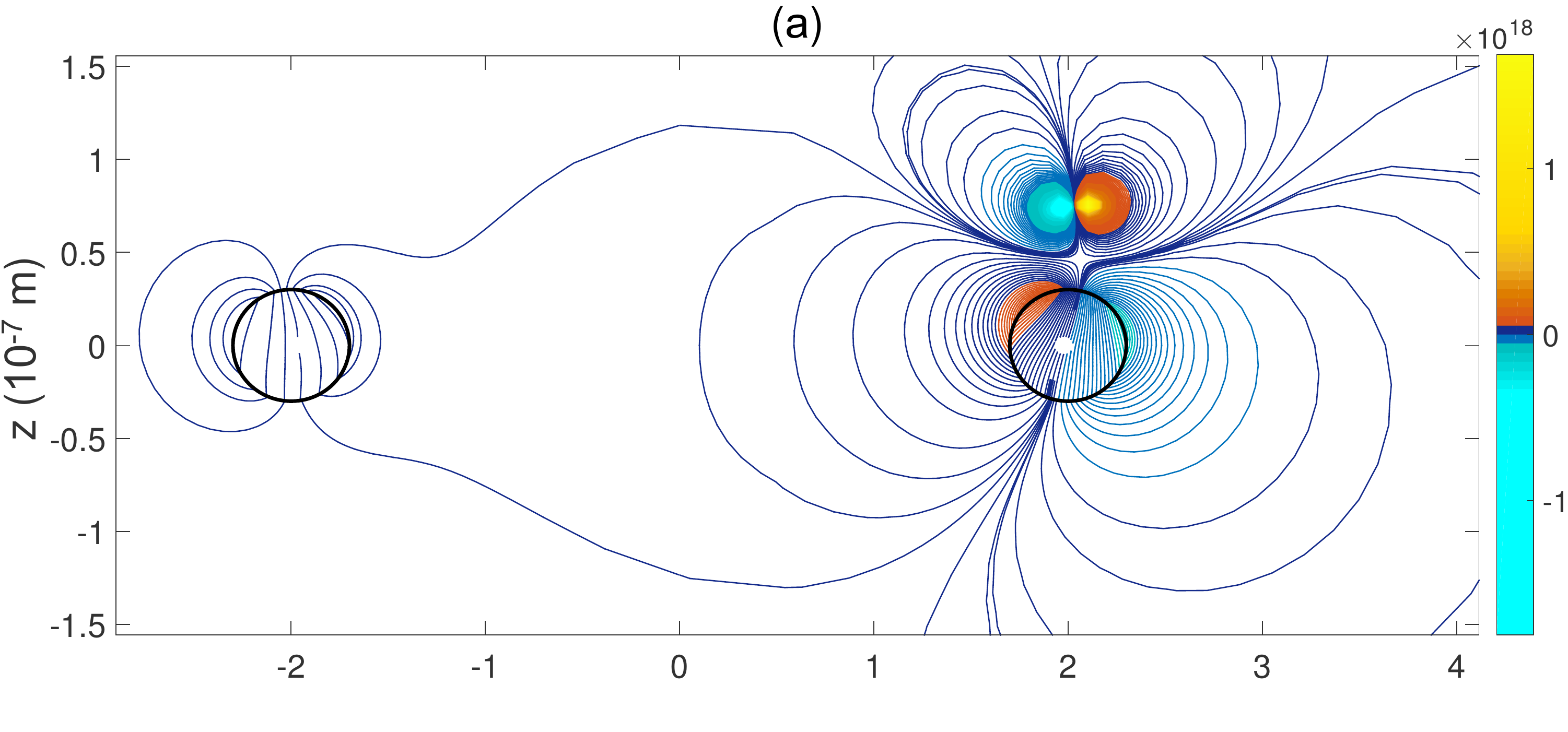}\\
       \includegraphics[width=3.4in]{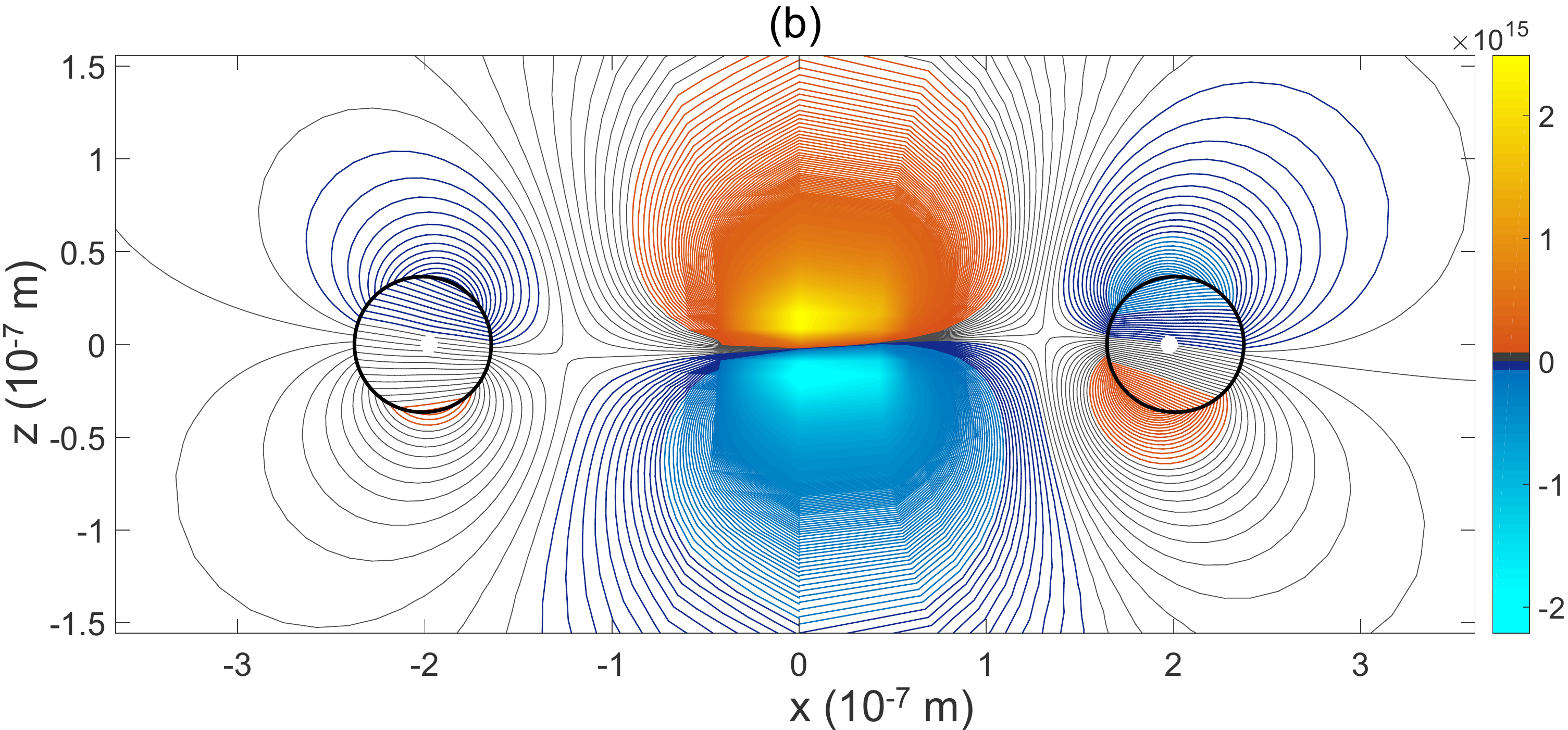}\\
     \caption{\bl{Quasi-static potential distribution representing the response of a metal nanoring to a nonuniform field of a dipolar emitter located (a) away from the ring, and (b) close to the origin.
The result is obtained via Eq.~\eqref{k_layer2} by summation of poloidal and toroidal modes for a solid ring with $a=\SI{0.1977}{\micro\meter}$,  $\cosh{\mu_1}=\text{(a)} \ 6.6667$, $\text{(b)} \ 5.5$, $\eps_1=-1.5$, and $\eps_2=1$.}}
\label{arb_charge}
\end{center}
\end{figure}
\begin{figure}[htp]
\begin{center}
     \includegraphics[width=3.4in]{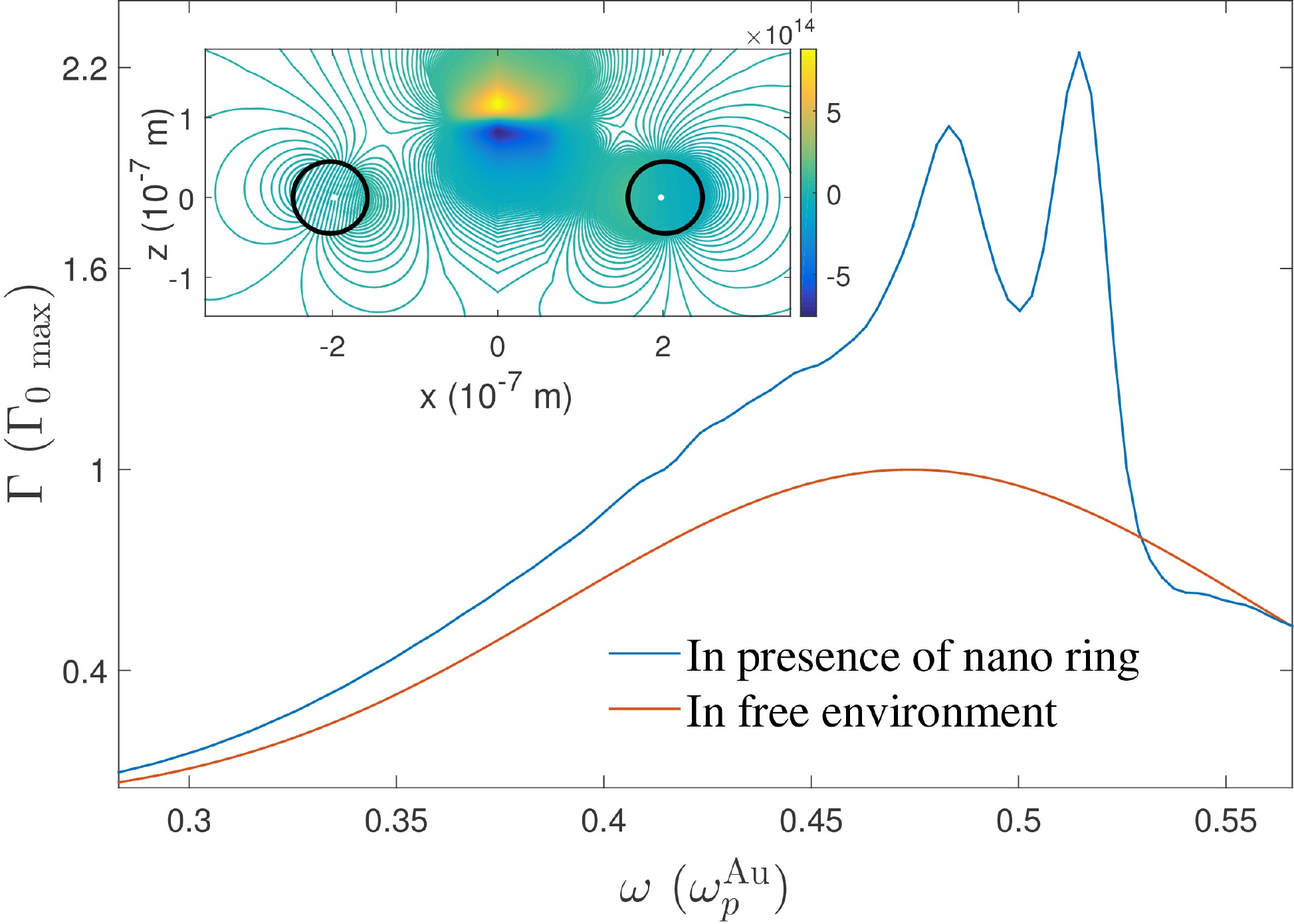} 
   		 \caption[]{\footnotesize  
\bl{Modification of  the spontaneous decay rate $\Gamma$ of a two-level quantum system due to the presence of a metal torus 100~nm below it.The higher decay rates corresponding to the surface modes of the ring in response  to the nonuniform field of the dipole with a moment $\bar{d}$ parallel to the symmetry axis of the gold ring
are clearly observable. The decay rate without the ring has been normalized with $\Gamma_{0\text{max}}$.The inset depicts the quasi-static potential distribution representing the response of a metal nanoring to a nonuniform field of a dipolar emitter located $\SI{100}{\nano\meter}$ above the $z=0$. The result, obtained via Eq.~\eqref{k_layer2} by summation of poloidal and toroidal modes for a solid ring with $a=\SI{0.1977}{\micro\meter}$,  $\cosh{\mu_1}=4.5$, $\eps_1=-1.5$, and $\eps_2=1$, may be used to calculate the dyadic Green's function.}}
     \label{ldos}
     \end{center}
\end{figure}
For example, setting ${\bf r}_0 $ to form a dipole parallel to the symmetry axis of the ring, then the partial ($z$) local density of states (LDOS), which is directly proportional to the spontaneous decay rate $\Gamma$, 
can be obtained from 
$$\text{LDOS}_z = 6\omega \Im \mathcal{G}_{zz}   /\pi^2 = \Im \left( \frac{6\epsilon c^2 \nabla_z \Phi}{p\omega\pi^2}   \right ) = \alpha \Gamma,$$ 
where $\alpha = 3\hbar/2\omega p^2$. Fig.~\ref{ldos} shows the modification of the radiative decay rate due to the presence of the gold ring.  The higher local density of states corresponding to the resonant response of the ring to the two polarization states of the exciting field (parallel and perpendicular to the symmetry axis of the ring) is clearly observable.  
Employing Eq.~\eqref{k_layer2} to obtain the response of the ring to an electric dipole with an arbitrary moment $\bar{d}$ we may thus obtain the partial local density of states.  
}
\subsubsection{\bl{Response of a vacuum bounded solid ring to uniform fields}}
In case of potential for a uniform field polarized along the $z$-axis, the corresponding potentials $\Phi_1$ and $\Phi_2$ for the regions inside and outside $\mu=\mu_1$ respectively, after solving second-order difference equation \eqref{threeterm_2} using the Green function with $k=1$ are given as
\begin{align}
\label{uniform_field_in}
\Phi_1  &=  f(\mu,\eta) \sum_{n=-\infty}^\infty
				\sum_{N=-\infty}^\infty
							d_{_{n,N}}Q_{n-\frac12}(\cosh \mu)
											 e^{in\eta},\\
											  & \hspace{50mm}\mu \ge \mu_{1},
\notag\\
\label{uniform_field_out}
\Phi_2 &=   f(\mu,\eta) \sum_{n=-\infty}^\infty
									\bigl[\bigl(\sum_{N=-\infty}^\infty d_{_{n,N}}
							- K_{n}\bigr)
							   P_{n-\frac12}(\cosh \mu)\notag\\
							&\times	\dfrac{Q_n^{1}}{P_n^{1}}
									+ K_{n}
										Q_{n-\frac12}(\cosh \mu)\bigr]
										   e^{in\eta}, \hspace{4mm}
											 0 \le \mu \le \mu_{1},
\end{align}
where \bl{$d_{_{n,N}}$ is as shown in Eq.~\eqref{dnN} but with $m=0$ considered} and $K_n$ is as defined in Eq.~\eqref{K2}.
\bl{
The toroidal and poloidal mode potential distribution obtained by considering relevant modes in Eqs.~\eqref{uniform_field_in}--\eqref{uniform_field_out} is further discussed in SM~\ref{G}.}
\subsection{Scattering cross section}
\begin{figure}[htp]
\centering
     \includegraphics[width=3.40in]{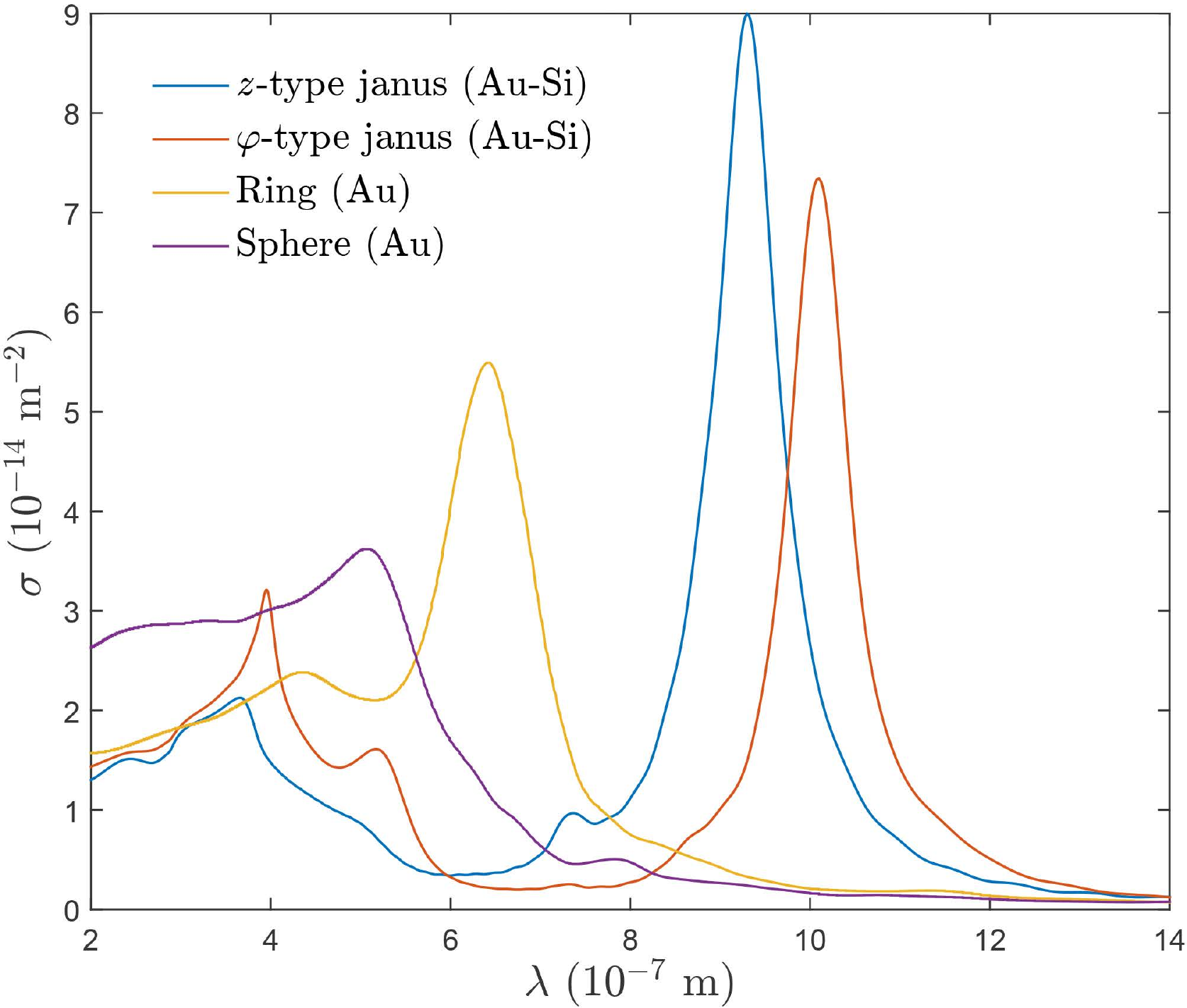} 
   		 \caption[]{\footnotesize 
		 \bl{Spectral dependence of plasmon excitation. Absorption cross sections were obtained for a gold ring $(r=25, R=50)$~nm as  compared to that of both  $z$- and $\varphi$-type gold/silicon janus rings and a gold sphere $(r=50)$~nm.
		 }}
		 \label{cross}
\end{figure}

To calculate the scattering cross section for a single ring, here we shall consider \bl{
the equations of a uniform field of an arbitrary polarization (subject to  the  conditions Eqs.~\eqref{k_bc1}, \eqref{k_bc2} with $k=1$)}. Similar procedure discussed in section III.A is followed and finally after solving the second-order difference equation Eq.~\eqref{threeterm_2} using the Green function with $k=1$, the corresponding potentials $\Phi_{1}$ and $\Phi_{2}$ for the regions inside and outside $\mu=\mu_1$ respectively, are obtained:
\begin{align}
 \label{xyz_in}
\Phi_{1} & =  f(\mu,\eta) \  \sum_{m=0}^1 \sum_{n=-\infty}^\infty
				\sum_{N=-\infty}^\infty
							d{_{m,n,N}}Q_{n-\frac12}^m(\cosh \mu) \notag\\	
											& \times	(\cos{m\varphi}+\beta\sin{m\varphi})
													 e^{in\eta}, 
											\	\mu \ge \mu_{1},\\
\label{point_charge_out_1}
\Phi_{2} & =   f(\mu,\eta) \  \sum_{m=0}^1 \sum_{n=-\infty}^\infty
									\bigl[\bigl(	\sum_{N=-\infty}^\infty d{_{m,n,N}}
							- K_{mn}\bigr)\notag\\
							& \times	P_{n-\frac12}^m(\cosh \mu)
								\dfrac{Q_n^{1}}{P_n^{1}}
									+ K_{mn}
										Q_{n-\frac12}^m(\cosh \mu)\bigr] \notag\\
										& \times (\cos{m\varphi}+\beta\sin{m\varphi})
													 e^{in\eta} ,
														\	0 \le \mu \le \mu_{1},
\end{align}
where $d{_{m,n,N}}$ is as defined in Eq.~\eqref{dnN} with
\bl{
\begin{align}\label{xyz_value}
K_{mn} & = \frac{\sqrt{2}a}{\pi}(\delta_{m0}+\delta_{m1})\mathcal{E}_{m+1} (-i)^{m-1} 
				(1+\delta_{m0}) \notag\\
				&\times (1+m\delta_{n0}) 
					\bigg(n^{m+1} - \frac14 m\bigg)
								\frac{\Gamma(n - m + \frac12)}{\Gamma(n + m+ \frac12)}.
\end{align}
The derivation of the applied potential 
\begin{align*}
\Phi_{ap}&=f(\mu,\eta) \sum_{m=0}^1 \sum_{n=-\infty}^\infty K_{mn}
				 Q_{n-\frac12}^m(\cosh \mu) e^{in\eta}  \\
				 	& \times (\cos{m\varphi}+\beta\sin{m\varphi}),
\end{align*}
in Eq.~\eqref{point_charge_out_1} is discussed in SM~\ref{B}.
}
The induced surface charge density on the ring in the limit of dipolar contribution ($m=0,1$) is therefore
\begin{align}
\sigma&=\sum_{m=0}^1 \sum_{n=-\infty}^\infty f(\mu_1,\eta) \sinh\mu_1(\cos{m\varphi}+\beta\sin{m\varphi})
						 e^{in\eta}\notag\\
&\times \bigl(\sum_{N=-\infty}^\infty d{_{m,n,N}}
							- K_{mn}\bigr)\bigl\{[Q+2Q'f^2(\mu_1,\eta)]\notag\\
&- 	\dfrac{Q_n^{1}}{P_n^{1}}[P+2P'f^2(\mu_1,\eta)]\bigr\}.
\end{align}
 which is connected to the volume charge density via
\begin{equation}
\rho=\sigma \frac{\delta(\mu-\mu_1)}{h_{\mu}},
\end{equation}
yielding  the induced dipole moment $\vec{p}$ 
\begin{equation}
\begin{cases}
p_x=\int\int\int h_\mu h_\eta h_\phi d_\mu d_\eta d_\phi \rho x,\\
p_z=\int\int\int h_\mu h_\eta h_\phi d_\mu d_\eta d_\phi \rho z,
\end{cases}
\end{equation}
where $x$, $z$ are Eqs.~\eqref{x_coeff}, \eqref{yz_coeff} in SM~\ref{B} and $h_\mu$, $h_\eta$, $h_\varphi$ are presented in Eq.~\eqref{scale_fac} in SM~\ref{C}.
The induced dipole gives rise to a scattered field such that the total scattering cross section can be calculated.  
Using this result, the simulated spectrum are shown in Fig.~\ref{cross} for a vacuum bounded  gold ring. For comparison,  the computationally obtained cross sections  for both types of janus nanorings and a metal nanosphere are presented.  
\section{Conclusion}
In conclusion, the introduction of a composite system of ring toroids to model multilayered metal, dielectric, metal-dielectric, and janus nanorings successfully produced the sought surface modes that may be exhibited by such a system. 
In particular,  the acquisition of the plasmon dispersion relations either analytically, up to a three-term difference equation including its associated continued fraction and infinite tridiagonal determinants, or numerically, visualized in the form of specific solution branches, was shown to produce the resonance spectra of the multilayered ring system. 
Specifically, using the simulated resonance values corresponding to the dominant (dipole) modes of a simple ring isolated in vacuum, reasonable agreement with spectral peak positions from the computationally determined absorption cross sections of the ring was demonstrated using FDTD. The agreement was demonstrated for several materials (gold, silver, aluminum and silicon) using experimental values of their dielectric functions. Furthermore, the analysis of the initial conditions for the obtained three-term difference equation and the consideration of its convergence properties, clarified the ambiguity of the reported surface modes in the literature. 
Therefore, exact quasi-static surface plasmon dispersion relations were calculated for a composite ring and can be simulated following Eq.~\eqref{Disp1} and Eq.~\eqref{Disp2} presented here \bl{for various configurations of current interest in plasmonics, nanophotonics, and metamaterial research. The agreement with the computationally obtained retarded dispersion relations in the case of a single ring, further implies that the presented results do offer working estimates of the resonance spectra of solid nanorings.}
Additionally, for multiple cases simulating the analytical results from a first order perturbation theory and visualizing and comparing the leading dispersion branches demonstrated the viability and accuracy of obtaining the exact dispersion relations via the presented results. \bl{Additionally, the perturbation solutions were instrumental for calculating the substrate induced plasmon shift, as well as for analytical observation of the energy splitting due to hybridization between the induced bonding and anti-bonding screening charges at the two interfaces of a toroidal shell.}
The presented formulation and modeling of the multilayered rings  also proved useful in obtaining the Green's function for rings with an arbitrary number of layers. Therefore, field distribution and scattering properties of composite nanorings may be studied when considering interaction with  arbitrarily polarized photons, which was shown to amount to modifying a source term appearing in the three-term difference equation. The correctness of the obtained Green's function was demonstrated for a single ring interacting with a uniform field with polarization parallel to the ring symmetry axis and perpendicular to it, as well as for a nonuniform field, resulting for example from \bl{electrons and dipoles. }
Therefore, following the results presented in Sec.~\ref{LDOS}, one may obtain the LDOS for applications requiring the determination of the spontaneous decay rate of an emitter, such as  a two level or four level quantum system in presence of multilayer nanorings and toroidal nanoshells. \bl{We demonstrated this viability by determining the partial LDOS and the corresponding spontaneous decay rate for the simpler system of a transition dipole within the nearfield of a gold ring.}
In addition to providing the resonance spectra, plasmon dispersion relations and fields, the presented results are of importance for use in validating computational models, in particular since retarded quantities are \bl{analytically} unavailable due to the inseparability of the wave equation in the toroidal geometry. Finally, the introduction of the two types of ring based janus particles may lead to new opportunities in nanofabrication, which could trigger novel applications in sensing, \bl{ quantum applications~\cite{salhi}, and metamaterials}.
In the presented view, in the case of $z$-type janus rings, for any non-negative integer $m$, one will have to establish a new orthogonality relation of the following kind
\begin{align*}
\int_{0}^{\infty} Q_{n-\frac12}^m(\cosh\mu)Q_{n'-\frac12}^m(\cosh\mu)&\sqrt{\cosh\mu-1}d\mu\\
&=C_{mn}\delta{nn'},
\end{align*}
for all non-negative integers $n$ and $n'$, where $C_{mn}$ only depends on $m$ and $n$. This orthogonality, which to the best of our knowledge is being reported for the first time, may need to hold in case of $\varphi$-type janus ring but now with fixed $n$, varying $m$ and weight function $\sqrt{\cosh\mu-\cos\eta}$ considered. The proof of this orthogonality will be the subject of a forthcoming article.
\section{Acknowledgements} 
This work was supported by the laboratory directed research and development (LDRD) fund at Oak Ridge National Laboratory (ORNL). ORNL is managed by UT-Battelle, LLC, for the US DOE under contract DE-AC05-00OR22725.


%
\clearpage
\setcounter{section}{0}
\setcounter{equation}{0}
\setcounter{figure}{0}
\section*{Supplemental Materials (SM)}

%
%

\section{}
\label{A}
Formulas in SM~\ref{A} hold for $n=0, \pm 1, \pm 2,\cdots$.
\begin{equation}
\small
\mathcal{C}_n = 
	\begin{bmatrix}\label{C_k}
 	C_n^2\\[0.3em]
	C_n^3\\[0.3em]
 	\vdots\\[0.3em]
	C_n^{k+1}\\[0.3em]
	\end{bmatrix}
	\hspace{2mm}
	\text{ and }
	\hspace{2mm}
\mathcal{D}_n = 
	\begin{bmatrix}
	 D_n^1\\[0.3em]
 	D_n^2\\[0.3em]
 	\vdots\\[0.3em]
	D_n^k\\[0.3em]
	\end{bmatrix},
\end{equation}
\begin{equation}\label{P_k}
\small
\mathbb{P}_n=
	\begin{bmatrix}
	-P_n^1 & & &0 \\[0.3em]
	P_n^2 & -P_n^2 &  &\\[0.3em]
	&  \ddots & \ddots & \\[0.3em]
	0&	&	P_n^k	&  -P_n^k 	\\[0.3em]
	\end{bmatrix},
\end{equation}
\begin{equation}\label{Q_k}
\small
\mathbb{Q}_n=
	\begin{bmatrix}
	-Q_n^1 & Q_n^1 & &0 \\[0.3em]
	& \ddots & \ddots &\\[0.3em]
	& &-Q_n^{k-1}&  Q_n^{k-1}\\[0.3em]
	0&	&	&  -Q_n^k	\\[0.3em]
	\end{bmatrix},
\end{equation}
\begin{equation}\label{P_k_1}
\small
\mathbb{P}'_n=
	\begin{bmatrix}
	-P_n^{'1} & & &0 \\[0.3em]
	P_n^{'2} & -P_n^{'2} &  &\\[0.3em]
	 & \ddots& \ddots   &  \\[0.3em]
	0 &	&P_n^{'k}& -P_n^{'k} \\[0.3em]
	\end{bmatrix},
\end{equation}
\begin{equation}\label{Q_k_1}
\small
\mathbb{Q}'_n=
	\begin{bmatrix}
	-Q_n^{'1} & Q_n^{'1} & &0 \\[0.3em]
	& \ddots & \ddots &\\[0.3em]
	& &-Q_n^{'k-1}&  Q_n^{'k-1}\\[0.3em]
	0&	&	&  -Q_n^{'k}	\\[0.3em]
	\end{bmatrix},
\end{equation}
\begin{equation}\label{D1}
\small
\mathrm{D}_\mu=
	\begin{bmatrix}
	2\ch \mu_1 &  & &0 \\[0.3em]
	& 2\ch \mu_2   & &\\[0.3em]
	&   & \ddots& \\[0.3em]
	0	&  & & 2\ch \mu_k	\\[0.3em]
	\end{bmatrix},
\end{equation}
\begin{equation}\label{E1_1}
\small
\mathrm{E}_1=
	\begin{bmatrix}
	\eps_1&  & &0 \\[0.3em]
	& \eps_2   & &\\[0.3em]
	&   & \ddots& \\[0.3em]
	0	&  & & \eps_k	\\[0.3em]
	\end{bmatrix}
\hspace{2mm}
\text{and}
\hspace{2mm}
\mathrm{E}_2=
	\begin{bmatrix}
	\eps_2&  & &0 \\[0.3em]
	& \eps_3   & &\\[0.3em]
	&   & \ddots& \\[0.3em]
	0	&  & & \eps_{k+1}	\\[0.3em]
	\end{bmatrix}.
\end{equation}
\section{}
\label{B0}
\begin{figure}[htp]
\centering
     \includegraphics[width=3.40in]{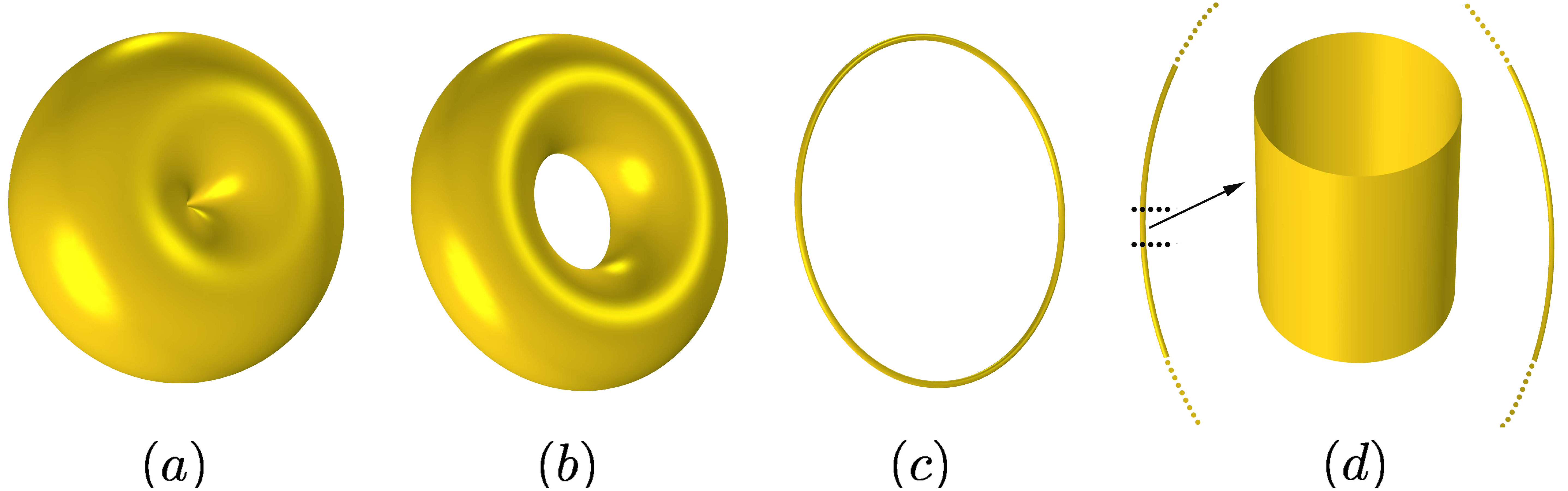} 
   		 \caption[]{\footnotesize Various limiting cases of a ring. (a) A small aspect ratio, $\cosh\mu_1=1.000001$, approximates a sphere where both $r$ and $R$ are infinite. (b) A normal ring is generated with a moderate aspect ratio $\cosh\mu_1=2$. (c) A large aspect ratio, $\cosh\mu_1=100$ generates a ring wire of radius $a$, \emph{i.e.}, $R\rightarrow a$ and $r\rightarrow 0$ as $\cosh\mu_1 \rightarrow \infty$. (d) The cylindrical limit is obtained similar to (c) but the minor radius, say $r=\SI{1}{\micro\meter}$, remains fixed, while the major radius $R$ and $a$ increase indefinitely as $\cosh\mu_1 \rightarrow \infty$. This simulates the case where the minor radius $r$ is much smaller than the major radius $R$ of a ring and hence a segment of the ring will not sense the  curvature and may therefore locally behave as a straight cylinder.}
		 \label{limit_ring}
\end{figure}

It follows from  Eq.~\eqref{eqn_ref} and Eq.~\eqref{k_layer_field1} that the general solution is a linear combinations of the modes\begin{align} \label{k_layer_mod_sep}
\phi_{mn} = f(\mu,\eta)  L_{n-\frac{1}{2}}^m( \mu) e^{in\eta} e^{im\varphi},
\end{align}
where $L_{n-\frac{1}{2}}^m( \mu)$ denotes either of the associated Legendre functions $Q_{n-\frac12}^m(\cosh \mu)$ or $P_{n-\frac12}^m(\cosh \mu)$. Letting $^*$ stand for the complex conjugation, we have
\begin{align}\label{k_layer_integral_1}
&\iiint
					 \phi_{mn}\phi_{m'n'}^* d\mathbf{r} \notag\\
					 	&=\int_{\mu=\mu'}^{\mu''} L_{n-\frac{1}{2}}^m( \mu) L_{n'-\frac{1}{2}}^{m'*}( \mu) d\mu \notag\\
					 		& \times \int_{\eta=0}^{2\pi} (\cosh\mu-\cos\eta) e^{i(n-n')\eta} d\eta 
					 		 \int_{\varphi=0}^{2\pi} e^{i(m-m')\varphi} d\varphi \notag\\
					 			&= I(m,n,n') \int_{\eta=0}^{2\pi} e^{i(n-n')\eta} d\eta \notag\\
					 				&-\dfrac{J(m,n,n')}{2}\int_{\eta=0}^{2\pi} \big(e^{i(n+1)\eta}+e^{i(n-1)\eta}\big)e^{-in'\eta} d\eta,
\end{align}
where
\begin{align*}
I(m,n,n') &= \int_{\mu=\mu'}^{\mu''} L_{n-\frac{1}{2}}^m( \mu) L_{n'-\frac{1}{2}}^{m*}( \mu) \cosh \mu \, d\mu,\\
J(m,n,n') &= \int_{\mu=\mu'}^{\mu''} L_{n-\frac{1}{2}}^m( \mu) L_{n'-\frac{1}{2}}^{m*}( \mu) \, d\mu.\\
\end{align*}
Using the orthogonality of $\{ e^{in\eta}\}$ in Eq.~\eqref{k_layer_integral_1}
\begin{align}\label{k_layer_integral_2}
&\iiint   \phi_{mn}\phi_{m'n'}^* d\mathbf{r} \notag\\
						&=I(m,n,n) -  \tfrac12 J(m,n,n+1) -  \tfrac12 J(m,n,n-1).
\end{align}
For the sake of the discussions, we note that for $\cosh\mu \rightarrow 1$, the behavior of the toroidal harmonics  resembles that of spherical harmonics as determined by using formulas $8.751-1$ and $8.751-2$ in Gradshteyn \emph{et al.}~\cite{Grad} On the other hand, when $\cosh\mu \rightarrow \infty$, the toroidal harmonics resemble cylindrical harmonics and are represented using modified Bessel functions of the first and second kind of order $n$ which are the solutions to Laplace's equation in cylindrical coordinates. Such cylindrical limit solution is discussed in Love~\cite{love:plasma}. Various limiting cases of a ring is shown in Fig.~\ref{limit_ring}.
\section{}
\label{B}
The potential for a uniform field polarized along $x-z$ axis in toroidal coordinates is given by:
\begin{equation}
\Phi_{ap} = - \mathcal{E}_1 z - \mathcal{E}_2 x.
\end{equation}
For a torus in the toroidal coordinate system $(\mu, \eta , \varphi),$ by Morse \& Feshbach~\cite{morse} on page 1304,
\begin{align}\label{expansion}	
\frac1{\sqrt{\ch \mu - \cos \eta}} & =
		\frac{\sqrt{2}}{\pi} \, \sum_{n=0}^\infty \Qn \cos n\eta \\ 
			& = \frac{\sqrt{2}}{2\pi} \sum_{n=-\infty}^\infty (1+\delta_{n0}) \Qn \notag\\\notag
			& \times e^{in\eta}.
\end{align}
Using the absolute convergence of the above series, differentiation w.r.t $\ch \mu$ yields
\begin{multline}\label{x_series}
 \frac{-1}{(\ch \mu - \cos \eta)^{\frac32}} \\ = 	
 \frac{\sqrt{2}}{\pi}
\sum_{n=-\infty}^\infty (1+\delta_{n0}) \frac{d}{d \ch \mu}\Qn  \, e^{in\eta}.
\end{multline}
Using 8.732, 8.734, and 8.736.4 from Gradshteyn \& Ryzhik~\cite{Grad}
\begin{equation*}
 \sqrt{z^2 - 1} \,  \frac{d}{dz} Q_{n-\frac12}(z) = 
	(n^2 - \frac14) \frac{\Gamma(n - \frac12)}{\Gamma(n + \frac32)}Q^1_{n-\frac12}(z).
\end{equation*}
Letting $z=\ch \mu$ in the last series Eq.~\eqref{x_series} and using the above relation together with the fact that 
\begin{equation}\label{x_coeff}
x= \dfrac{a \sh \mu}{\ch \mu - \cos \eta} \cos \varphi ,
\end{equation}
we obtain the desired expansion						
\begin{align}\label{x_label}
x&= 
	\frac{\sqrt{2}a}{\pi} \sqrt{\ch \mu - \cos \eta} 
		\sum_{n=-\infty}^\infty -(1+\delta_{n0}) (n^2 - \frac14) \notag\\ 
			&\times
				\frac{\Gamma(n - \frac12)}{\Gamma(n + \frac32)} Q^1_{n-\frac12}(\ch \mu) e^{in\eta} \cos \varphi.
\end{align}
From the  known expansion~\cite{love:jmp} for $z$,  we have
\begin{equation*}
 z=\frac{-2\sqrt{2}a i}{\pi}  \sqrt{\ch \mu - \cos \eta} \sum_{n=-\infty}^\infty n \Qn e^{in\eta}.
\end{equation*}
Thus,
\begin{align*}
 \Phi_{ap} &= - \mathcal{E}_1 z - \mathcal{E}_2 x \notag\\
 &= f(\mu,\eta)\sum_{n=-\infty}^\infty \sum_{m=0}^1 K_{mn} Q_{n-\frac12}^m(\cosh \mu)
						e^{in\eta}  \cos{m\varphi},
\end{align*}
where
$K_{mn}$ is as given in Eq.~\eqref{K3}.

Furthermore we have
\begin{equation}\label{yz_coeff}
y= \dfrac{a \sh \mu}{\ch \mu - \cos \eta} \sin \varphi, \hspace{5mm} z= \dfrac{a\sin\eta}{\ch \mu - \cos \eta}.
\end{equation}
Similar to Eq.~\eqref{x_label}, we now have
\begin{align}\label{y_label}
y&= 
	\frac{\sqrt{2}a}{\pi} \sqrt{\ch \mu - \cos \eta} 
		\sum_{n=-\infty}^\infty -(1+\delta_{n0}) (n^2 - \frac14) \notag\\ 
			&\times
				\frac{\Gamma(n - \frac12)}{\Gamma(n + \frac32)} Q^1_{n-\frac12}(\ch \mu) e^{in\eta} \sin \varphi.
\end{align}
The potential for a uniform field along $x-y-z$ axis in toroidal coordinates is given by:
\begin{align}
\Phi_{ap} &= - \mathcal{E}_1 z - \mathcal{E}_2 x-\mathcal{E}_3 y \notag\\\notag
&= f(\mu, \eta) \sum_{n=-\infty}^\infty \sum_{m=0}^1
K_{mn}Q_{n-\frac12}^m(\cosh \mu) e^{in\eta} \notag\\
& \times (\cos{m\varphi}+\beta\sin{m\varphi}),\label{xyz_field}
\end{align}
where
$K_{mn}$ is as given in Eq.~\eqref{xyz_value} with the assumption that $\mathcal{E}_3=\beta\mathcal{E}_2$.
Using the above obtained expression for the applied potential $\Phi_{ap}$, the superposed potential for a uniform field of arbitrary polarization can be written as
\begin{align}\label{xyz_superposed1}
\Phi_{1} & =  f(\mu,\eta)\sum_{m=0}^1 \sum_{n=-\infty}^\infty
							D_{mn}^1Q_{n-\frac12}^m(\cosh \mu) \notag\\
								& \times	(\cos{m\varphi}+\beta\sin{m\varphi})
									 e^{in\eta}, \	\mu \ge \mu_{1},\\
			\label{xyz_superposed2}
\Phi_{2} & =   f(\mu,\eta)\sum_{m=0}^1 \sum_{n=-\infty}^\infty
									\bigl[C_{mn}^2 P_{n-\frac12}^m(\cosh \mu)
										\notag \\
									& + K_{mn}
										Q_{n-\frac12}^m(\cosh \mu)
										   \bigr](\cos{m\varphi}+\beta\sin{m\varphi})
										   		e^{in\eta}, 										 
\end{align}
for $0 \le \mu \le \mu_{1}$.
\section{}
\label{C}
The scale factors for the toroidal coordinates are
\begin{align}\label{scale_fac}
h_{\mu}=h_{\eta}=\dfrac{a}{\cosh\mu-\cos\eta},\hspace{4mm} h_{\varphi}=\dfrac{a\sinh\mu}{\cosh\mu-\cos\eta}.
\end{align}
Using the chain rule on the partial derivative terms involved in Eq.~\eqref{bcc_janus} gives
\begin{align}\label{diff_yz}
\frac{\partial \Phi_{\text{in}}^i}{\partial z}&= \frac{\partial \Phi_{\text{in}}^i}{\partial \eta}
	\frac{\partial \eta}{\partial z} + 
		\frac{\partial \Phi_{\text{in}}^i}{\partial \mu}
			\frac{\partial \mu}{\partial z},  \hspace{2mm} i=1,2, \notag\\
		\frac{\partial \Phi_{\text{in}}^i}{\partial y}&= \frac{\partial \Phi_{\text{in}}^i}{\partial \mu}
			\frac{\partial \mu}{\partial y}+
				\frac{\partial \Phi_{\text{in}}^i}{\partial \eta}
					\frac{\partial \eta}{\partial y} + 
						\frac{\partial \Phi_{\text{in}}^i}{\partial \varphi}
							\frac{\partial \varphi}{\partial y}, \hspace{2mm}
								i=1,2. \notag\\
\end{align}
Furthermore, differentiating $x$, $y$, $z$ terms in Eqs.~\eqref{x_coeff}, \eqref{yz_coeff} gives
\begin{align}\label{diff_tor}
\frac{\partial x}{\partial \mu}&=\frac{a\cos\varphi(1-\cosh\mu\cos\eta)}{(\cosh\mu-\cos\eta)^2}, \notag\\
\frac{\partial y}{\partial \mu}&=\frac{a\sin\varphi(1-\cosh\mu\cos\eta)}{(\cosh\mu-\cos\eta)^2},  \notag\\
\frac{\partial z}{\partial \mu}&=\frac{-a\sin\eta\sinh\mu}{(\cosh\mu-\cos\eta)^2}, \notag\\
\frac{\partial x}{\partial \eta}&=\frac{-a\sinh\mu\cos\varphi\sin\eta}{(\cosh\mu-\cos\eta)^2},  \notag\\
\frac{\partial y}{\partial \eta}&=\frac{-a\sinh\mu\sin\varphi\sin\eta}{(\cosh\mu-\cos\eta)^2}, \notag\\
\frac{\partial z}{\partial \eta}&=\frac{a(\cosh\mu\cos\eta-1)}{(\cosh\mu-\cos\eta)^2}, \notag\\
\frac{\partial x}{\partial \varphi}&=\frac{-a\sinh\mu\sin\varphi}{(\cosh\mu-\cos\eta)},  \notag\\
\frac{\partial y}{\partial \varphi}&=\frac{a\sinh\mu\cos\varphi}{(\cosh\mu-\cos\eta)}, \hspace{2mm}
\frac{\partial z}{\partial \varphi}=0. \notag\\
\end{align} 
\section{}
\label{D}
Consider the system
\begin{equation}\label{infty}
\begin{bmatrix}
1 & 1 & \\[0.3em]
1 & 1 & 1 & \\[0.3em]
 & 1 & 1 & 1 & \\[0.3em]
 &  & \ddots & \ddots & \ddots & \\[0.3em]
\end{bmatrix}
\begin{bmatrix}
f_0 \\[0.4em] f_1 \\[0.4em] f_2 \\[0.4em] \vdots
\end{bmatrix}
=
\begin{bmatrix}
0 \\[0.4em] 0 \\[0.4em] 0 \\[0.4em] \vdots
\end{bmatrix}.
\end{equation}
The infinite determinant of the above matrix does not exist, but yet the system has non-trivial solutions: $f_0=1, f_1=-1, f_2=0, f_3=1, \cdots$.
\section{}
\label{E}
It is instructive to consider some general points regarding the character of the charge density in the absence of applied fields.
Here, we shall consider a solid ring without external sources,  and therefore the charge density (confined to the surfaces) should integrate to zero. Thus, the charge density $\rho$ must satisfy Eq.~\eqref{charge_zero}.
We will now discuss the potential satisfying the Poisson equation
\begin{align}\label{poisson}
\rho (\mu,\eta,\varphi) = - \epsilon_0 \nabla^2 \Phi(\mu,\eta,\varphi).
\end{align}
The discussion related to the Poisson equation where distributed charges of density $q$ may be present in a region $\mu > \mu_1$ can be found elsewhere~\cite{zotter}. 
In case of a solid ring, the electrostatic potential given in Eq.~\eqref{eqn_ref} becomes
\begin{align}\label{ring_ref}
\Phi(\mu,\eta,\varphi)  & =    \Theta (\mu - \mu_1)\Phi_1(\mu,\eta,\varphi) \notag\\
								& + \Theta (\mu_1 - \mu) \Phi_2(\mu,\eta,\varphi),
\end{align}
and considering Eq.~\eqref{k_layer_field1} in general terms, Eq.~\eqref{ring_ref} can be arranged as a linear combination of the product solutions
\begin{align}\label{ring_rearrange}
\Phi(\mu,\eta,\varphi)  & =   f(\mu,\eta) \notag\\
								&	\sum_{m,n} \mathcal{A}_{mn}U_{mn}(\mu)T_n(\eta)V_m(\varphi),
\end{align}
where $\mathcal{A}_{mn}$ is the corresponding amplitudes, $U_{mn}(\mu)$ is an associated Legendre function $Q_{n-\frac12}^m(\cosh\mu)$ or $P_{n-\frac12}^m(\cosh\mu)$, $T_n(\eta)$ is $\sin{n\eta}$ or $\cos{n\eta}$, and $V_m(\varphi)$ is $\sin{m\varphi}$ or $\cos{m\varphi}$. One makes the potential continuous across the boundary $\mu=\mu_1$ by writing
\begin{align}\label{ring_bound}
U_{mn}(\mu)&=\varepsilon_1 Q_{n-\frac12}^m(\cosh\mu)\Theta (\mu - \mu_1)\notag\\
				&+ \varepsilon_2 P_{n-\frac12}^m(\cosh\mu)\Theta (\mu_1 - \mu),
\end{align}
and 
\begin{align}\label{ring_delta}
\dfrac{d}{d\mu}\Theta(\mu)=\delta(\mu),
\end{align}
where $\delta(\mu)$ is the Dirac delta function.
The  Laplacian of $\Phi$ in toroidal coordinate system is
\begin{align}\label{laplace}
\nabla^2\Phi&=\dfrac{(\cosh\mu-\cos\eta)^3}{a^2\sinh\mu}
				\bigg[\dfrac{\partial}{\partial \mu}\bigg(\dfrac{\sinh\mu}{\cosh\mu-\cos\eta}\dfrac{\partial \Phi}{\partial \mu}\bigg)\notag\\
					&+\dfrac{\partial}{\partial \eta}\bigg(\dfrac{\sinh\mu}{\cosh\mu-\cos\eta}\dfrac{\partial \Phi}{\partial \eta}\bigg)\notag\\
						&+\dfrac{1}{\sinh\mu(\cosh\mu-\cos\eta)}\dfrac{\partial^2 \Phi}{\partial \varphi^2}\bigg].
\end{align}
We shall insert the basis in Eq.~\eqref{ring_rearrange} which is
\begin{align}\label{basis}
\Phi(\mu,\eta,\varphi)=f(\mu,\eta)U_{mn}(\mu)T_n(\eta)V_m(\varphi),
\end{align}
into Eq.~\eqref{laplace} and obtain~\cite{weis} the corresponding Laplace equation
\begin{align}\label{laplace_1}
\nabla^2\Phi&=\dfrac{1}{4}\sinh^2\mu+\cosh\mu\sinh\mu\dfrac{U'(\mu)}{U(\mu)}+\sinh^2\mu\dfrac{U''(\mu)}{U(\mu)}\notag\\
			&	+\sinh^2\mu\dfrac{T''(\eta)}{T(\eta)}+\dfrac{V''(\varphi)}{V(\varphi)}=0.
\end{align}
An alternative separation~\cite{andrew} of Laplace's equation in toroidal coordinates can be obtained by transforming Eq.~\eqref{laplace} into
\begin{align}\label{laplace_alt_1}
\nabla^2\Phi&=\dfrac{(\cosh\mu-\cos\eta)^3}{a^2\sinh\mu}
				\bigg[\dfrac{\partial}{\partial \mu}\bigg(r\dfrac{\partial \Phi}{\partial \mu}\bigg)
					+\dfrac{\partial}{\partial \eta}\bigg(r\dfrac{\partial \Phi}{\partial \eta}\bigg)\notag\\
						&+\dfrac{1}{\sinh^2\mu}r\dfrac{\partial^2 \Phi}{\partial \varphi^2}\bigg],
\end{align}
where $r$ is the distance from the $z$-axis given by 
\begin{align}
r=\sqrt{x^2+y^2}=\dfrac{a\sinh\mu}{\cosh\mu-\cos\eta},
\end{align}
and $x$ and $y$ are as given in Eqs.~\eqref{x_coeff} and \eqref{yz_coeff} in SM~\ref{B}. 
Further substituting~\cite{bye} $U=\Phi\sqrt{r}$ into Eq.~\eqref{laplace_alt_1}, one would obtain the Laplace equation 
\begin{align}\label{laplace_alt_2}
\nabla^2\Phi&=\dfrac{(\cosh\mu-\cos\eta)^2}{a^2\sqrt{r}}
				\bigg[\dfrac{\partial^2 U}{\partial \mu^2}+\dfrac{\partial^2 U}{\partial \eta^2} 
						+\dfrac{1}{\sinh^2\mu} \notag\\
						& \times	\bigg(\dfrac{\partial^2 U}{\partial \varphi^2}+\dfrac{1}{4}U\bigg)\bigg]=0,
\end{align}
The solutions of the above Laplace equation form a complete basis~\cite{andrew} and are given by   
\begin{align}\label{basis_alt}
\Phi(\mu,\eta,\varphi)=\sqrt{a/r}U_{pq}(\mu)T_q(\eta)V_p(\varphi),
\end{align}
where $U_{pq}(\mu)$ is $Q_{p-\frac12}^q(\coth\mu)$ or $P_{p-\frac12}^q(\coth\mu)$, $T_q(\eta)$ is $\sin{q\eta}$ or $\cos{q\eta}$, and $V_p(\varphi)$ is $\sin{p\varphi}$ or $\cos{p\varphi}$. This basis is more appropriate for the situations where the boundary conditions do not depend on $\eta$. In such cases  $q=0$ is considered and the solution involves the Legendre functions  $Q_{p-\frac12}(\coth\mu)$ or $P_{p-\frac12}(\coth\mu)$. The basis in Eq.~\eqref{basis} is particularly convenient for axially symmetric situations, in which case we set $m=0$ and the solution involves the Legendre functions  $Q_{n-\frac12}(\cosh\mu)$ or $P_{n-\frac12}(\cosh\mu)$. However the equivalence of the two approaches to separation of the variables is discussed in SM B in Andrews~\cite{andrew}. 

The charge density $\rho (\mu,\eta,\varphi)$ corresponding to the potential of Eq.~\eqref{ring_rearrange} is given by Eq.~\eqref{charge_density}
and $W_{mn}(\mu_1)$ for each fixed $m$ is
\begin{align}
W_{mn}(\mu_1)&=\varepsilon_1D_n^1\bigg[f_1Q_n^{'1}+\dfrac{Q_n^1}{2f_1}\bigg]\sinh\mu_1\notag\\
				&	-\varepsilon_2C_n^2\bigg[f_1P_n^{'1}+\dfrac{P_n^1}{2f_1}\bigg]\sinh\mu_1.
\end{align}
Substituting $\rho$ given by Eq.~\eqref{charge_density} into Eq.~\eqref{charge_zero} and evaluating the integral with respect to $\mu$, one obtains
\begin{align}\label{int_1}
\int_{\varphi=0}^{2\pi} \int_{\eta=0}^{2\pi} \dfrac{\sinh\mu_1}{\cosh\mu_1-\cos\eta} \rho_1(\mu_1,\eta,\varphi) d\eta d\varphi,
\end{align}
where
\begin{align*}
\rho_1(\mu_1,\eta,\varphi)=\epsilon_0\sum_{m,n}W_{mn}(\mu_1)T_n(\eta)V_m(\varphi).
\end{align*}
Evaluating the integral in Eq.~\eqref{int_1} with respect to $\varphi$, one obtains
\begin{align}\label{int_2}
\int_{\eta=0}^{2\pi} \dfrac{\sinh\mu_1}{\cosh\mu_1-\cos\eta} \rho_1(\mu_1,\eta) d\eta,
\end{align}
where
\begin{align*}
\rho_1(\mu_1,\eta)=2\pi\epsilon_0\sum_{n}W_{0n}(\mu_1)T_n(\eta).
\end{align*}
That is, to infer that a solid ring without external sources is axially symmetric. Simplifying integral expression in Eq.~\eqref{int_2}, we obtain
\begin{align}\label{int_simplify}
2\pi\epsilon_0\sinh^2{\mu_1}&\sum_{n}\varepsilon_1D_n^1\bigg[Q'_{n-\frac12}(\cosh\mu_1)\int_{\eta=0}^{2\pi}\dfrac{1}{f_1}\notag\\
					&	+Q_{n-\frac12}(\cosh\mu_1)\int_{\eta=0}^{2\pi}\dfrac{1}{2f_1^3}\bigg]\notag\\
					&      -\varepsilon_2C_n^2\bigg[P'_{n-\frac12}(\cosh\mu_1)\int_{\eta=0}^{2\pi}\dfrac{1}{f_1}\notag\\
					&	+P_{n-\frac12}(\cosh\mu_1)\int_{\eta=0}^{2\pi}\dfrac{1}{2f_1^3}\bigg]T_n(\eta) d\eta.
\end{align}
One can now substitute Eq.~\eqref{expansion} in SM~\ref{B} into Eq.~\eqref{int_simplify} and together with the orthogonality property of $T_n(\eta)$ on $0\le\eta\le 2\pi$, the following sufficient condition  
\begin{align}\label{int_cond}
\varepsilon_1 D_n^1Q_{n-\frac12}(\cosh\mu_1)= \varepsilon_2 C_n^2P_{n-\frac12}(\cosh\mu_1), 
\end{align}
needs to hold in order to satisfy Eq.~\eqref{charge_zero} for $n=0,1,\cdots$.
As a final remark, we note that Eq.~\eqref{int_cond} further signifies the fact that for a given aspect ratio $\cosh\mu_1$, there are no roots for the dispersion relation Eqs.~\eqref{Disp1} and \eqref{Disp2} when $m=0$ is considered. This can be observed from the fact that the left hand side and the right hand side of the equation obtained by applying boundary condition Eq.~\eqref{k_bc2} for $k=1$ cancel each other out with Eq.~\eqref{int_cond} considered at $m=0$ mode. Notice that the superscripts in Eq.~\eqref{int_cond} indicate the $k$ layered index, in this case, $k=1$ and not to be confused with $m$ modes.
\section{}
\label{E1}
\begin{figure}[H]
  \centering
    \includegraphics[width=8cm]{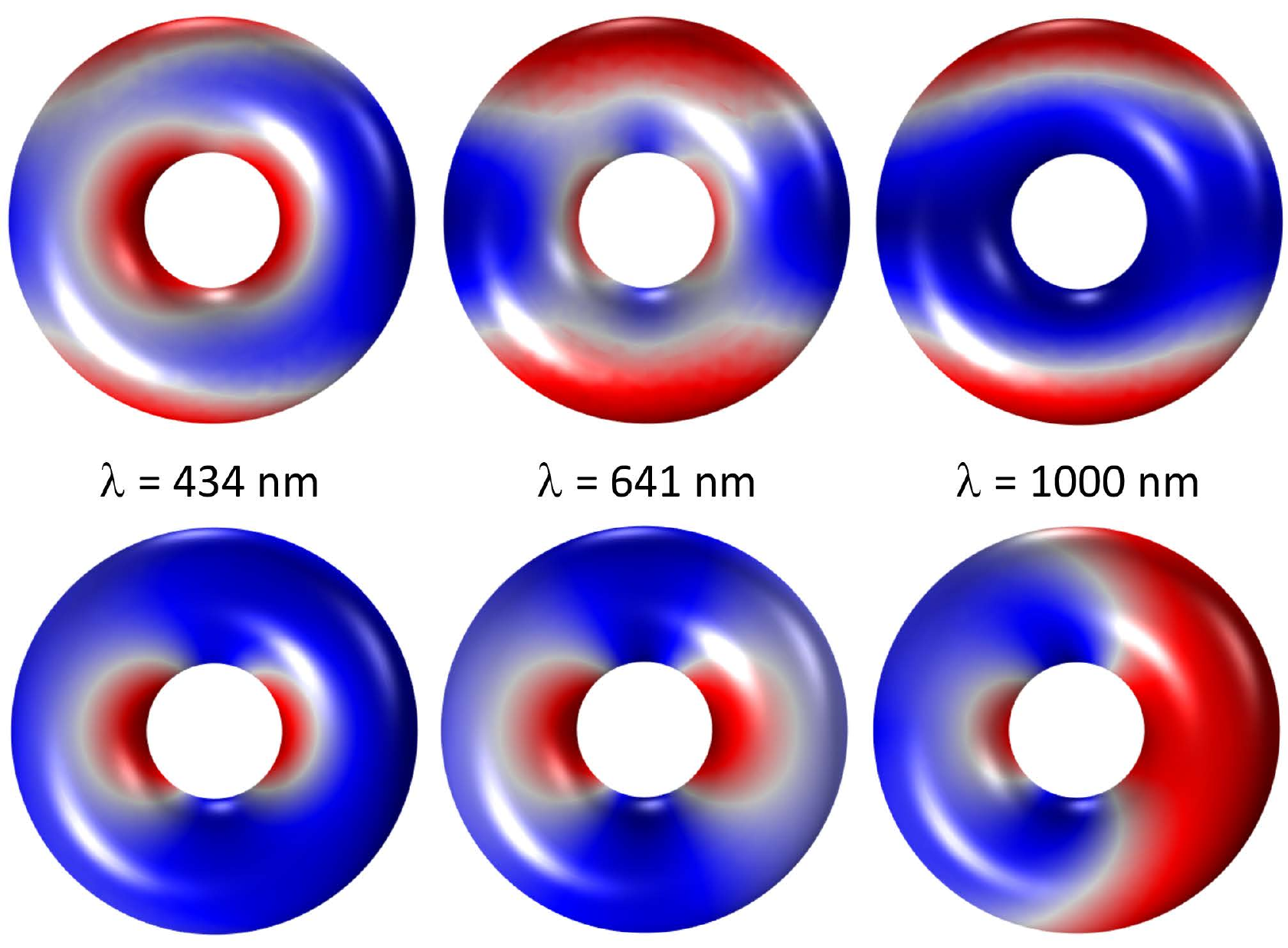} 
      \caption{Retarded response of a gold nanoring to linearly polarized photons of wavelength $\lambda$. The norm of the engendered electric field (top), and the induced current density (bottom) are visualized.}
  \label{ring-eigenmodes}
\end{figure}

 Employing the FEM method, in Fig.~\ref{ring-eigenmodes} we present the  computationally determined transient response of a gold nanoring to linearly polarized incident photons.
\section{}
\label{F}
In case of a toroidal shell we assume $\varepsilon_v=1$ and $\varepsilon_c=\varepsilon$ in Eq.~\eqref{pert_det}.
The determinant in Eq.~\eqref{pert_det} generates a second order polynomial in $\varepsilon$:
\begin{align}\label{pert_poly}
&\varepsilon^2\left(\dfrac{Dh_{u_2}Df_{u_1}}{f_{u_2}}-\dfrac{Df_{u_2}Dh_{u_1}}{h_{u_2}}\right) \notag\\
& +\varepsilon\bigg(\dfrac{Dh_{u_1}Dh_{u_2}}{h_{u_2}}-\dfrac{2Df_{u_1}Dh_{u_2}}{f_{u_2}}
				+\dfrac{Df_{u_1}Df_{u_2}h_{u_1}}{f_{u_1}h_{u_2}}\bigg)  \notag\\
					& +\dfrac{Df_{u_1}Dh_{u_2}}{f_{u_2}}
							-\dfrac{Df_{u_1}Dh_{u_2}h_{u_1}}{f_{u_1}h_{u_2}}=0, \notag\\
\end{align}
with
\begin{align}\label{pert_coeff}
\mathcal{A}&= \dfrac{\dfrac{Dh_{u_1}Dh_{u_2}}{h_{u_2}}-\dfrac{2Df_{u_1}Dh_{u_2}}{f_{u_2}}
				+\dfrac{Df_{u_1}Df_{u_2}h_{u_1}}{f_{u_1}h_{u_2}}}
						{\dfrac{Dh_{u_2}Df_{u_1}}{f_{u_2}}-\dfrac{Df_{u_2}Dh_{u_1}}{h_{u_2}}}, \notag\\
\mathcal{B}&=\dfrac{\dfrac{Df_{u_1}Dh_{u_2}}{f_{u_2}}
							-\dfrac{Df_{u_1}Dh_{u_2}h_{u_1}}{f_{u_1}h_{u_2}}}
								{\dfrac{Dh_{u_2}Df_{u_1}}{f_{u_2}}-\dfrac{Df_{u_2}Dh_{u_1}}{h_{u_2}}}. \notag\\			
\end{align}
\section{}
\label{H}
\begin{figure}[H]
\begin{center}
     \includegraphics[width=3.4in]{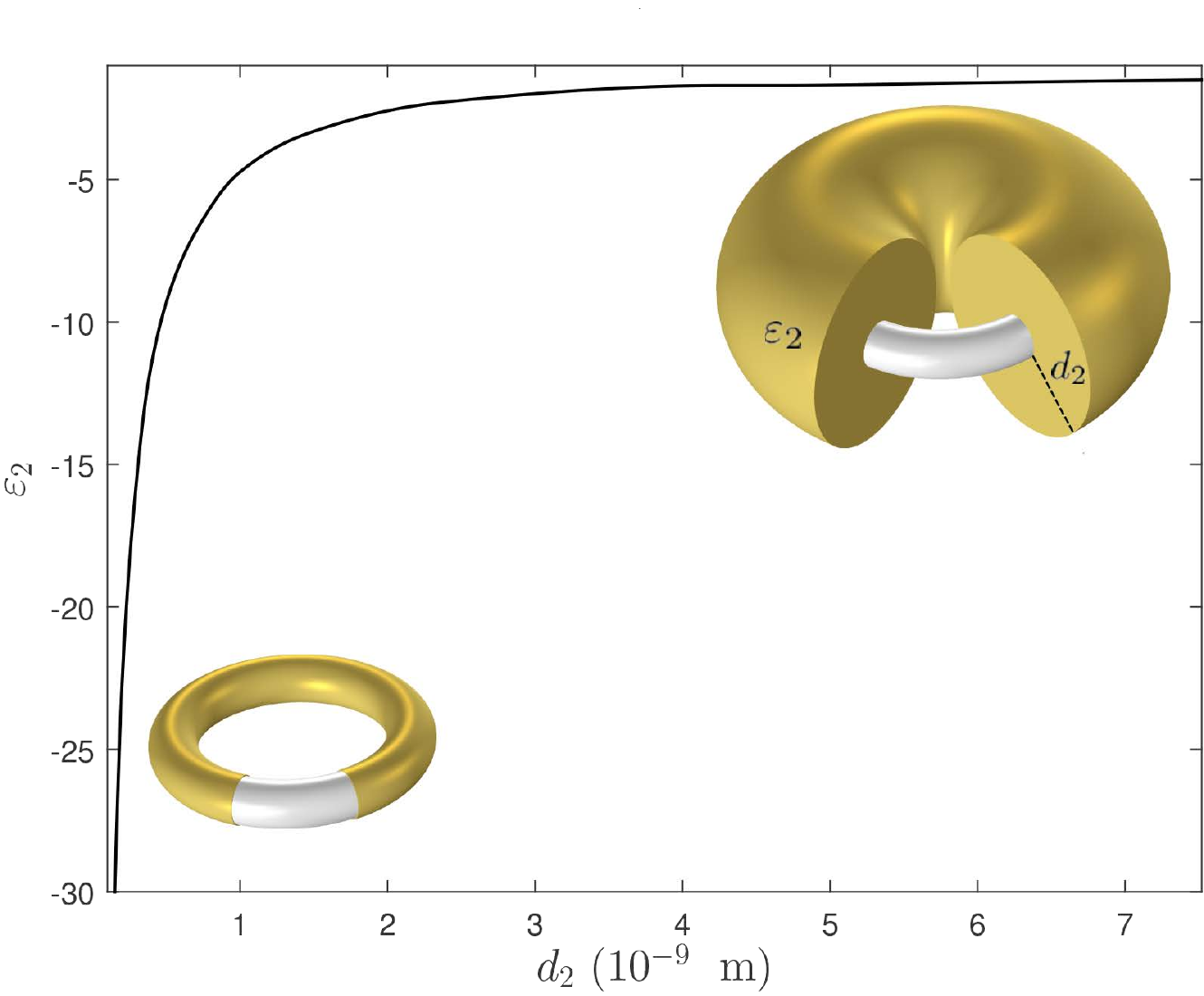} 
     
     \vspace{1cm}
     
     \includegraphics[width=3.4in]{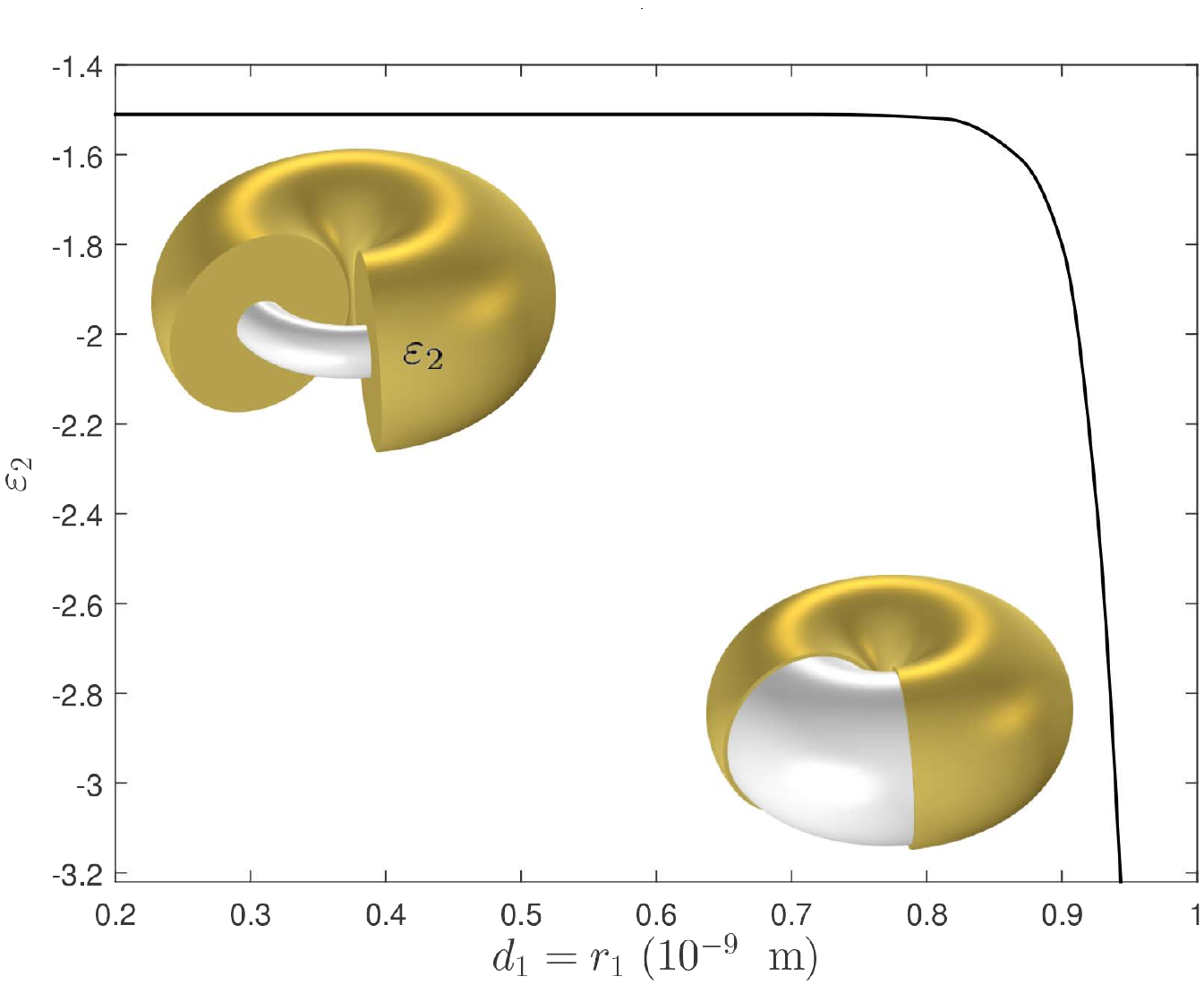} 
   		 \caption[]{\footnotesize (a) Single shell in the limit $r_2 \to R$: A single shell with fixed $R=\SI{10}{\nano\meter}$, $\cosh\mu_1=5$, $r_1=d_1=R/\cosh\mu_1=\SI{2}{\nano\meter}$, is considered. The outer aspect ratio is varied between $1.05 \le\cosh\mu_2 \le 4.99$. The $x-$axis represents $d_2=r_2-r_1$. 
(b) Single shell as $r_1 \to 0$: A single shell with fixed $R=\SI{1}{\nano\meter}$, $\cosh\mu_2=1.05$, $r_2=R/\cosh\mu_2=\SI{0.95}{\nano\meter}$, is considered. The inner aspect ratio is varied between $1.06 \le\cosh\mu_1 \le 5$. The $x-$axis represents $d_1=r_1$. The curves shown are plotted for $m=4$, $n=0$, $\eps_1=\eps_3=1$ using the first continued fraction Eq.~\eqref{Disp1} with $k=2$ and $\eps_2$ as its solution.}
     \label{SingleShell1}
\end{center}
\end{figure}

We briefly discuss plasmon dispersion in metallic shells with dielectric core. Instructive modes include the resonance values $\eps_2$ of a single shell (in  vacuum)  as a function of the outer aspect ratio, assuming  $\eps_1=1$ as shown in Fig.~\ref{SingleShell1}(a), where  the inner aspect ratio is fixed at $\cosh\mu_1=5$.
The $x$-axis represents the shell thickness $d_2$ which is $d_2=r_2-r_1$ while the $y$-axis denote the roots of the dispersion relation Eq.~\eqref{Disp1} in the form of a dielectric function $\eps_2$ for a toroidal surface mode of $m=4$ and $n=0$. Similarly, in Fig.~\ref{SingleShell1}(b) the outer aspect ratio is fixed at $\cosh\mu_2=1.05$ while the inner varied. 
The $x$-axis represents the inner ring radius $d_1=r_1$. 
\begin{figure}[H]
\begin{center}
     \includegraphics[width=3.35in]{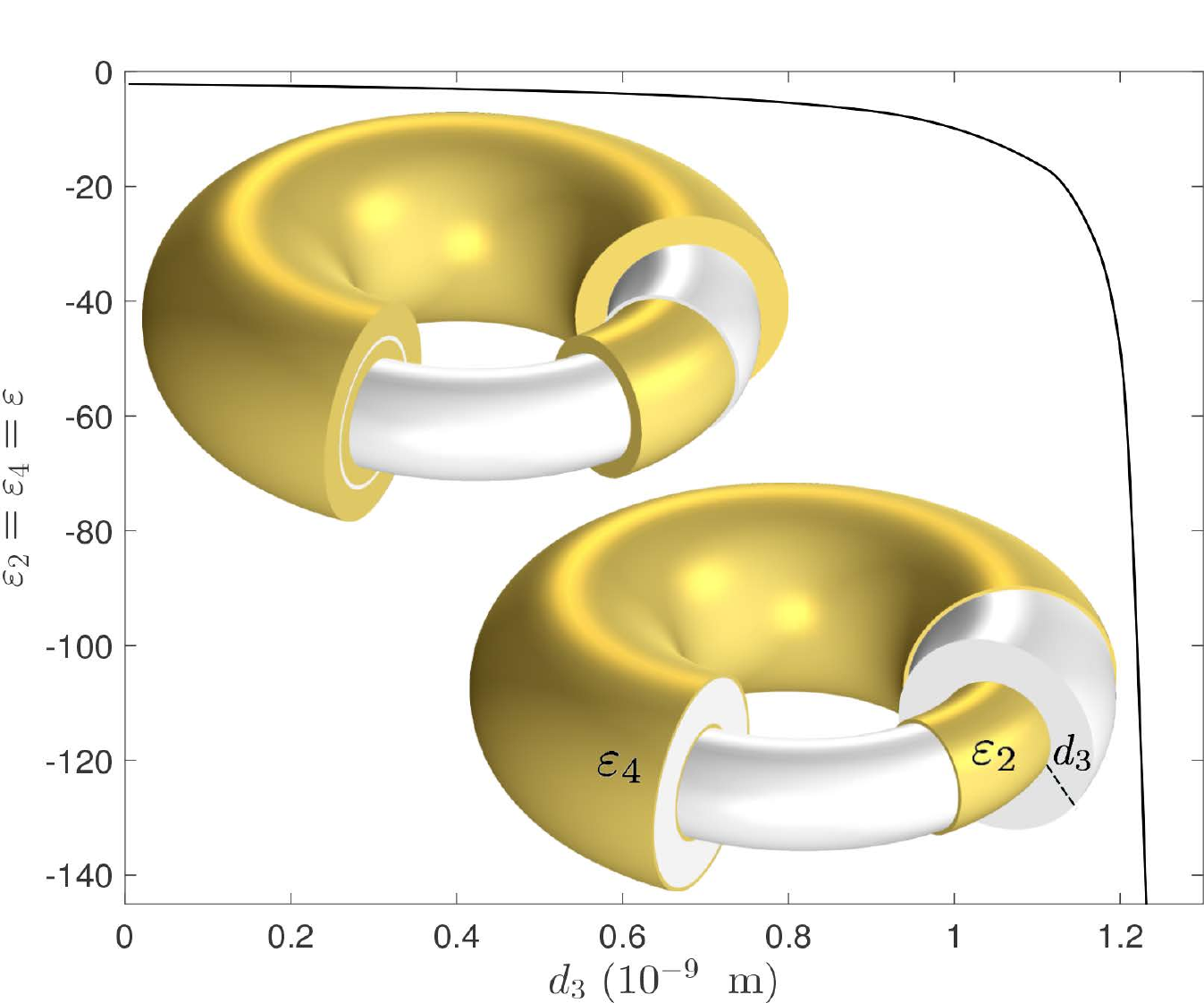} 
     
      \vspace{1cm}
     
     \includegraphics[width=3.4in]{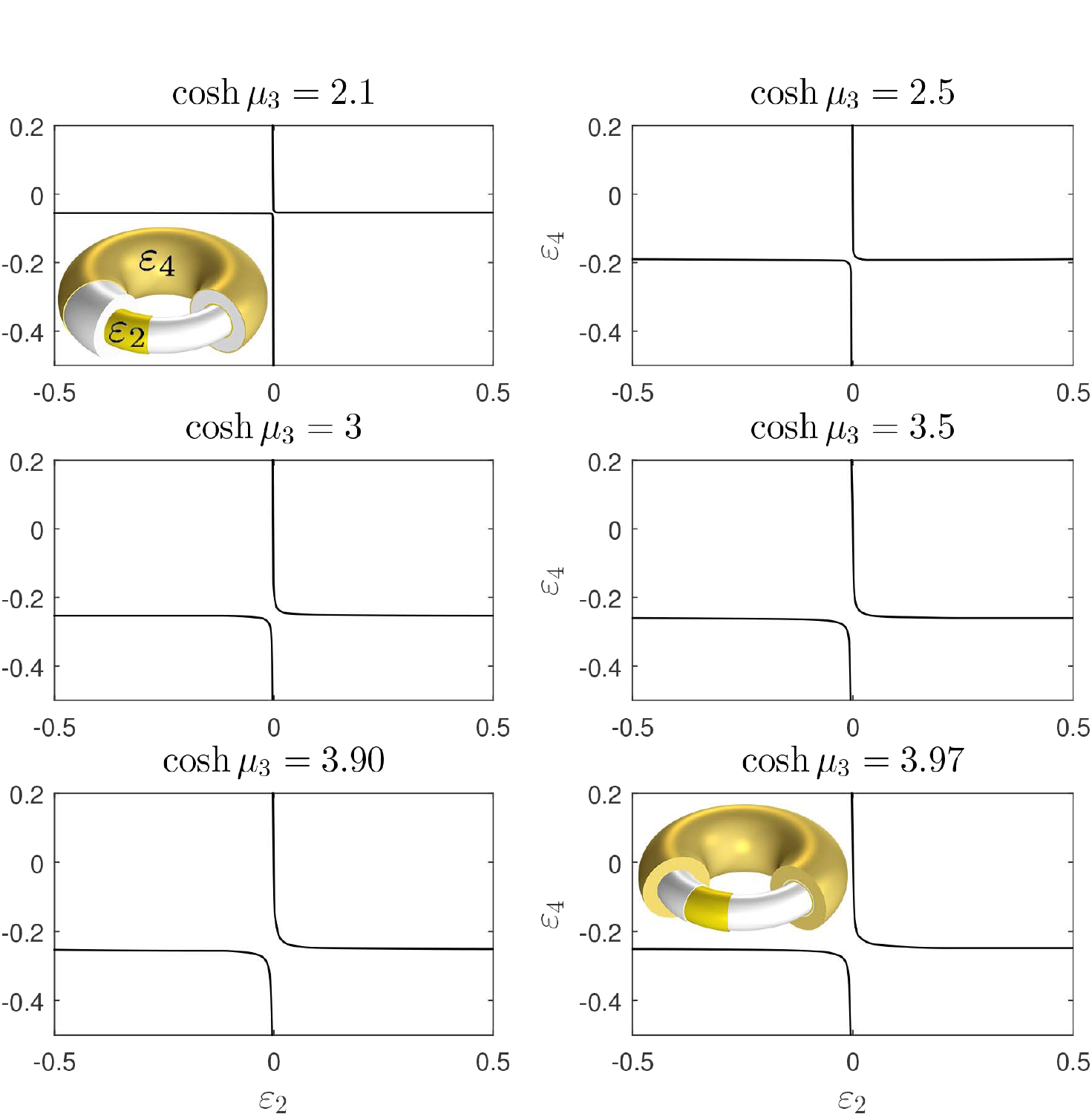} 
   		 \caption[]{\footnotesize (a) Tuning the resonant response of a metal double shell. A four layered torus with similar $\eps_2$, $\eps_4$ shells, fixed $R=\SI{5}{\nano\meter}$, $\cosh\mu_1=3.99$, $\cosh\mu_4=2$, $\eps_1=\eps_3=\eps_5=1$, and $r_1=d_1=R/\cosh\mu_1=\SI{1.253}{\nano\meter}$ with  inner aspect ratios $3 \le\cosh\mu_2\le 3.98$, and $2.01 \le\cosh\mu_3\le 2.99$ yielding $d_3=r_3-r_2=R\bigg(\dfrac{1}{\cosh\mu_3}-\dfrac{1}{\cosh\mu_2}\bigg)$. 
The eigenvalues are obtained from Eq.~\eqref{Disp1} for  $m=4$, $n=0$, $k=4$, and same values of $\eps_2$, $\eps_4$ as its solution.
(b) A four layered torus shell including positive $\eps$ for fixed aspect ratios  $\cosh\mu_1=4$, $\cosh\mu_2=3.98$, $\cosh\mu_4=2$, and  $\eps_1=\eps_3=\eps_5=1$. The eigenvalues are obtained from Eq.~\eqref{Disp1} with $k=4$, $m=4$, and $n=0$.}
   		 \label{fourlayer2}
\end{center}
\end{figure}
We allow the inner radius $r_1$ to approach 0 in order to establish its relation with $\eps_2$ values satisfying the dispersion relation Eq.~\eqref{Disp1} for a toroidal surface mode of $m=4$ and $n=0$ with $k=2$.
As noticed in Fig.~\ref{SingleShell1}(b), smaller ranges of $\eps_2$ values with negative magnitudes tend to satisfy the dispersion relation Eq.~\eqref{Disp1} when the distance $d_1$ approaches towards zero. But when $d_1$ approaches $R=\SI{1}{\nano\meter}$ as shown in the plot, the $\eps_2$ values satisfying the dispersion relation Eq.~\eqref{Disp1} tend to gradually increase. 
This phenomenon is opposite to the one described in Fig.~\ref{SingleShell1}(a) wherein the $\eps_2$ values gradually increase for smaller shell size thickness $d_2$.

Using the dispersion relation Eq.~\eqref{Disp1} with $k=4$ for the toroidal mode $m=4$ and $n=0$, other effects such as  the dependence on the distance or layer thickness, as shown in Fig.~\ref{fourlayer2}(a), or obtaining positive values satisfying Eq.~\eqref{Disp1}, as shown in Fig.~\ref{fourlayer2}(b) may be studied.
\section{}
\label{G}
\begin{figure}[H]
  \centering
  \begin{tabular}{c}
  \hspace{0.05mm} 
  \includegraphics[width =3.315in]{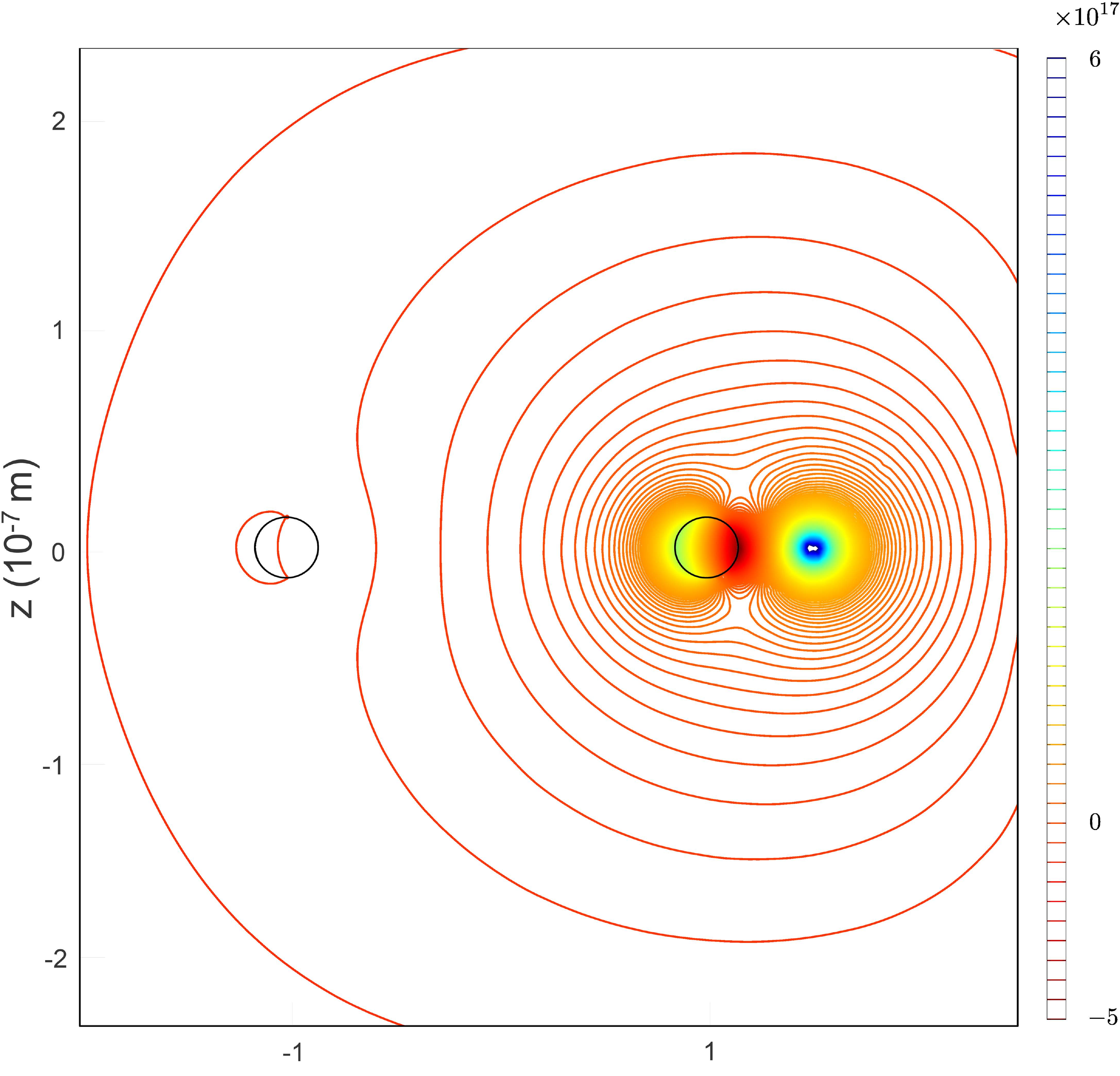} \\\\
     \includegraphics[width =3.325in]{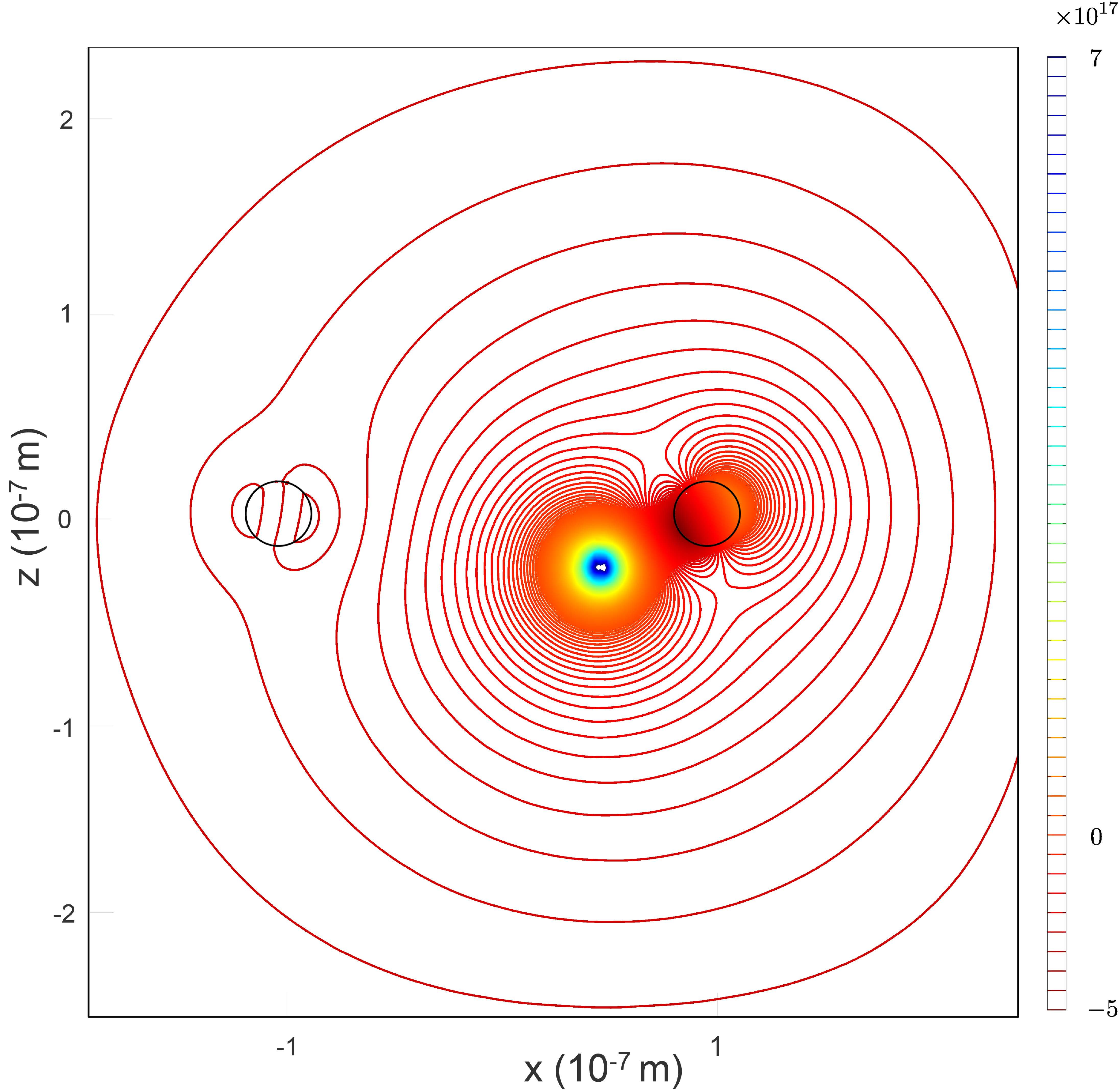}
  \end{tabular}
  \caption[]{\footnotesize Potential distribution representing the response of a metal nanoring to a nonuniform field of an electron located at $\cosh{\mu_0}=2.8$, $\eta_0=0$ (top) and $-150$ (bottom), $\varphi_0=0$ via Eqs.~\eqref{point_charge_in} and \eqref{point_charge_out}
  by summation of poloidal and toroidal modes for a solid ring with $a=\SI{0.0980}{\micro\meter}$,  $\cosh{\mu_1}=6.6667$, $\eps_1=-1.5$, and $\eps_2=1$.}
  \label{point_charge}
\end{figure}

We briefly discuss the procedure used in plotting the non-uniform and uniform fields. To plot the non-uniform field, one may convert Eq.~\eqref{point_charge_in} into real Fourier series. A real Fourier series function in general can be expressed as
\begin{equation}\label{fourier1}
f(x)=\frac{a_0}{2}+\sum_{n=1}^\infty a_n\cos{nx}+\sum_{n=1}^\infty b_n\sin{nx}.
\end{equation}
Expressing $\cos{nx}$ and $\sin{nx}$ in exponential form, we may rewrite Eq.~\eqref{fourier1} as
\begin{equation}\label{fourier2}
f(x)=\sum_{n=-\infty}^\infty c_ne^{inx},
\end{equation}
where $c_n=\frac{1}{2}(a_n-ib_n)$ and $c_{-n}=\frac{1}{2}(a_n+ib_n)$ for $n>0$ and $c_0=\frac{1}{2}a_0$. Hence $a_n$ and $b_n$ in Eq.~\eqref{fourier1} are given by $a_n=c_n+c_{-n}$ and $b_n=i(c_n-c_{-n})$.
We now convert Eq.~\eqref{point_charge_in} into real Fourier series with respect to $\eta$. Thus, considering $c_n$ to be
\begin{equation}
c_n=\sum_{m=0}^\infty\sum_{N=-\infty}^\infty d{_{m,n,N}}Q_{n-\frac12}^m(\cosh \mu)\cos{m(\varphi-\varphi_0)},
\end{equation}
together with the following relations
\begin{align}
d{_{m,n,N}}&=d{_{m,-n,-N}}e^{-2iN\eta_0},\notag\\
d{_{m,n,-N}}&=d{_{m,-n,N}}e^{2iN\eta_0},\notag\\
\end{align}
we obtain
\begin{align}
a_n&=c_n+c_{-n} \notag\\
		&=\sum_{N=0}^\infty\big(1+e^{2iN\eta_0}\big)\notag\\
		&\times\sum_{m=0}^\infty\big[d{_{m,n,N}}Q_{n-\frac12}^m(\cosh \mu)\cos{m(\varphi-\varphi_0)}\big] \notag\\
		& + \sum_{N=1}^\infty\big(1+e^{-2iN\eta_0}\big)\notag\\
		& \times\sum_{m=0}^\infty\big[d{_{m,n,-N}}Q_{n-\frac12}^m(\cosh \mu)\cos{m(\varphi-\varphi_0)}\big],
\end{align}
\begin{align}
b_n&=i(c_n-c_{-n}) \notag\\
		&=i\bigl\{\sum_{N=0}^\infty\big(1-e^{2iN\eta_0}\big)\notag\\
		&\times\sum_{m=0}^\infty\bigl[d{_{m,n,N}}Q_{n-\frac12}^m(\cosh \mu)\cos{m(\varphi-\varphi_0)}\bigr] \notag\\
		& + \sum_{N=1}^\infty\bigl(1-e^{-2iN\eta_0}\bigr)\notag\\
		&\times\sum_{m=0}^\infty\bigl[d{_{m,n,-N}}Q_{n-\frac12}^m(\cosh \mu)\cos{m(\varphi-\varphi_0)}\bigr]\bigr\}.
\end{align}
In the same fashion we can convert Eq.~\eqref{point_charge_out} into real Fourier series by considering $c_n$ to be
\begin{align}
c_n & = \sum_{m=0}^\infty 
					\bigl[\bigl(	\sum_{N=-\infty}^\infty d{_{m,n,N}}
							- K_{mn}\bigr)
								P_{n-\frac12}^m(\cosh \mu)
							  		\dfrac{Q_n^{1}}{P_n^{1}}\notag\\
								  &	+ K_{mn}
										Q_{n-\frac12}^m(\cosh \mu)\bigr] 
										  \cos{m(\varphi-\varphi_0)},
\end{align}
and using the fact that
\begin{equation*}
K_{mn}=K'_{mn}e^{-in\eta_0},
\end{equation*}
one can compute $a_n$, $b_n$ at the respective toroidal, poloidal modes (considering other modes to be irrelevant) and then use Eq.~\eqref{fourier1} to obtain the potential distribution $\Phi_{1}$, $\Phi_{2}$ of the poloidal and toroidal modes induced by a nearby electron. Similar procedure as above but using Eq.~\eqref{k_layer2} accordingly one can obtain the potential distribution at the respective toroidal, poloidal modes induced by a dipole.
Considering all other modes via summation of $m$ and $n$, results in the total potential distribution as shown in Fig.~\ref{point_charge}.
To plot the uniform field, We convert Eq.~\eqref{uniform_field_in} into real Fourier series with respect to $\eta$ by considering $c_n$ to be
\begin{equation}
c_n=\sum_{N=-\infty}^\infty d_{_{n,N}}Q_{n-\frac12}(\cosh \mu),
\end{equation}
together with the following relation
\begin{equation}
d_{_{-n,-N}}=-d_{_{n,N}},
\end{equation}
\begin{figure}[H]
\begin{center}
     \includegraphics[width=3.4in]{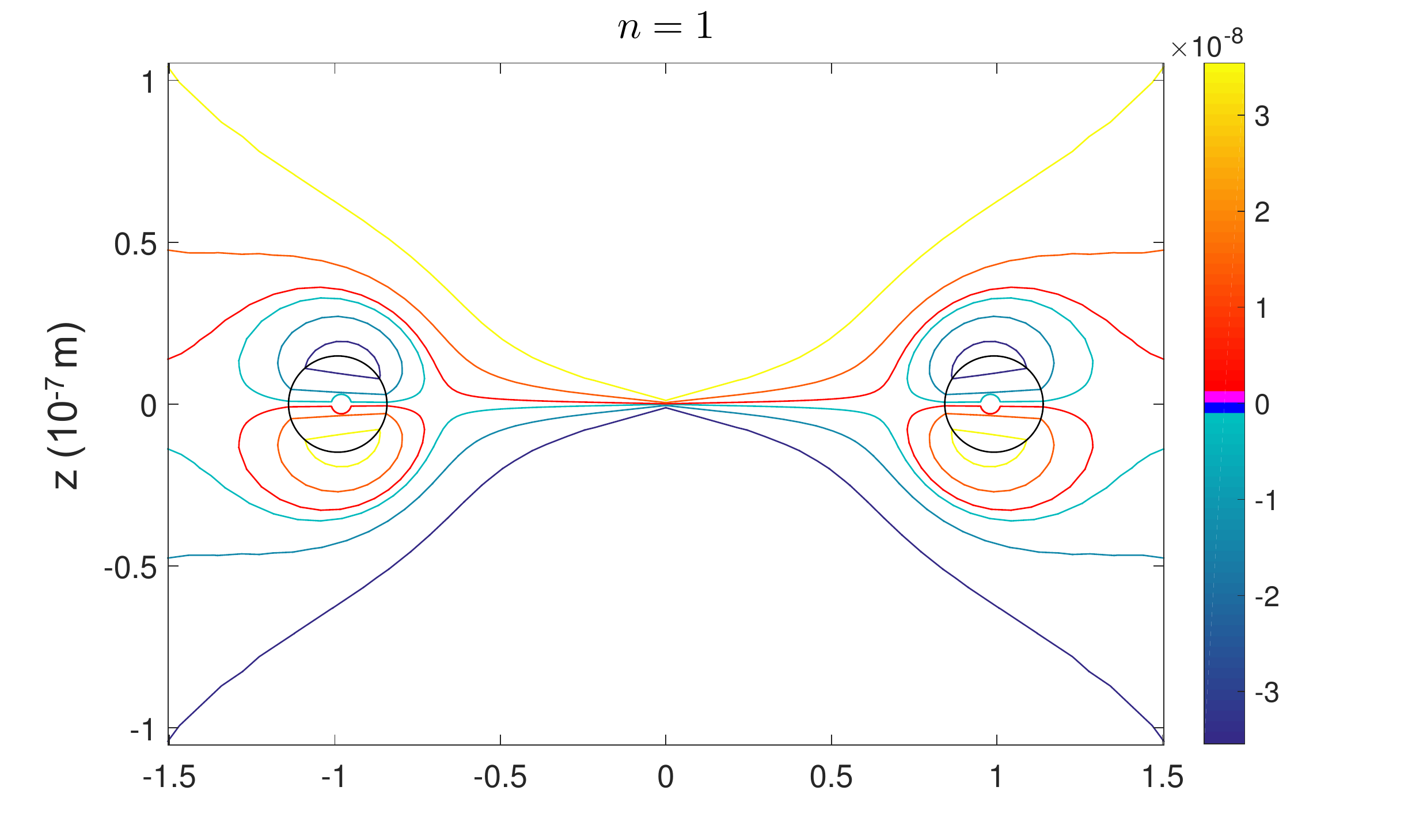}\\
     \includegraphics[width=3.4in]{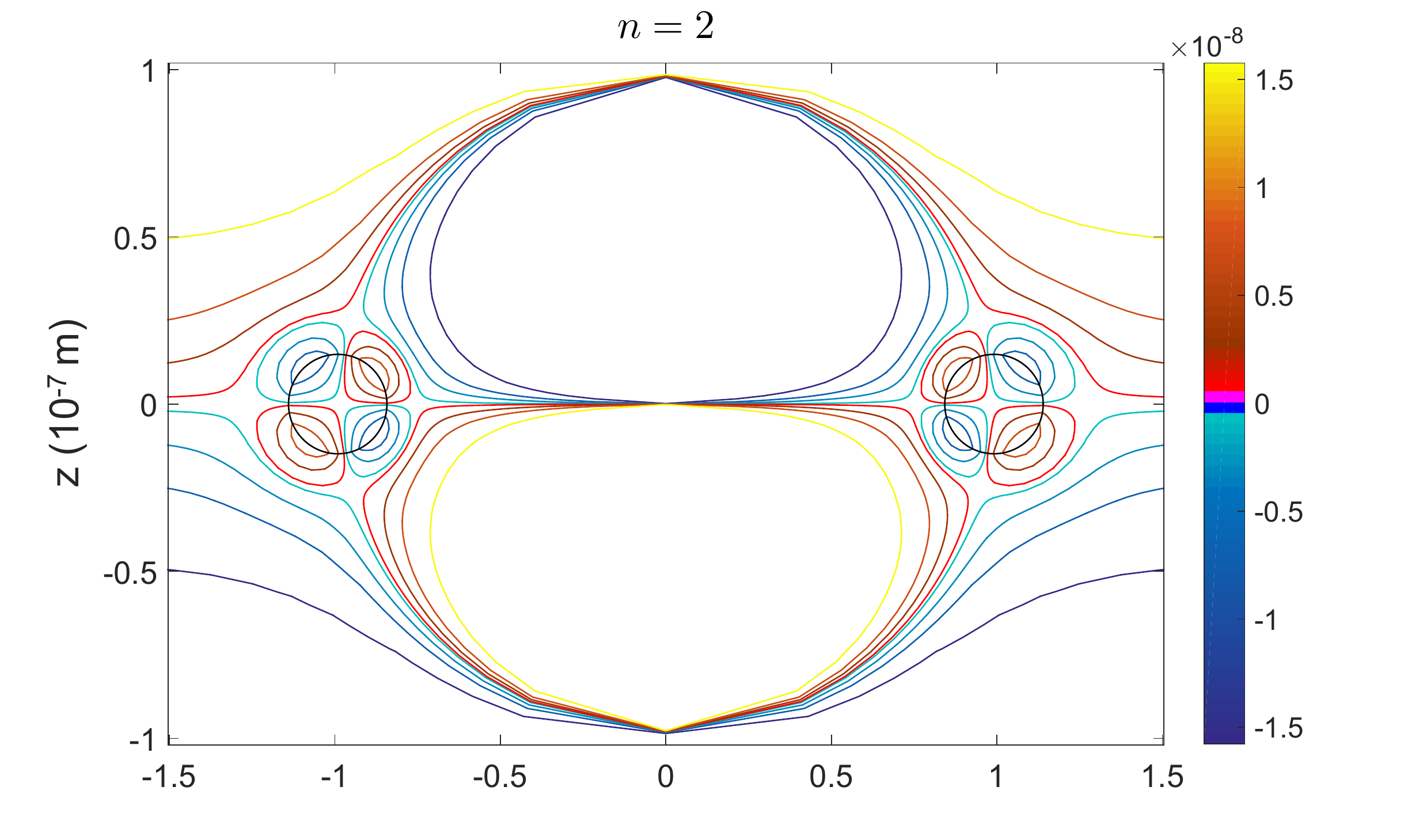}\\
      \includegraphics[width=3.4in]{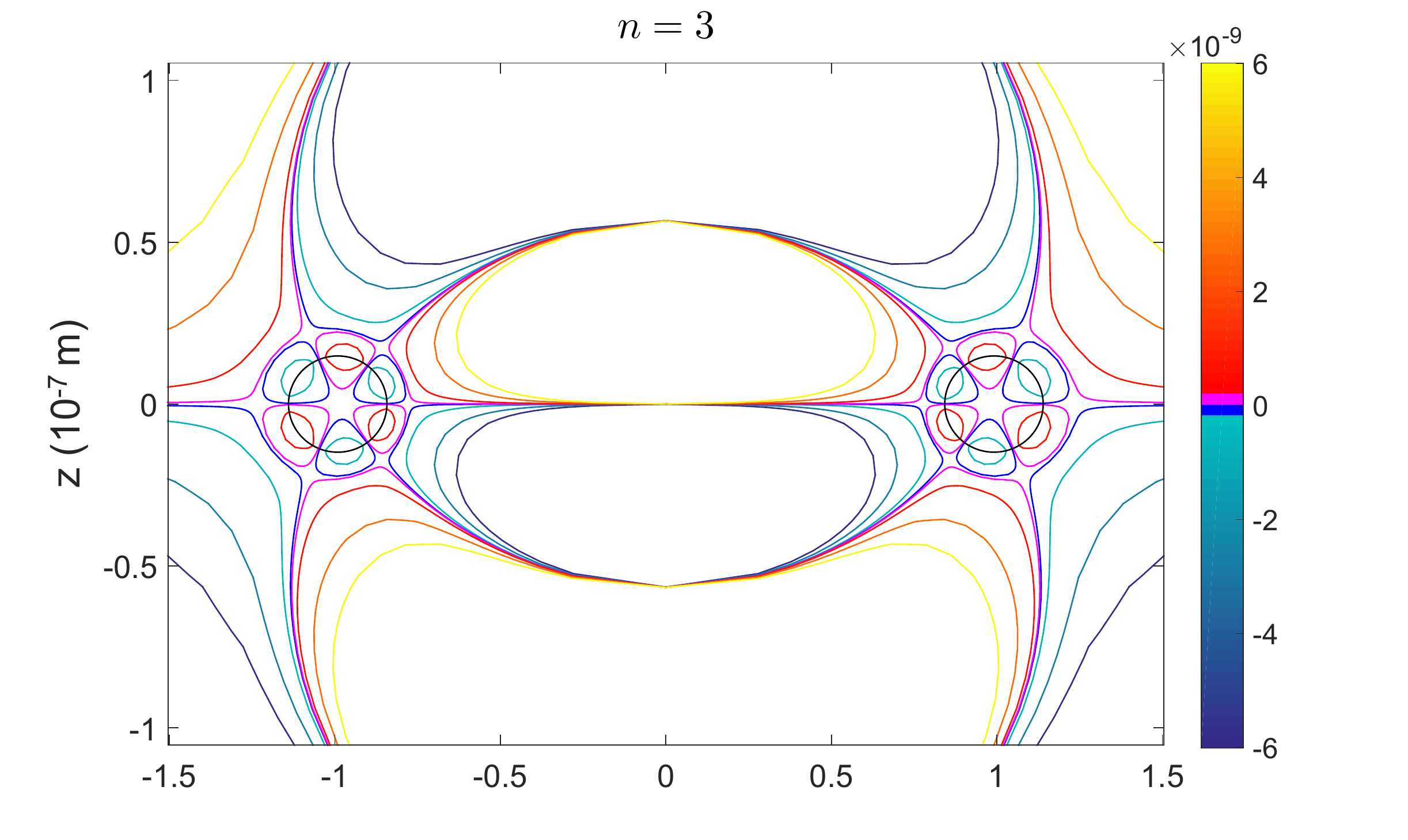}\\
       \includegraphics[width=3.4in]{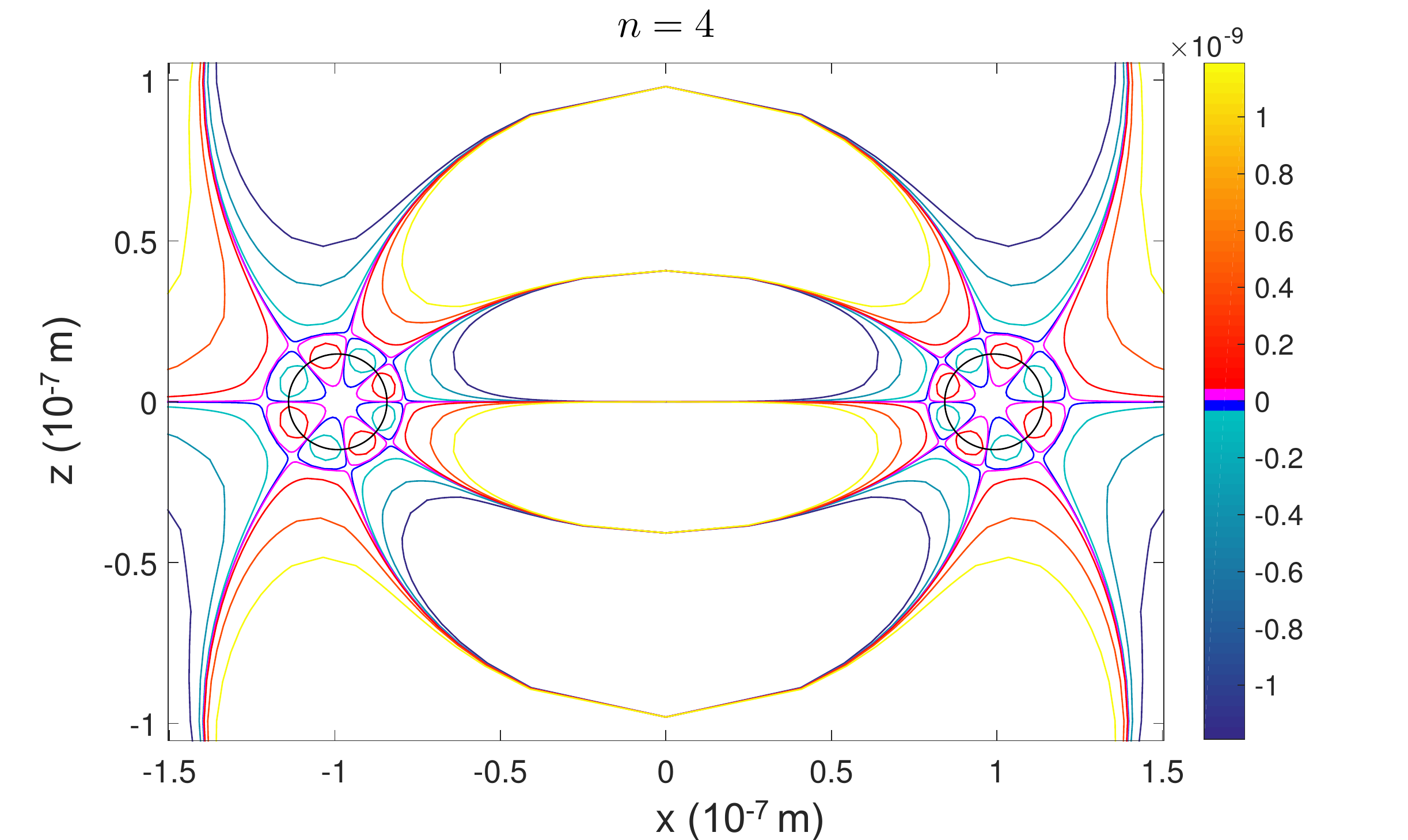}
     \caption[]{\footnotesize
		 	Poloidal eigenmodes of a vacuum bounded solid ring. Simulation was made for a ring with $a=\SI{0.0980}{\micro\meter}$,  $\cosh{\mu_1}=6.6667$, $\eps_1=-1.5$, and $\eps_2=1$.   The visualization was made for angles  $\varphi=0,\pi$ and $0 < \eta < 2\pi$.
The column displays the $m=0$ and $n=1,2,3,4$ modal potential distributions corresponding the response to a uniform field polarized parallel to the symmetry axis of the ring.}
     \label{uni_field}
   	\end{center}
\end{figure}
we obtain
\begin{align}
a_n&=c_n+c_{-n}=0,
\\
b_n&=i(c_n-c_{-n}) \notag\\
		&=i\big[2d_{_{n,0}}Q_{n-\frac12}(\cosh \mu)\notag\\
		&+\sum_{N=1}^\infty(2d_{_{n,N}}+2d_{_{n,-N}})Q_{n-\frac12}(\cosh \mu)\big].
\end{align}
In similar fashion we convert Eq.~\eqref{uniform_field_out} into real Fourier series by considering $c_n$ to be
\begin{align}
c_n & =   			\bigl(\sum_{N=-\infty}^\infty d_{_{n,N}}
							- K_{n}\bigr)
							   P_{n-\frac12}(\cosh \mu)
								\dfrac{Q_n^{1}}{P_n^{1}} \notag\\
								&	+ K_{n}
										Q_{n-\frac12}(\cosh \mu),
\end{align}
and using the fact that $K_{n}=-K_{-n}$,
one can use Eq.~\eqref{fourier1} to plot $\Phi_1$ and $\Phi_2$ at the respective toroidal and poloidal modes as shown in Fig.~\ref{uni_field}. The coloring scheme across the two cross-sections in case of uniform field in Fig.~\ref{uni_field} depends on the evenness/oddness of $n$-mode.  When odd, the color scheme differs at $\varphi=0$ or $\pi$ noticing the opposite loops on the ring cross-section and vice-versa. The total potential is then obtained by summation of $n$ modes as in Eqs.~\eqref{uniform_field_in}--\eqref{uniform_field_out} is shown in Fig.~\ref{uniform_comsol}.
\begin{figure}[H]
\begin{center}
     \includegraphics[width=3.4in]{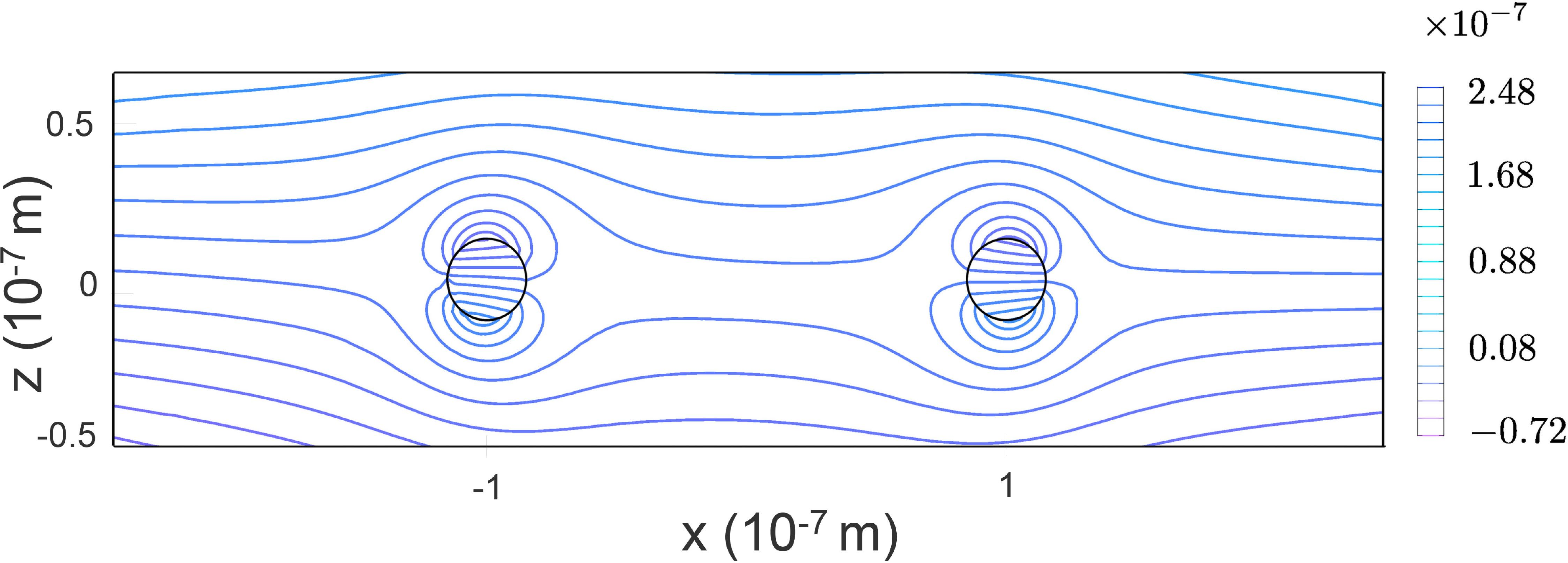} 
   		 \caption[]{\footnotesize 
		 	Potential distribution representing the response of a metal nanoring to a uniform field polarized parallel to the ring symmetry axis via Eqs.~\eqref{uniform_field_in} and \eqref{uniform_field_out}. The response is obtained from the summation of poloidal modes for a solid ring with $a=\SI{0.0980}{\micro\meter}$,  $\cosh{\mu_1}=6.6667$, $\eps_1=-1.5$, and $\eps_2=1$. 
		 	}
     \label{uniform_comsol}
     \end{center}
\end{figure}


\end{document}